\documentclass[sigconf]{acmart}

\usepackage{enumitem}
\usepackage{multirow}
\usepackage{makecell}
\usepackage{colortbl}
\usepackage{array}
\usepackage{arydshln}
\usepackage{subcaption}
\usepackage{booktabs}
\usepackage[most]{tcolorbox}

\AtBeginDocument{%
  }

\copyrightyear{2026}
\acmYear{2026}
\setcopyright{cc}
\setcctype{by}
\acmConference[ICMR '26]{International Conference on Multimedia Retrieval}{June 16--19, 2026}{Amsterdam, Netherlands}
\acmBooktitle{International Conference on Multimedia Retrieval (ICMR '26), June 16--19, 2026, Amsterdam, Netherlands}
\acmDOI{10.1145/3805622.3810742}
\acmISBN{979-8-4007-2617-0/2026/06}

\begin{document}

\title{ComMark: Covert and Robust Black-Box Model Watermarking with Compressed Samples}

\author{Yunfei Yang}
\affiliation{%
  \institution{Institute of Information Engineering, Chinese Academy of Sciences}
  \institution{State Key Laboratory of Cyberspace Security Defense}
  \institution{School of Cyber Security, University of Chinese Academy of Sciences}
  \city{Beijing}
  \country{China}}
\email{yangyunfei@iie.ac.cn}

\author{Xiaojun Chen}
\authornote{Corresponding author. This is an extended version of the paper accepted by the 16th ACM International Conference on Multimedia Retrieval (ICMR 2026).}
\affiliation{%
  \institution{Institute of Information Engineering, Chinese Academy of Sciences}
  \institution{State Key Laboratory of Cyberspace Security Defense}
  \institution{School of Cyber Security, University of Chinese Academy of Sciences}
  \city{Beijing}
  \country{China}}
\email{chenxiaojun@iie.ac.cn}

\author{Zhendong Zhao}
\affiliation{%
  \institution{Institute of Information Engineering, Chinese Academy of Sciences}
  \institution{State Key Laboratory of Cyberspace Security Defense}
  \institution{School of Cyber Security, University of Chinese Academy of Sciences}
  \city{Beijing}
  \country{China}}
\email{zhaozhendong@iie.ac.cn}

\author{Yu Zhou}
\affiliation{%
  \institution{College of Computer Science, Nankai University}
  \city{Tianjin}
  \country{China}}
\email{yzhou@nankai.edu.cn}

\author{Xiaoyan Gu}
\affiliation{%
  \institution{Institute of Information Engineering, Chinese Academy of Sciences}
  \institution{State Key Laboratory of Cyberspace Security Defense}
  \institution{School of Cyber Security, University of Chinese Academy of Sciences}
  \city{Beijing}
  \country{China}}
\email{guxiaoyan@iie.ac.cn}

\author{Juan Cao}
\affiliation{%
  \institution{Institute of Computing Technology, Chinese Academy of Sciences}
  \city{Beijing}
  \country{China}}
\email{caojuan@ict.ac.cn}


\begin{abstract}
The rapid advancement of deep learning has turned models into highly valuable assets due to their reliance on massive data and costly training processes. However, these models are increasingly vulnerable to leakage and theft, highlighting the critical need for robust intellectual property protection. Model watermarking has emerged as an effective solution, with black-box watermarking gaining significant attention for its practicality and flexibility. Nonetheless, existing black-box methods often fail to better balance covertness (hiding the watermark to prevent detection and forgery) and robustness (ensuring the watermark resists removal)—two essential properties for real-world copyright verification. In this paper, we propose ComMark, a novel black-box model watermarking framework that leverages frequency-domain transformations to generate compressed, covert, and attack-resistant watermark samples by filtering out high-frequency information. To further enhance watermark robustness, our method incorporates simulated attack scenarios and a similarity loss during training. Comprehensive evaluations across diverse datasets and architectures demonstrate that ComMark achieves state-of-the-art performance in both covertness and robustness. Furthermore, we extend its applicability beyond image recognition to tasks including speech recognition, sentiment analysis, image generation, image captioning, and video recognition, underscoring its versatility and broad applicability.
\end{abstract}

\ccsdesc[500]{Security and privacy~Social aspects of security and privacy}
\ccsdesc[500]{Information systems~Multimedia information systems}
\ccsdesc[500]{Computing methodologies~Artificial intelligence}

\keywords{Deep Learning, Copyright Protection, Model Watermarking}


\maketitle

\section{Introduction}
Deep neural networks \cite{sze2017efficient,samek2021explaining,alam2020survey} have been widely applied in fields such as face recognition \cite{laishram2025toward}, autonomous driving \cite{chen2024end}, and content generation \cite{cao2025survey}. Building state-of-the-art models demands significant investment in large-scale data collection and annotation, algorithm design, and computational resources. These models embody valuable intellectual property but are vulnerable to unauthorized reproduction, dissemination, and derivation \cite{tan2023deep}, highlighting the urgent need for effective protection.

Model watermarking \cite{boenisch2021systematic,wan2022comprehensive,nie2024deep,zhang2021deep,zhang2020model}, which embeds hidden information into model parameters or outputs, has emerged as a promising solution. Techniques are generally classified into white-box and black-box schemes. White-box methods \cite{wang2021riga,yan2023rethinking,lv2023robustness} require full access to model internals during verification, which is often impractical in real-world settings. In contrast, black-box watermarking \cite{zhu2024reliable,jia2021entangled,tan2023deep} only requires querying the model with crafted samples and inferring ownership from predictions, making it more practical and widely adopted.

Despite their potential, existing black-box methods struggle to balance covertness and robustness. Some \cite{zhang2018protecting,jia2021entangled,lv2024mea} offer strong robustness against removal attacks but lack visual stealth, making watermark samples easily detectable or forgeable. Others \cite{li2019prove,kim2023margin,xi2024invisible} achieve higher covertness, with samples nearly indistinguishable from clean data, but sacrifice robustness. Designing black-box watermarks that ensure both remains an open challenge.

\begin{figure}[t]
\centering
\includegraphics[width=0.48\textwidth]{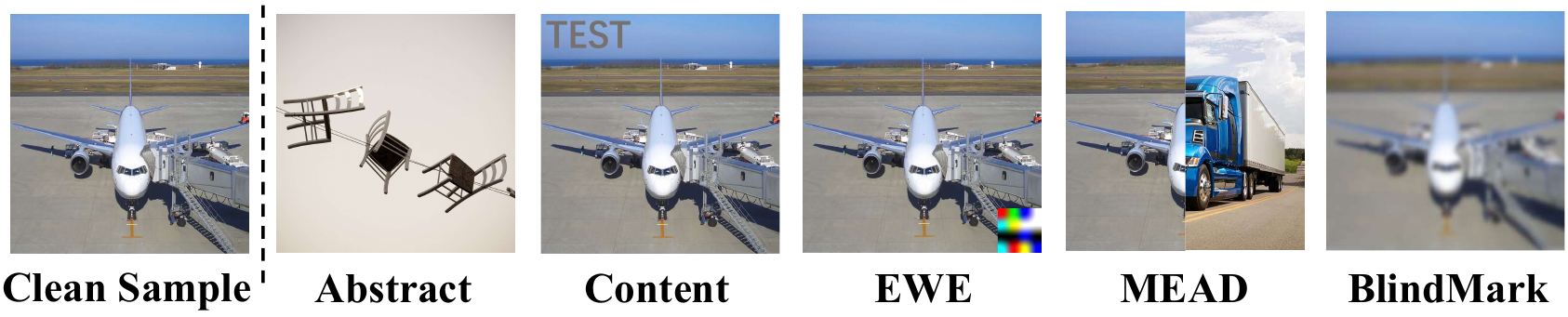}
\caption{Examples of watermark samples from existing black-box methods including Abstract \cite{adi2018turning}, Content \cite{zhang2018protecting}, EWE \cite{jia2021entangled}, MEAD \cite{lv2024mea} and BlindMark \cite{li2019prove}.}
\label{fig: examples of watermark samples}
\end{figure}

We identify a key limitation: most prior methods construct watermarks in the spatial domain at pixel level, often embedding features in only part of the sample (see Figure \ref{fig: examples of watermark samples}). These designs are both fragile and conspicuous, rendering them vulnerable to cropping, scaling \cite{wu2025robust}, JPEG compression \cite{nie2024fedcrmw}, model extraction \cite{pang2025modelshield}, and forgery \cite{li2023plmmark}, potentially leading to verification failures. To address this, we propose embedding covert watermark signals across the entire sample.

\begin{figure}[t]
\centering
\includegraphics[width=0.45\textwidth]{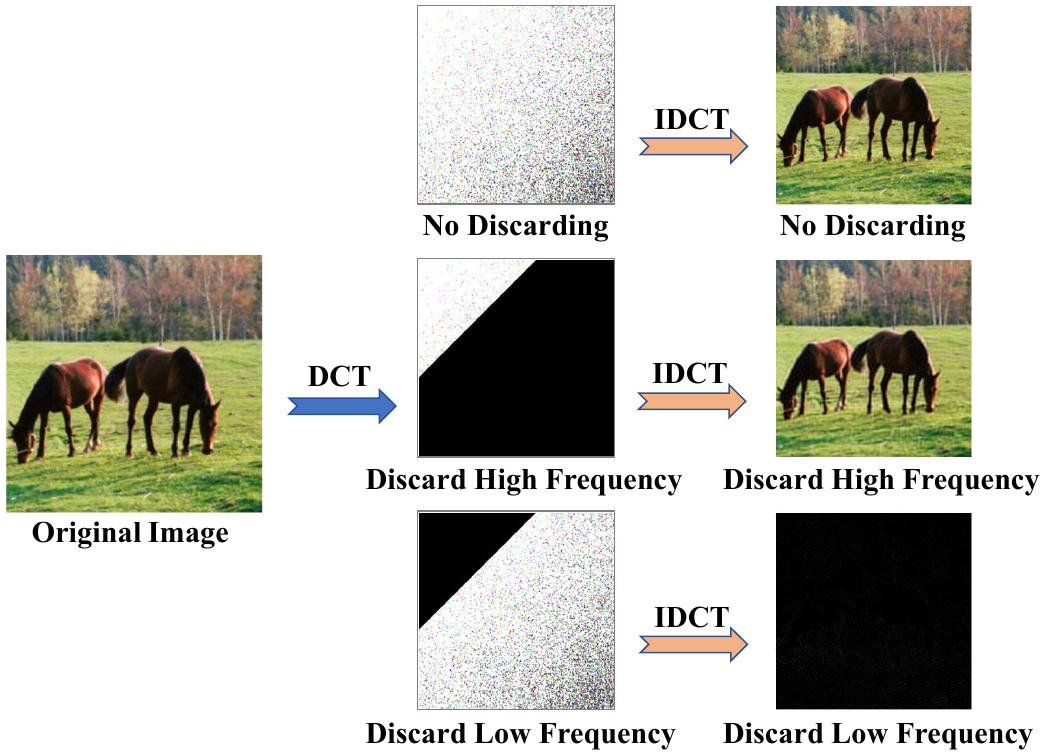}
\caption{The effect of visual differences in the spatial domain resulting from different components being discarded in the frequency domain. The DCT and IDCT are the discrete cosine transform and its inverse transform, respectively.}
\label{fig: effects of DCT}
\end{figure}

Frequency-domain transformations \cite{long2022frequency,zhong2022detecting,hou2023stealthy,kong2023efficient,yue2023invisible} decompose spatial features into frequencies, enabling global processing (Figure \ref{fig: effects of DCT}). Leveraging this, we first transform training samples into the frequency domain. Inspired by lossy image compression \cite{bhowmik2022lost,cunha2023lossy,behme2023art,cai2024make,duan2024conditional,yang2023everyone}, we avoid embedding explicit signals to prevent reverse inspection. Instead, we discard high-frequency components via quantization—imperceptible to human vision—and reconstruct the samples. These globally removed components create watermark samples whose trigger condition is compression behavior rather than pixel-level patterns, enabling both high covertness and robustness. To further enhance robustness and prevent false triggering, we simulate common input preprocessing attacks for joint training and introduce a similarity loss to cluster watermark samples with the same label in feature space, reinforcing their link to the watermark target label. Our main contributions are:
\begin{itemize}[itemsep=0pt,topsep=0pt,parsep=0pt]
    \item \textbf{Novel Approach for Watermark Sample Construction.} We introduce a new approach to constructing watermark samples in frequency domain, using global compression behavior as watermark triggering signal. Compared to prior methods, this is more covert, and its global distribution property enhances the watermark robustness.
    \item \textbf{Covert and Robust Watermarking Framework.} We propose ComMark, a black-box watermarking framework that balances better covertness and robustness. Our framework includes frequency-domain compression for constructing watermark samples, attack simulation strategy to boost robustness and prevent false triggering, and similarity loss to improve feature-level robustness. Our source code is available at \cite{SourceCode}.
    \item \textbf{Comprehensive Performance Evaluation.} We compare ComMark with state-of-the-art black-box watermarking techniques on multiple datasets. Our results demonstrate its superior performance across effectiveness, covertness, and robustness. Additionally, we test various watermark triggering conditions, confirming no false triggering. Extending our approach to audio, video, and text modalities yields consistently strong performance, showcasing its generalizability.
\end{itemize}

\section{Related Work}

Black-box watermarking methods \cite{chen2019deepmarks,namba2019robust,lv2022ssl} verify ownership without accessing model parameters, relying on whether watermark samples elicit target predictions. The earliest approach, Abstract \cite{adi2018turning}, was inspired by backdoor learning and used out-of-distribution (OOD) images as triggers. Content \cite{zhang2018protecting} embedded patterns (e.g., text) into clean samples and linked them to target labels. MEAD \cite{lv2024mea} generated composite samples by combining two task categories, enhancing watermark persistence in stolen models. While these methods offer robustness, their watermarks are often visually obvious, enabling adversaries to bypass verification. To improve covertness, BlindMark \cite{li2019prove} used an encoder-discriminator framework to embed fixed OOD images into clean data. MAB \cite{kim2023margin} assigned random labels to clean samples and applied projected gradient ascent to push them away from decision boundaries, creating imperceptible watermarks. However, such methods often lack robustness.

In contrast, we propose a novel watermarking technique inspired by JPEG compression \cite{wallace1991jpeg,chang2002steganographic,al2013jpeg,alkholidi2007new,duan2024conditional}. By quantizing frequency-domain features to remove high-frequency components, we generate global, hard-to-remove triggers. These modifications remain imperceptible in the spatial domain, achieving both strong covertness and robustness.

\begin{figure*}[t]
\centering
\includegraphics[width=0.98\textwidth]{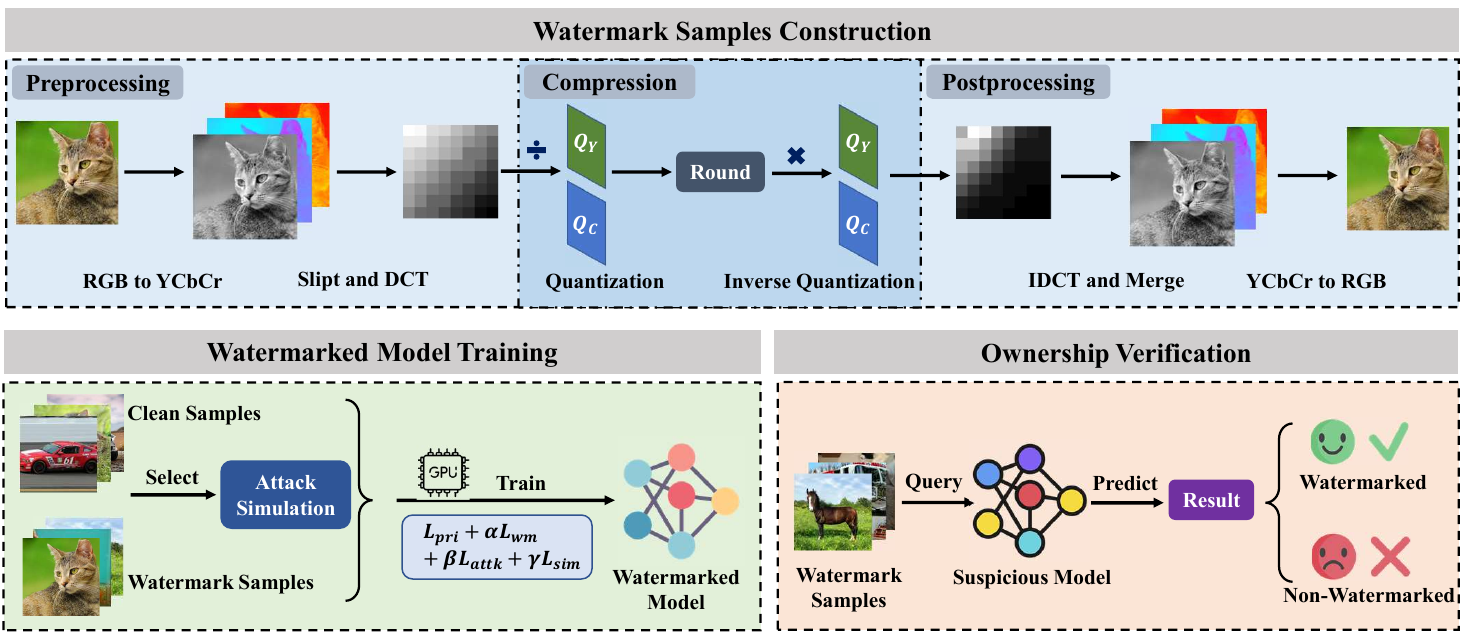}
\caption{An overview of our ComMark method.}
\label{fig: overview of method}
\end{figure*}

\section{Proposed Method}
\subsection{Threat Model}
\textbf{Adversary}: After obtaining the victim model via replication or extraction, the adversary attempts to invalidate watermark verification. This may involve removal attacks (e.g., finetuning, pruning) or preprocessing the defender’s verification samples through cropping, scaling, rotation, or JPEG compression to hinder ownership verification. Another potential threat is that the adversary could embed their own forged watermark into the model or discover samples that cause false triggering, potentially leading to disputes over ownership.

\textbf{Defender}: The defender, as the model owner, controls the training process and embeds the watermark before deployment. Given unknown future attacks, the goal is to craft a watermark that is effective, imperceptible, and robust. Verification is conducted in a black-box setting, where the defender only accesses the model’s predictions via an API controlled by the adversary.

\subsection{Overview}

Our method, illustrated in Figure \ref{fig: overview of method}, proceeds as follows: the defender selects a subset of clean samples and compresses them via Watermark Samples Construction module, assigning them a predefined target label to form the watermark set. During Watermarked Model Training, the model is trained on both clean and watermark data, with a portion of data undergoing adversarial processing each epoch. Finally, in Ownership Verification, defender queries the suspicious model with watermark test samples and, if the target class prediction rate exceeds a threshold, attributes the ownership.

\subsection{Watermark Samples Construction}
Previous watermarking methods typically constructed watermark samples in the spatial domain, where watermark triggers relied on fixed pixel-level feature patterns (often localized in specific regions). This reliance made such watermarking methods susceptible to detection or removal. To address these issues, inspired by JPEG compression algorithm, we propose constructing watermark samples in the frequency domain, leveraging quantization-based compression behavior as the triggering pattern. This novel black-box watermarking approach ensures superior covertness and robustness. Our construction process involves three stages: preprocessing, compression, and postprocessing.

\noindent\textbf{Preprocessing.}
We first convert input images from RGB to YCbCr, separating luminance (Y) from chrominance (Cb, Cr) components, as human vision is more sensitive to luminance. This separation allows selective compression.

Images are resized to $H\times H$ (with $H$ divisible by 8) and split into $8\times8$ blocks. Each block undergoes a 2D Discrete Cosine Transform (DCT) \cite{ahmed2006discrete}, converting it into frequency domain:
\begin{equation}
\small
\begin{aligned}
F(&u,v)\!=\!\alpha(u)\alpha(v)\!\sum\limits_{i=0}^{N-1}\sum\limits_{j=0}^{N-1}\!f(i,j)\text{cos}\big[\!\frac{(\!2i\!+\!1\!)u\pi}{2N}\!\big] \text{cos}\big[\!\frac{(\!2j\!+\!1\!)v\pi}{2N}\!\big] \\
&\alpha(u)=
\begin{cases}
\sqrt{1/N},~u=0 \\
\sqrt{2/N},~u\neq0
\end{cases}
~~~~~~~~~\alpha(v)=
\begin{cases}
\sqrt{1/N},~v=0 \\
\sqrt{2/N},~v\neq0
\end{cases}
\end{aligned}
\end{equation}
where $F(u,v)$ represents the intensity at position $(u,v)$ in frequency domain, and $f(i,j)$ represents the pixel value at position $(i,j)$ in spatial domain. DCT is used over DFT \cite{sundararajan2001discrete} due to its superior energy compaction, making it well-suited for compression.

\noindent\textbf{Compression.}
In the frequency domain, we apply quantization to suppress high-frequency components, which correspond to fine details less visible to the human eye. Low-frequency components, which carry primary visual content, are largely retained. Different quantization tables are used for Y and Cb/Cr channels to reflect human visual sensitivity. This results in globally distributed, covert watermark signals. The quantization and inverse quantization processes are expressed as:
\begin{equation}
F_q(u,v)=
\begin{cases}
Round[\frac{F(u,v)_{Y}}{Q_Y(u,v)}]\times Q_Y(u,v) \\[5pt]
Round[\frac{F(u,v)_{CbCr}}{Q_C(u,v)}]\times Q_C(u,v)
\end{cases}
\end{equation}
where $Round(\cdot)$ is the rounding operation. $Q_Y$ and $Q_C$ are the quantization tables for luminance and chrominance components, respectively. They are adapted based on the compression quality factor using the standard JPEG quantization table $Q_S$ \cite{kornblum2008using}, calculated as:
\begin{equation}
Q(i,j)=
\begin{cases}
\frac{50\times Q_S(i,j)}{Factor},&Factor\leq50 \\[5pt]
\frac{(200-2\times Factor)\times Q_S(i,j)}{100},&Factor>50
\end{cases}
\end{equation}
where $Q$ represents $Q_Y$ or $Q_C$, and $Factor$ is the quality factor ranging from 0 to 100.

\noindent\textbf{Postprocessing.}
After quantization-based compression, we first perform inverse DCT (IDCT) on each block to return to spatial domain:
\begin{equation}
\small
f_q(i,j)\!=\!\alpha(u)\alpha(v)\!\sum\limits_{u=0}^{N-1}\sum\limits_{v=0}^{N-1}\!F_q(u,v)\text{cos}\big[\!\frac{(\!2i\!+\!1\!)u\pi}{2N}\!\big] \text{cos}\big[\!\frac{(\!2j\!+\!1\!)v\pi}{2N}\!\big]
\end{equation}
where the symbols retain the same meanings as in DCT.

Next, we reassemble the blocks into a full $H\times H$ image, and convert the image back to RGB space.

Following above procedures, we can generate a sufficient number of watermark training and verification samples for subsequent use.

\subsection{Watermarked Model Training}
\noindent\textbf{Composition of Training Dataset.}
We randomly sample part of the original training data, relabel it to a predefined randomly selected watermark target label $y_t^w$, and form the watermark dataset $D_w$. The rest constitutes the primary task dataset $D_p$. To enhance robustness and reduce the risk of false activation from benign augmentations, we simulate attacks during training. Specifically, in each epoch, we first randomly select a portion $p$ from both $D_p$ and $D_w$, obtaining $D'_p$ and $D'_w$:
\begin{equation}
\small
D'_p\!=\!Sample_{rand}(D_p,p\!\cdot\!|D_p|),~D'_w\!=\!Sample_{rand}(D_w,p\!\cdot\!|D_w|)
\end{equation}
where $Sample_{rand}(D,n)$ represents randomly sampling $n$ samples from dataset $D$ without replacement.

Each replicated dataset is then partitioned into $k$ non-overlapping subsets. The $i$-th subset is subjected to attack $A_i$ as follows:
\begin{equation}
\begin{aligned}
D_{p}^{A_i}=ApplyAttack(A_i,Split(D'_p,i,k)) \\
D_{w}^{A_i}=ApplyAttack(A_i,Split(D'_w,i,k))
\end{aligned}
\end{equation}
where $Split(D,i,k)\!=\!\{x_j\!\in \!D|j\!\equiv \!i(\text{mod}~k)\}$ denotes partitioning $D$ into $k$ subsets and returning the $i$-th subset. $ApplyAttack(A,D)$ applies attack $A$ to dataset $D$, returning the attacked version. In this work, we utilize 10 common data preprocessing attacks (i.e., $k\!=\!10$), including cropping, rotation, scaling, Gaussian noise, Gaussian blur, brightness change, image quantization, JPEG2000 compression, WEBP compression, and color quantization. For each epoch, all the attacked data together form the attacked dataset $D_a$.

\noindent\textbf{Design of Training Loss.}
We first apply the regular classification loss on the three training sub-datasets, i.e.,
\begin{equation}
\begin{aligned}
L_{pri}=\sum\nolimits_{x_i^p\in D_p}\mathcal{L}(f_\theta(x_i^p),y_i^p) \\
L_{wm}=\sum\nolimits_{x_i^w\in D_w}\mathcal{L}(f_\theta(x_i^w),y_t^w) \\
L_{attk}=\sum\nolimits_{x_i^a\in D_a}\mathcal{L}(f_\theta(x_i^a),y_i^a)
\end{aligned}
\end{equation}
where $f_\theta$ is a model parameterized by $\theta$ and $\mathcal{L}$ is cross-entropy loss.

To further improve the correlation between watermark samples and the target label, we introduce a similarity loss $L_{sim}$ that pulls same-label samples closer and pushes different-label samples apart. We adopt contrastive loss \cite{hadsell2006dimensionality} to achieve this:
\begin{equation}
L_{sim}\!=\!\frac{1}{2N}\!\sum\limits_{i=1}^{N}[y_i\!\cdot d_i^2+(1\!-\!y_i)\cdot\text{max}(margin\!-\!d_i,0)^2]
\end{equation}
where $d_i$ is the Euclidean distance of the $i$-th sample pair, and $y_i = 1$ if the labels match, 0 otherwise. The $margin$ defines the minimum distance between sample features of different labels.

Thus, the total training loss for watermarked model is:
\begin{equation}
L=L_{pri}+\alpha L_{wm}+\beta L_{attk}+\gamma L_{sim}
\end{equation}
where $\alpha$, $\beta$, and $\gamma$ are coefficients that balance these losses.

\subsection{Ownership Verification}
To verify ownership, the defender queries the suspicious model with watermark test samples in watermark verification dataset $D_v$, all labeled with the target class. The watermark success rate $Acc_{wm}$ is then computed. If $Acc_{wm}$ exceeds a predefined threshold, ownership is confirmed. A higher $Acc_{wm}$ indicates stronger ownership evidence under black-box scenarios.

\section{Experiments}

\begin{table}[t]
    \vspace{-0.0cm}
    \centering
    \caption{Comparison of effectiveness and harmlessness (\%).}
    \label{tab: Comparison of effectiveness and harmlessness}
    \resizebox{0.48\textwidth}{!}{
    \begin{tabular}{lcccccccc}
        \toprule
        \multirow{2}{*}{\textbf{Method}} & \multicolumn{2}{c}{\textbf{GTSRB}} & \multicolumn{2}{c}{\textbf{CIFAR10}} & \multicolumn{2}{c}{\textbf{CIFAR100}} & \multicolumn{2}{c}{\textbf{VGGFace}} \\
        \cmidrule(lr){2-3} \cmidrule(lr){4-5} \cmidrule(lr){6-7} \cmidrule(lr){8-9}
        & Acc$\uparrow$ & WSR$\uparrow$ & Acc$\uparrow$ & WSR$\uparrow$ & Acc$\uparrow$ & WSR$\uparrow$ & Acc$\uparrow$ & WSR$\uparrow$ \\
        \midrule
        Clean Model	& 95.59 & 2.32 & 89.79 & 9.48 & 66.09 & 1.03 & 52.37 & 0.96 \\
        \hdashline
        Abstract & 93.23 & \textbf{100.00} & \underline{88.53} & \textbf{100.00} & \textbf{65.19} & \textbf{100.00} & \textbf{51.93} & \textbf{100.00} \\
        Content & 93.50 & 99.29 & 88.22 & \underline{99.96} & 63.41 & \underline{99.93} & 49.99 & \underline{99.99} \\
        MEAD & \underline{94.03} & 99.72 & \textbf{88.89} & \textbf{100.00} & 65.04 & \textbf{100.00} & 51.66 & 98.00 \\
        Noise & 92.18 & 99.76 & 87.73 & 99.92 & 64.58 & 99.50 & 50.38 & 99.91 \\
        BlindMark & 92.30 & 94.80 & 87.70 & 99.80 & 64.51 & 97.80 & 51.59 & 99.80 \\
        MAB	& 93.74 & \textbf{100.00} & 87.84 & \textbf{100.00} & 64.39 & \textbf{100.00} & 51.75 & \textbf{100.00} \\
        \textbf{ComMark} & \cellcolor{gray!20}\textbf{94.06} & \cellcolor{gray!20}\underline{99.83} & \cellcolor{gray!20}88.12 & \cellcolor{gray!20}\textbf{100.00} & \cellcolor{gray!20}\underline{65.06} & \cellcolor{gray!20}\textbf{100.00} & \cellcolor{gray!20}\underline{51.79} & \cellcolor{gray!20}99.71 \\
        \bottomrule
    \end{tabular}
    }
\end{table}

\begin{table*}[t]
    \vspace{-0.0cm}
    \centering
    \footnotesize
    \caption{Comparison of robustness (\%) against soft and hard label settings for three popular model extraction attacks. These results are obtained through testing on stolen models. The performance of victim model is presented in Table \ref{tab: Comparison of effectiveness and harmlessness}.}
    \label{tab: Comparison of robustness against three popular extraction attacks}
    \resizebox{\textwidth}{!}{
    \begin{tabular}{
        m{1.15cm}
        m{1.22cm}
        m{0.8cm}<{\centering}
        m{0.8cm}<{\centering}
        m{0.8cm}<{\centering}
        m{0.8cm}<{\centering}
        m{0.8cm}<{\centering}
        m{0.8cm}<{\centering}
        m{0.8cm}<{\centering}
        m{0.8cm}<{\centering}
        m{0.8cm}<{\centering}
        m{0.8cm}<{\centering}
        m{0.8cm}<{\centering}
        m{0.8cm}<{\centering}}
        \toprule
        \multirow{3}{*}{\textbf{Dataset}} & \multirow{3}{*}{\textbf{Method}} & \multicolumn{4}{c}{\textbf{Distillation}} & \multicolumn{4}{c}{\textbf{JBDA}} & \multicolumn{4}{c}{\textbf{Knockoff}} \\
        \cmidrule(lr){3-6} \cmidrule(lr){7-10} \cmidrule(lr){11-14}
        & & \multicolumn{2}{c}{Soft Label} & \multicolumn{2}{c}{Hard Label} & \multicolumn{2}{c}{Soft Label} & \multicolumn{2}{c}{Hard Label} & \multicolumn{2}{c}{Soft Label} & \multicolumn{2}{c}{Hard Label} \\
        \cmidrule(lr){3-4} \cmidrule(lr){5-6} \cmidrule(lr){7-8} \cmidrule(lr){9-10} \cmidrule(lr){11-12} \cmidrule(lr){13-14}
        & & \scriptsize Acc & \scriptsize WSR $\uparrow$ & \scriptsize Acc & \scriptsize WSR $\uparrow$ & \scriptsize Acc & \scriptsize WSR $\uparrow$ & \scriptsize Acc & \scriptsize WSR $\uparrow$ & \scriptsize Acc & \scriptsize WSR $\uparrow$ & \scriptsize Acc & \scriptsize WSR $\uparrow$ \\
        \midrule
        \multirow{7}{*}{GTSRB} & Abstract & 65.50 & 28.00 & 86.17 &	9.00 & 88.88 & 31.00 & 75.01 & 6.00 & 79.45 & \textbf{100.00} & 67.56 & 88.00 \\
        & Content & 50.71 & \underline{94.80} & 47.05 & 87.34 & 76.43 & \textbf{99.81} & 65.44 & 88.31 & 81.72 & \underline{99.99} & 18.15 & \underline{99.96} \\
        & MEAD & 78.73 & 4.03 & 64.83 & 0.28 & 83.17 & 26.53 & 72.75 & 3.47 & 89.69 & 98.06 & 67.43 & 85.00 \\
        & Noise	& 3.75 & 93.77 & 3.74 & \underline{92.32} & 3.73 & 98.71 & 4.13 & \underline{96.33} & 4.54 & 99.94 & 3.56 & \textbf{100.00} \\
        & BlindMark	& 56.83 & 6.40 & 53.82 & 4.60 & 72.61 & 10.60 & 67.62 & 6.80 & 76.63 & 18.40 & 48.89 & 9.60 \\
        & MAB & 60.38 & 50.51 & 55.34 & 46.43 & 84.70 & 76.79 & 83.55 & 68.11 & 78.59 & 82.40 & 63.99 & 64.80 \\
        & \textbf{ComMark} & \cellcolor{gray!20}25.33 & \cellcolor{gray!20}\textbf{98.99} & \cellcolor{gray!20}30.10 & \cellcolor{gray!20}\textbf{98.54} & \cellcolor{gray!20}28.31 & \cellcolor{gray!20}\underline{99.09} & \cellcolor{gray!20}20.09 & \cellcolor{gray!20}\textbf{98.73} & \cellcolor{gray!20}4.69 & \cellcolor{gray!20}\textbf{100.00} & \cellcolor{gray!20}3.74 & \cellcolor{gray!20}\textbf{100.00} \\
        \midrule
        \multirow{7}{*}{CIFAR10} & Abstract & 52.08 & 57.00 & 39.68 & 11.00 & 41.93 & 37.00 & 47.05 & 19.00 & 43.31 & 74.00 & 30.58 & 40.00 \\
        & Content & 68.85 & 15.36 & 40.55 & 19.06 & 34.55 & 12.65 & 50.14 & 20.77 & 53.13 & 21.09 & 39.71 & 15.12 \\
        & MEAD & 13.59 & \underline{94.08} & 20.58 & \underline{57.90} & 36.70 & \underline{92.85} & 24.10 & \underline{79.90} & 35.74 & \underline{96.88} & 16.72 & \underline{89.75} \\
        & Noise	& 49.50 & 8.37 & 30.00 & 19.60 & 58.71 & 12.03 & 40.12 & 23.92 & 57.67 & 15.47 & 30.26 & 42.77 \\
        & BlindMark	& 26.95 & 11.20 & 41.79 & 14.20 & 49.31 & 14.60 & 40.02 & 14.40 & 32.33 & 18.20 & 33.53 & 12.80 \\
        & MAB & 59.51 & 13.40 & 37.05 & 12.20 & 42.99 & 13.20 & 38.16 & 13.60 & 46.79 & 12.60 & 29.04 & 11.40 \\
        & \textbf{ComMark} & \cellcolor{gray!20}62.66 & \cellcolor{gray!20}\textbf{94.51} & \cellcolor{gray!20}45.71 & \cellcolor{gray!20}\textbf{84.62} & \cellcolor{gray!20}48.88 & \cellcolor{gray!20}\textbf{96.21} & \cellcolor{gray!20}40.30 & \cellcolor{gray!20}\textbf{93.03} & \cellcolor{gray!20}72.32 & \cellcolor{gray!20}\textbf{98.93} & \cellcolor{gray!20}53.80 & \cellcolor{gray!20}\textbf{94.45} \\
        \midrule
        \multirow{7}{*}{CIFAR100} & Abstract & 20.71 & 60.00 & 1.00 & 10.00 & 12.99 & 37.00 & 12.68 & 24.00 & 22.54 & 84.00 & 15.44 & 55.00 \\
        & Content & 18.00 & 0.05 & 2.59 & 0.04 & 11.75 & 0.00 & 15.44 & 0.18 & 18.99 & 96.22 & 16.74 & 54.49 \\
        & MEAD & 13.26 & \underline{86.12} & 1.80 & \underline{69.00} & 14.78 & \underline{91.00} & 9.93 & \underline{73.92} & 20.57 & \underline{99.27} & 10.79 & \underline{93.08} \\
        & Noise	& 13.93 & 0.00 & 1.00 & 0.00 & 14.69 & 0.01 & 12.33 & 0.33 & 16.17 & 0.02 & 10.08 & 0.02 \\
        & BlindMark	& 13.02 & 3.40 & 5.45 &	2.20 & 16.11 & 2.00 & 8.73 & 1.00 & 29.32 & 26.40 & 15.94 & 4.20 \\
        & MAB & 18.07 & 1.80 & 2.41 & 0.60 & 20.29 & 2.60 & 12.26 & 3.80 & 15.18 & 2.20 & 10.20 & 1.60 \\
        & \textbf{ComMark} & \cellcolor{gray!20}22.91 & \cellcolor{gray!20}\textbf{91.61} & \cellcolor{gray!20}0.98 & \cellcolor{gray!20}\textbf{75.88} & \cellcolor{gray!20}24.67 & \cellcolor{gray!20}\textbf{92.16} & \cellcolor{gray!20}18.09 & \cellcolor{gray!20}\textbf{80.30} & \cellcolor{gray!20}48.15 & \cellcolor{gray!20}\textbf{99.61} & \cellcolor{gray!20}30.84 & \cellcolor{gray!20}\textbf{98.64} \\
        \midrule
        \multirow{7}{*}{VGGFace} & Abstract	& 7.46 & 21.00 & 3.63 & 9.00 & 12.43 & 15.00 & 4.92 & 12.00 & 9.48 & 95.00 & 2.77 & 71.00 \\
        & Content & 8.09 & \textbf{99.84} & 1.80 & \textbf{87.28} & 10.21 & \underline{96.99} & 4.25 & 72.04 & 10.85 & \underline{99.19} & 2.38 & 98.98 \\
        & MEAD & 8.18 & 3.00 & 3.25 & 4.00 & 10.00 & 1.00 & 6.50 & 2.00 & 10.42 & 87.00 & 3.73 & 52.00 \\
        & Noise	& 1.04 & 89.98 & 1.00 & 69.18 & 1.01 & 95.49 & 1.00 & \underline{78.90} & 2.85 & 99.07 & 1.00 & \underline{99.12} \\
        & BlindMark	& 10.13 & 5.20 & 2.39 & 0.20 & 6.85 & 3.60 & 5.68 & 3.40 & 3.21 & 30.60 & 3.22 & 29.20 \\
        & MAB & 10.64 & 27.54 & 4.28 & 8.56 & 7.17 & 19.25 & 4.57 & 16.84 & 10.02 & 24.87 & 3.45 & 7.75 \\
        & \textbf{ComMark} & \cellcolor{gray!20}12.75 & \cellcolor{gray!20}\underline{96.20} & \cellcolor{gray!20}1.09 & \cellcolor{gray!20}\underline{80.24} & \cellcolor{gray!20}6.50 & \cellcolor{gray!20}\textbf{98.94} & \cellcolor{gray!20}13.66 & \cellcolor{gray!20}\textbf{81.87} & \cellcolor{gray!20}15.63 & \cellcolor{gray!20}\textbf{99.83} & \cellcolor{gray!20}3.61 & \cellcolor{gray!20}\textbf{99.84} \\
        \bottomrule
    \end{tabular}
    }
\end{table*}

\subsection{Experimental Setup}
\noindent\textbf{Tasks, Datasets, and Models.}
We evaluate ComMark on three tasks across four datasets: traffic sign recognition (GTSRB \cite{stallkamp2011german}), object classification (CIFAR10, CIFAR100 \cite{krizhevsky2009learning}), and face recognition (VGGFace \cite{parkhi2015deep}). ResNet18 \cite{he2016deep} is used for GTSRB and CIFAR10, and ResNet34 for CIFAR100 and VGGFace.

\noindent\textbf{Watermark Methods for Comparison.}
These methods include Abstract \cite{adi2018turning}, Content \cite{zhang2018protecting}, Noise \cite{zhang2018protecting}, MEAD \cite{lv2024mea}, BlindMark \cite{li2019prove}, and MAB \cite{kim2023margin}. Their detailed introductions are as follows:

$\bullet$ Abstract \cite{adi2018turning}: Uses abstract art images as watermark samples.

$\bullet$ Content \cite{zhang2018protecting}: Adds meaningful text (e.g., "TEST") to normal samples to form watermark samples.

$\bullet$ Noise \cite{zhang2018protecting}: Adds Gaussian noise with a specific distribution to normal samples to construct watermark samples.

$\bullet$ MEAD \cite{lv2024mea}: Selects two categories of normal samples from primary task and then concatenates them to form watermark samples.

$\bullet$ BlindMark \cite{li2019prove}: Encodes meaningful images (e.g., copyright logos) into normal images as watermark samples.

$\bullet$ MAB \cite{kim2023margin}: Randomly selects normal images and changes their labels to random labels as watermark samples, with training using projected gradient ascent to increase the distance between these samples and the decision boundary.

\noindent\textbf{Evaluation Metrics.}
Effectiveness and robustness are measured by: (1) Accuracy (Acc): the classification accuracy on primary test sets; (2) Watermark Success Rate (WSR): the percentage of watermark samples classified as the target label.

To assess covertness, we use PSNR \cite{huynh2008scope}, SSIM \cite{wang2004image}, and LPIPS \cite{zhang2018unreasonable} to quantify similarity between watermark and clean images.

\begin{figure*}[t]
    \centering
    \vspace{-0.0cm}
    \begin{subfigure}{0.24\textwidth}
        \centering
        \includegraphics[width=1.0\textwidth]{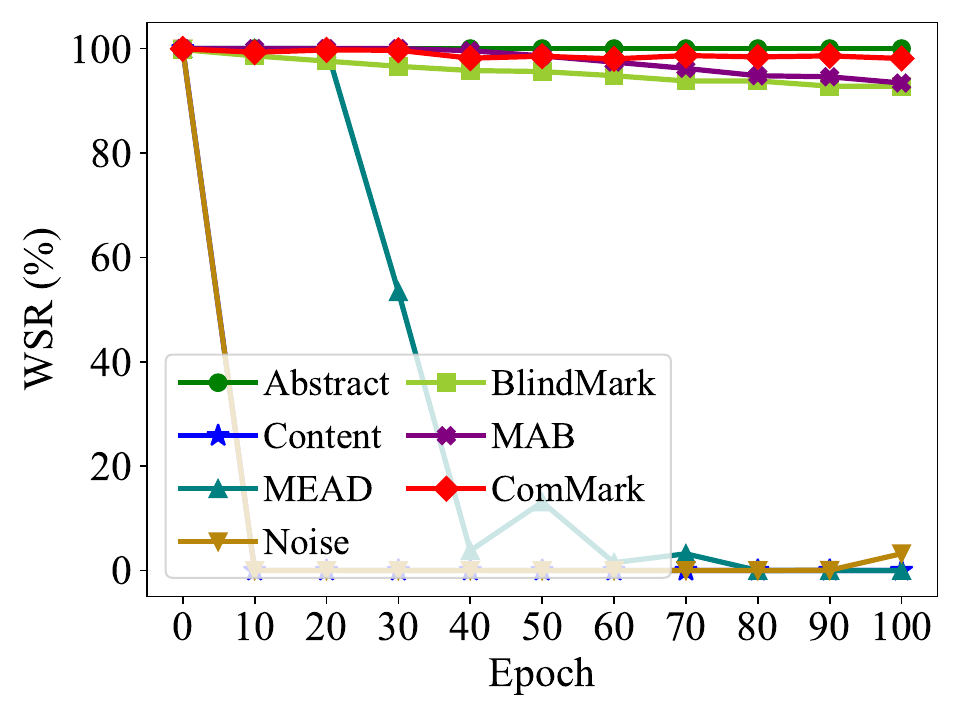}
        \caption{Model Finetuning Attack}
        \label{fig: finetune_cifar10_wsr}
    \end{subfigure}
    \begin{subfigure}{0.24\textwidth}
        \centering
        \includegraphics[width=1.0\textwidth]{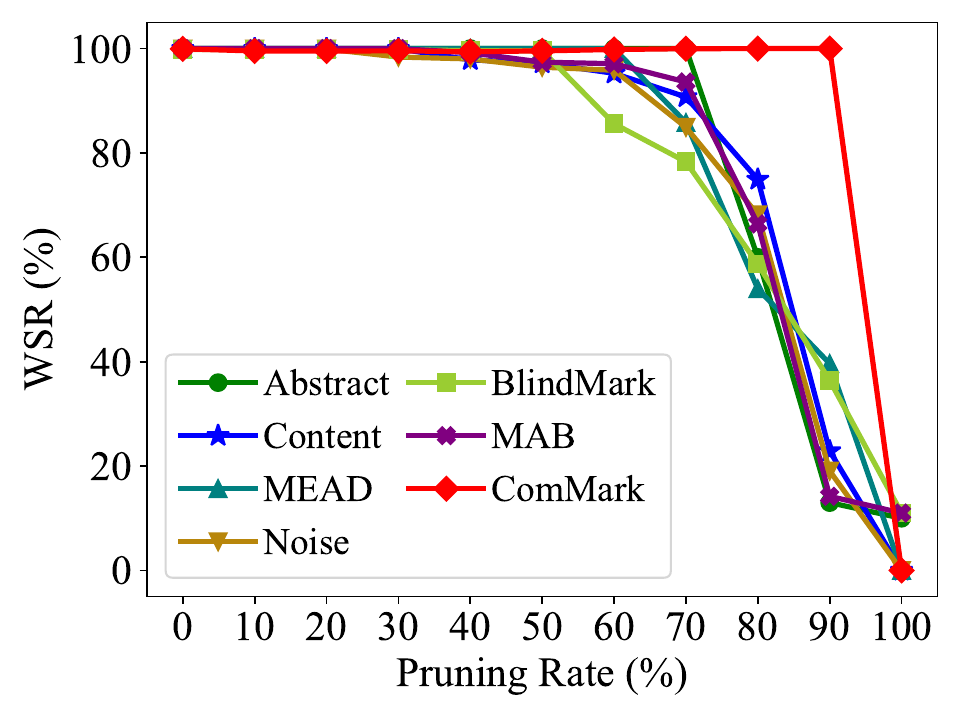}
        \caption{Model Pruning Attack}
        \label{fig: prune_cifar10_wsr}
    \end{subfigure}
    \begin{subfigure}{0.24\textwidth}
        \centering
        \includegraphics[width=1.0\textwidth]{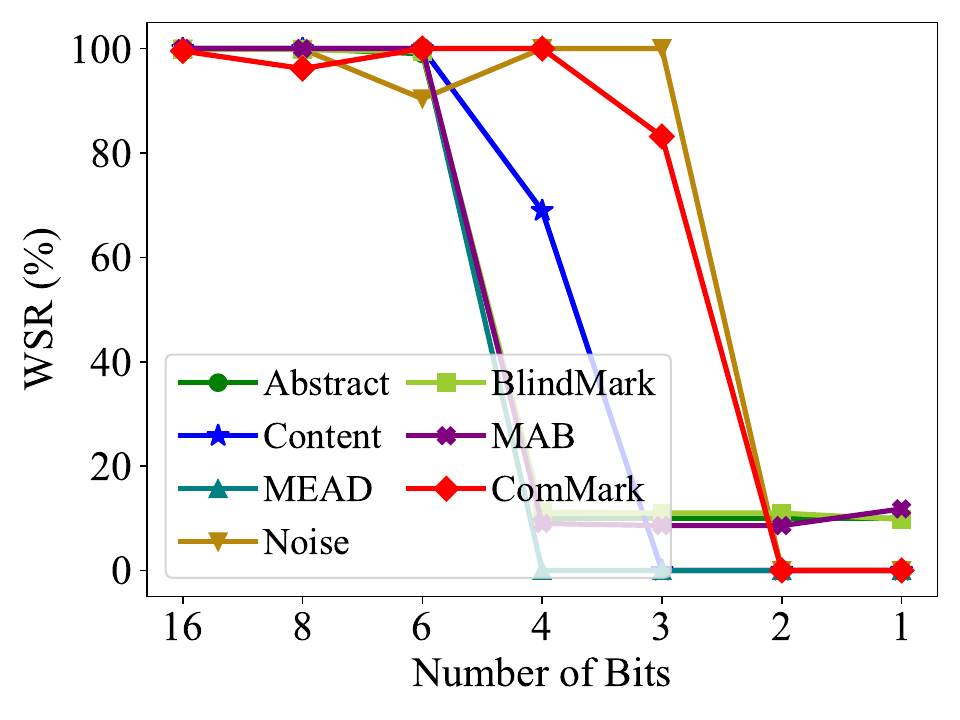}
        \caption{Model Quantization Attack}
        \label{fig: quantization_cifar10_wsr}
    \end{subfigure}
    \caption{Comparison of robustness against watermark removal attacks. (On CIFAR10)}
    \label{fig: robustness against removal attacks on cifar10}
    \vspace{-0.0cm}
\end{figure*}

\begin{figure*}[t]
    \centering
    \begin{subfigure}{0.24\textwidth}
        \centering
        \includegraphics[width=1.0\textwidth]{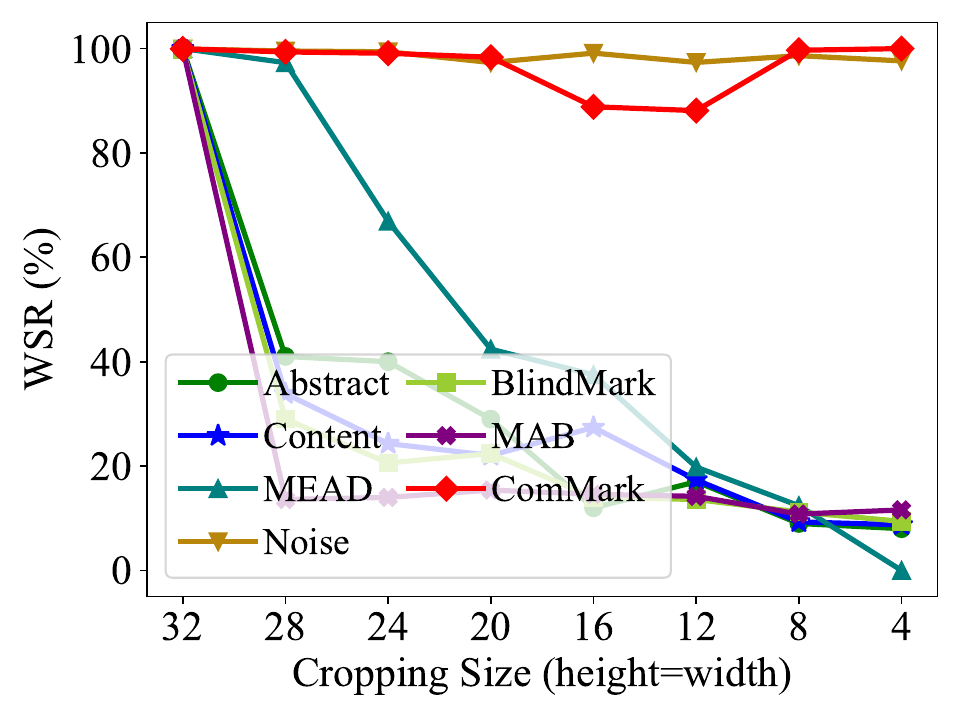}
        \caption{Cropping}
        \label{fig: crop_cifar10_wsr}
    \end{subfigure}
    \begin{subfigure}{0.24\textwidth}
        \centering
        \includegraphics[width=1.0\textwidth]{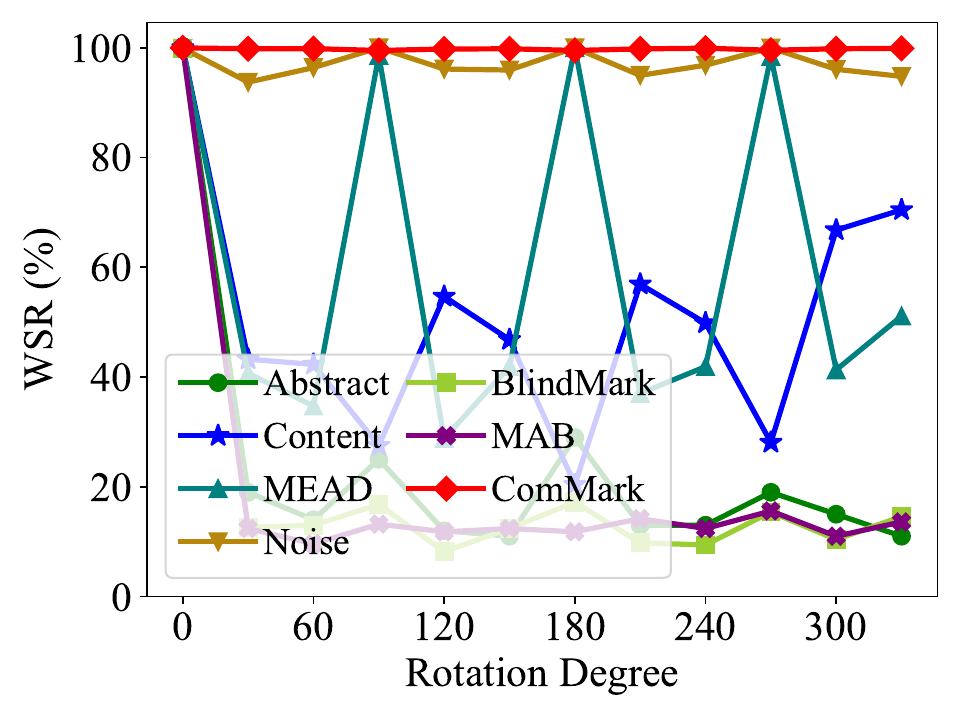}
        \caption{Rotation}
        \label{fig: rotate_cifar10_wsr}
    \end{subfigure}
    \begin{subfigure}{0.24\textwidth}
        \centering
        \includegraphics[width=1.0\textwidth]{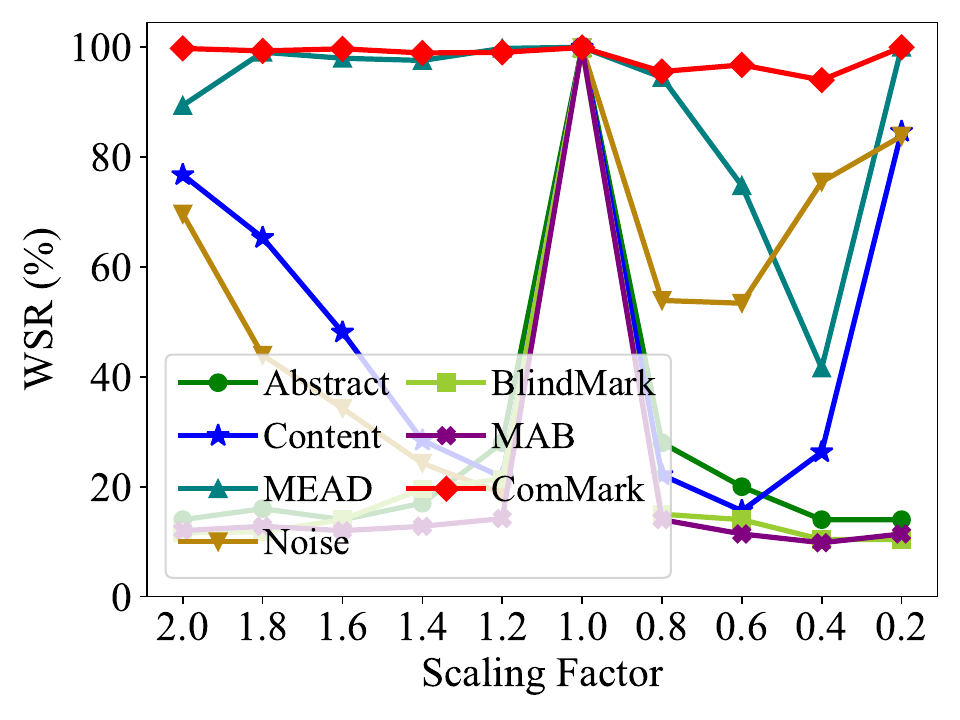}
        \caption{Scaling}
        \label{fig: scale_cifar10_wsr}
    \end{subfigure}
    \begin{subfigure}{0.24\textwidth}
        \centering
        \includegraphics[width=1.0\textwidth]{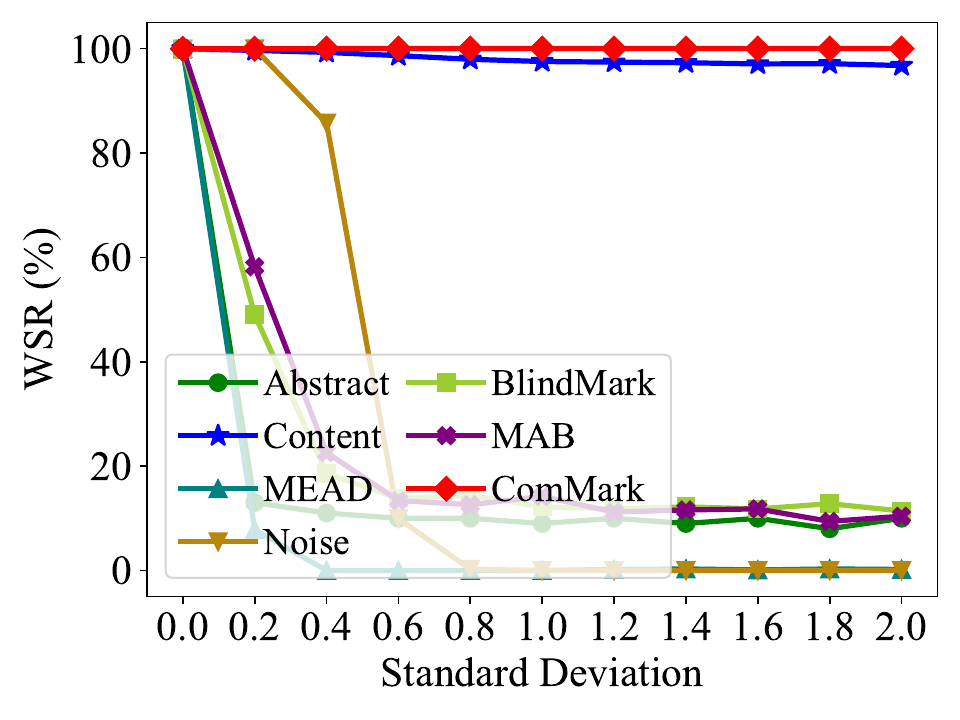}
        \caption{Gaussian Noise}
        \label{fig: gaussian_noise_cifar10_wsr}
    \end{subfigure}
    \vspace{0.2cm}

    \begin{subfigure}{0.24\textwidth}
        \centering
        \includegraphics[width=1.0\textwidth]{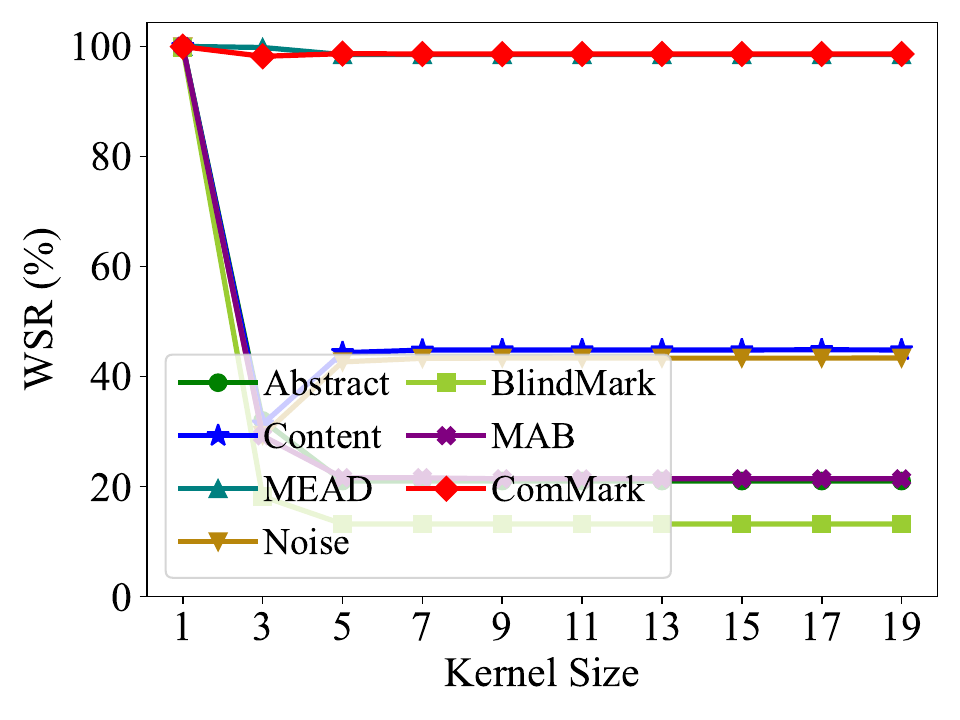}
        \caption{Gaussian Blur}
        \label{fig: gaussian_blur_cifar10_wsr}
    \end{subfigure}
    \begin{subfigure}{0.24\textwidth}
        \centering
        \includegraphics[width=1.0\textwidth]{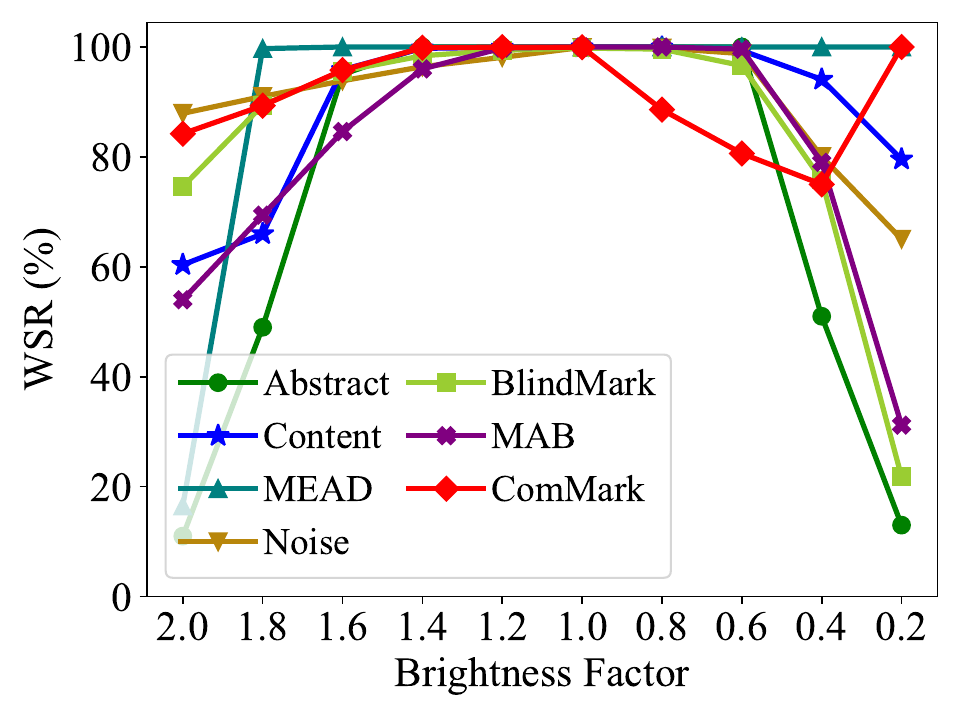}
        \caption{Brightness Change}
        \label{fig: brightness_change_cifar10_wsr}
    \end{subfigure}
    \begin{subfigure}{0.24\textwidth}
        \centering
        \includegraphics[width=1.0\textwidth]{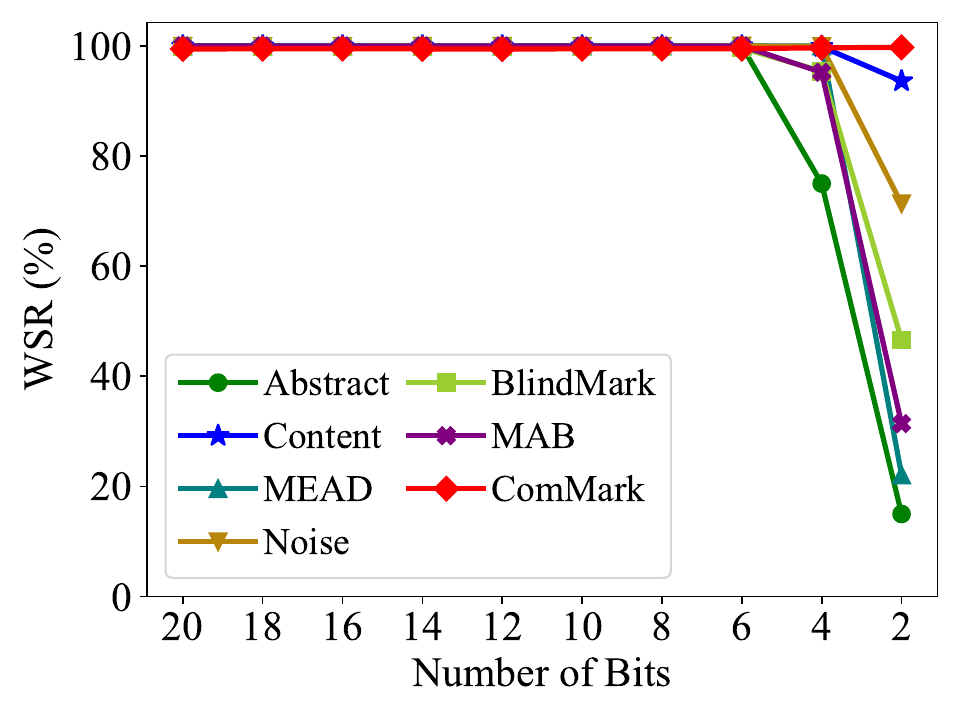}
        \caption{Image Quantization}
        \label{fig: image_quantization_cifar10_wsr}
    \end{subfigure}
    \begin{subfigure}{0.24\textwidth}
        \centering
        \includegraphics[width=1.0\textwidth]{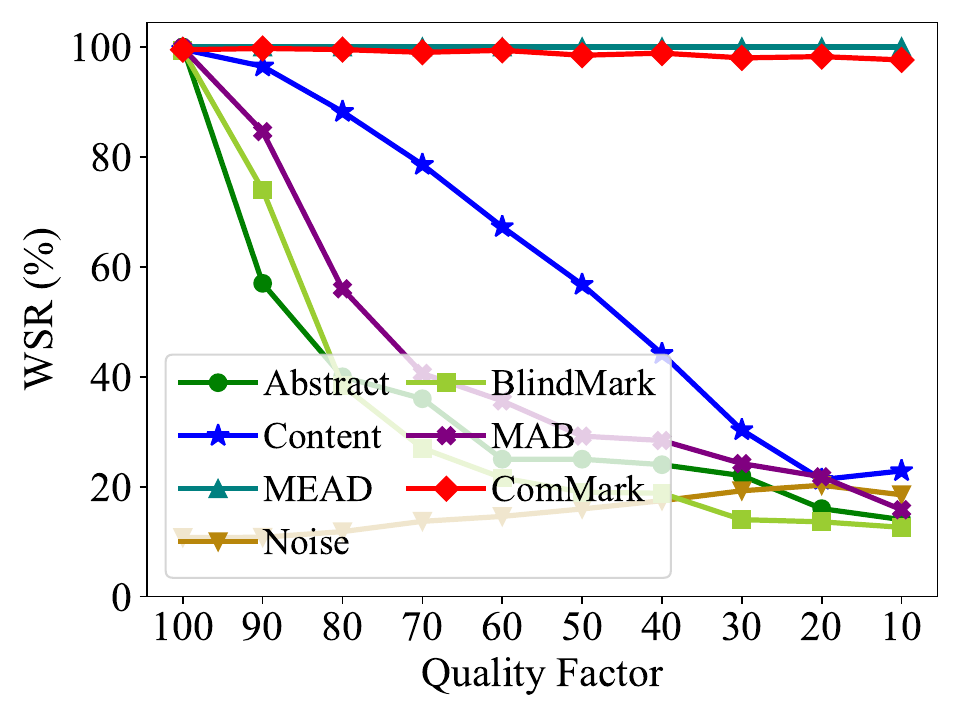}
        \caption{JPEG Compression}
        \label{fig: jpeg_compression_cifar10_wsr}
    \end{subfigure}
    \vspace{0.2cm}

    \begin{subfigure}{0.24\textwidth}
        \centering
        \includegraphics[width=1.0\textwidth]{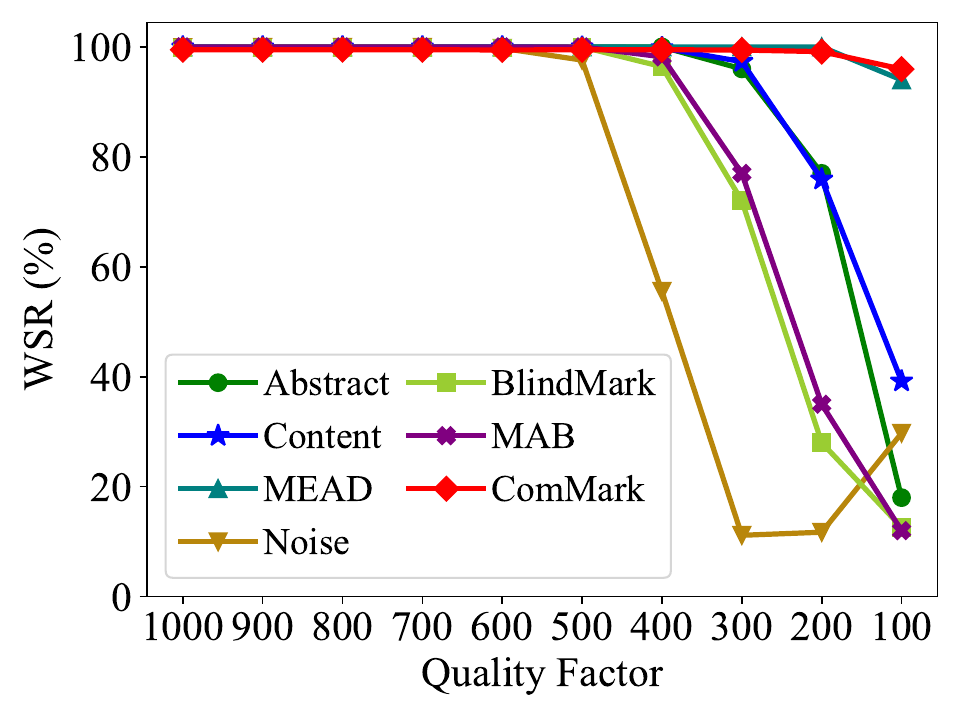}
        \caption{JPEG2000 Compression}
        \label{fig: jpeg2000_compression_cifar10_wsr}
    \end{subfigure}
    \begin{subfigure}{0.24\textwidth}
        \centering
        \includegraphics[width=1.0\textwidth]{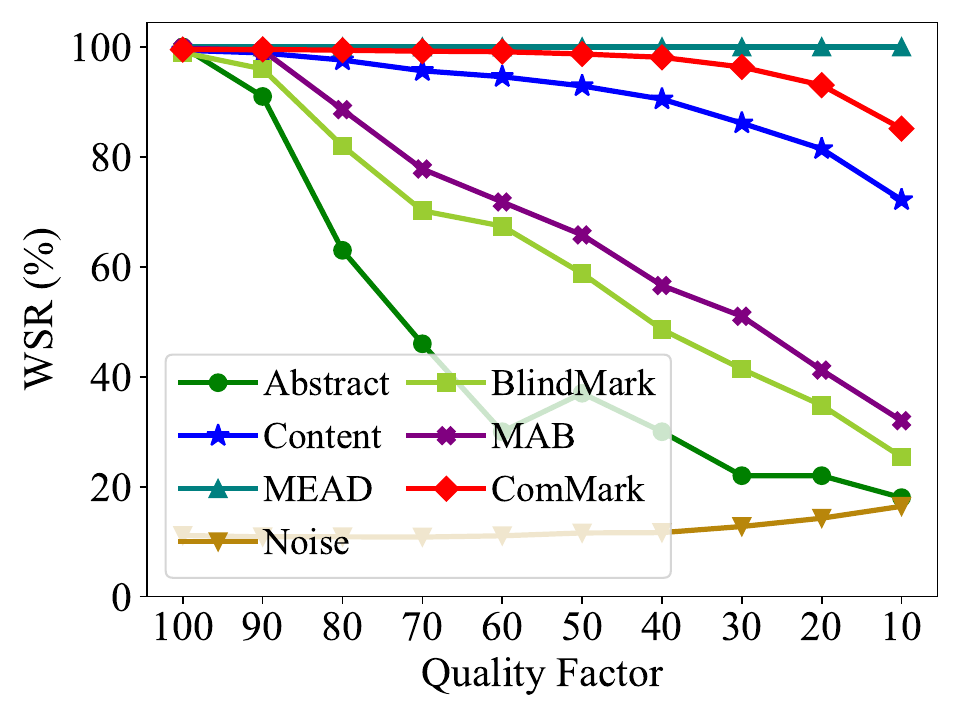}
        \caption{WEBP Compression}
        \label{fig: webp_compression_cifar10_wsr}
    \end{subfigure}
    \begin{subfigure}{0.24\textwidth}
        \centering
        \includegraphics[width=1.0\textwidth]{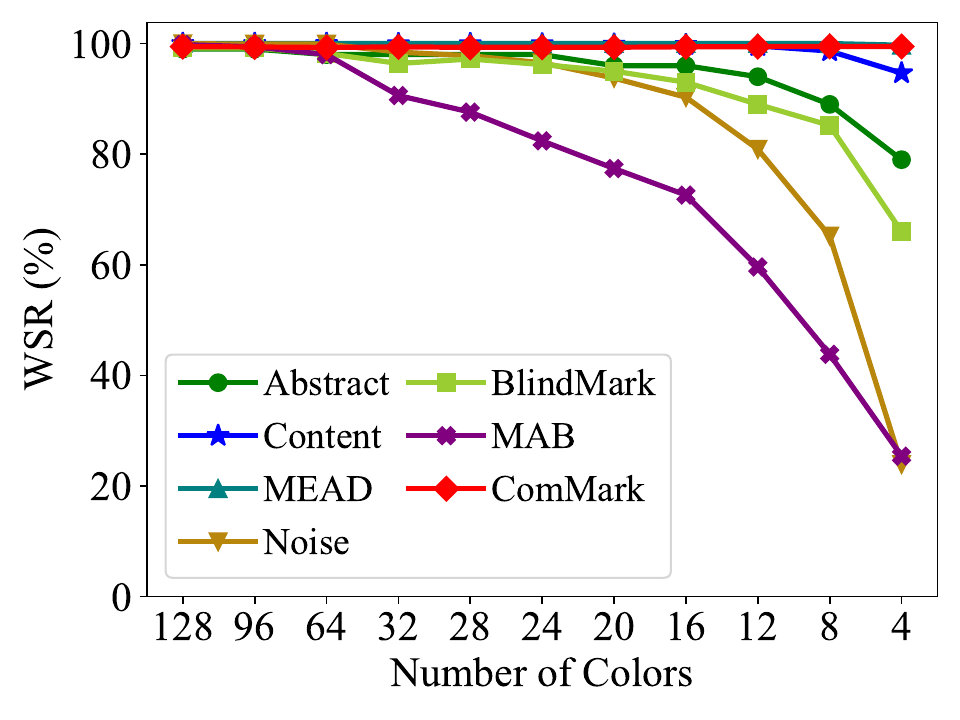}
        \caption{Color Quantization}
        \label{fig: color_quantization_cifar10_wsr}
    \end{subfigure}
    \vspace{-0.2cm}
    \caption{Comparison of robustness against watermark evasion attacks. (On CIFAR10)}
    \label{fig: robustness against evasion attacks on cifar10}
\end{figure*}

\noindent\textbf{Implementation Details.}
Watermark samples comprise 10\% of training data, with a compression quality factor of 90. We train for 100 epochs using Adam (initial learning rate 0.001, decayed by 0.1 every 30 epochs), and set loss weights to $\alpha\!=\!1.0$, $\beta\!=\!1.0$, $\gamma\!=\!0.1$. $margin$ is 1.0 in $L_{sim}$. Each epoch includes 50\% adversarially processed samples. Competing methods and attacks are tuned based on original papers for fair comparison. To eliminate the error influence of randomness, the results reported in our evaluation are the average of five repeated experiments with different random seeds.

\subsection{Effectiveness and Harmlessness}
In this section, we evaluate the effectiveness and harmlessness of various methods. Table \ref{tab: Comparison of effectiveness and harmlessness} shows our method achieves WSR comparable to SOTA methods on protected model, with rates of 99.83\%, 100.00\%, 100.00\%, and 99.71\% on GTSRB, CIFAR10, CIFAR100, and VGGFace datasets, respectively. The primary task accuracy drops by only 1.53\%, 1.67\%, 1.03\%, and 0.58\% on these datasets, confirming minimal harm to benign users. While prior methods also achieve similar effectiveness and harmlessness, we highlights that better robustness and covertness of black-box watermarks remain critical challenges.

\subsection{Robustness against Extraction Attacks}
Model extraction is known to be highly effective in removing watermarks \cite{tan2023deep,jia2021entangled,lv2024mea,tang2023exposing}, since extracted models often fail to capture watermark triggers that are not tied to the main task. We evaluate ComMark against popular extraction attacks: Distillation \cite{hinton2015distilling}, JBDA \cite{papernot2017practical}, Knockoff \cite{orekondy2019knockoff}, and their hard-label variants. As shown in Table \ref{tab: Comparison of robustness against three popular extraction attacks}, ComMark consistently achieves top watermark success rates, with few exceptions.

Unlike most methods that only perform well in specific settings, ComMark and MEAD show consistent robustness across attacks and datasets. ComMark's resilience stems from embedding global, covert triggers through JPEG-inspired compression and enforcing feature-space alignment via similarity loss. Further evaluations under Cross-Dataset, Cross-Architecture, and combined attacks (Tables \ref{tab: Comparison of robustness against cross-dataset extraction attacks}-\ref{tab: Comparison of robustness against cross-dataset and cross-architecture extraction attacks} of Appendix) confirm ComMark’s superior performance.

\begin{figure*}[t]
\vspace{-0.0cm}
\centering
\includegraphics[width=0.99\textwidth]{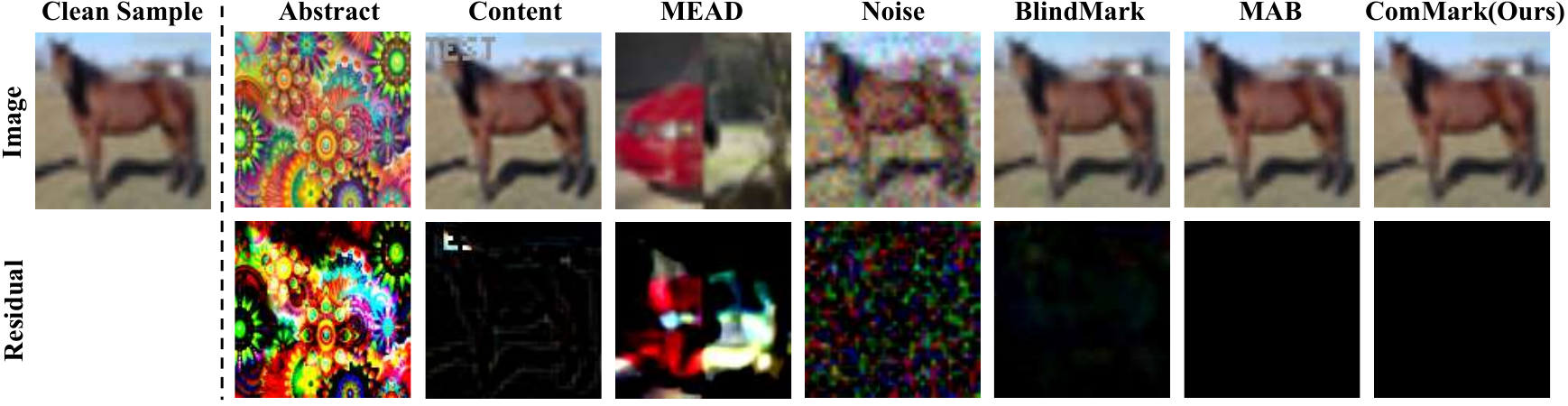}
\caption{Visual comparison of covertness of different watermarking methods. (On CIFAR10)}
\label{fig: visual comparison of covertness on cifar10}
\vspace{-0.0cm}
\end{figure*}

\begin{table*}[t]
    \centering
    \footnotesize
    \caption{Quantitative comparison of covertness.}
    \label{tab: Quantitative comparison of covertness}
    \resizebox{\textwidth}{!}{
    \begin{tabular}{
        m{1.6cm}
        m{0.9cm}<{\centering}
        m{0.9cm}<{\centering}
        m{0.9cm}<{\centering}
        m{0.9cm}<{\centering}
        m{0.9cm}<{\centering}
        m{0.9cm}<{\centering}
        m{0.9cm}<{\centering}
        m{0.9cm}<{\centering}
        m{0.9cm}<{\centering}
        m{0.9cm}<{\centering}
        m{0.9cm}<{\centering}
        m{0.9cm}<{\centering}}
        \toprule
        \multirow{2}{*}{\textbf{Method}} & \multicolumn{3}{c}{\textbf{GTSRB}} & \multicolumn{3}{c}{\textbf{CIFAR10}} & \multicolumn{3}{c}{\textbf{CIFAR100}} & \multicolumn{3}{c}{\textbf{VGGFace}} \\
        \cmidrule(lr){2-4} \cmidrule(lr){5-7} \cmidrule(lr){8-10} \cmidrule(lr){11-13}
        & \scriptsize PSNR $\uparrow$ & \scriptsize SSIM $\uparrow$ & \scriptsize LPIPS $\downarrow$ & \scriptsize PSNR $\uparrow$ & \scriptsize SSIM $\uparrow$ & \scriptsize LPIPS $\downarrow$ & \scriptsize PSNR $\uparrow$ & \scriptsize SSIM $\uparrow$ & \scriptsize LPIPS $\downarrow$ & \scriptsize PSNR $\uparrow$ & \scriptsize SSIM $\uparrow$ & \scriptsize LPIPS $\downarrow$\\
        \midrule
        Abstract & 9.93 & 0.0733 & 0.7087 & 10.92 & 0.0224 & 0.6141 & 10.77 & 0.0635 & 0.7107 & 10.29 & 0.0330 & 0.7421 \\
        Content	& 22.37 & 0.9699 & 0.0769 & 26.48 & 0.9898 & 0.0421 & 23.42 & 0.9859 & 0.0580 & 25.46 & 0.9842 & 0.0289 \\
        MEAD & 8.21 & 0.1322 & 0.2297 & 8.65 & 0.0735 & 0.6171 & 10.02 & 0.0435 & 0.6573 & 9.29 & 0.1371 & 0.5426 \\
        Noise & 20.39 & 0.3741 & 0.4478 & 23.70 & 0.6104 & 0.4396 & 23.93 & 0.5934 & 0.4049 & 20.44 & 0.3647 & 0.3842 \\
        BlindMark & 31.24 & 0.9904 & 0.0091 & 36.68 & 0.9963 & 0.0079 & 23.86 & 0.9407 & 0.1285 & 33.22 & 0.9888 & 0.0115 \\
        MAB & \textbf{inf} & \textbf{1.0000} & \textbf{0.0000} & \textbf{inf} & \textbf{1.0000} & \textbf{0.0000} & \textbf{inf} & \textbf{1.0000} & \textbf{0.0000} & \textbf{inf} & \textbf{1.0000} & \textbf{0.0000} \\
        \textbf{ComMark} & \cellcolor{gray!20}\underline{43.19} & \cellcolor{gray!20}\underline{0.9976} & \cellcolor{gray!20}\underline{0.0006} & \cellcolor{gray!20}\underline{42.85} & \cellcolor{gray!20}\underline{0.9984} & \cellcolor{gray!20}\underline{0.0009} & \cellcolor{gray!20}\underline{42.76} & \cellcolor{gray!20}\underline{0.9963} & \cellcolor{gray!20}\underline{0.0010} & \cellcolor{gray!20}\underline{43.04} & \cellcolor{gray!20}\underline{0.9986} & \cellcolor{gray!20}\underline{0.0007} \\
        \bottomrule
    \end{tabular}
    }
\end{table*}

\subsection{Robustness against Removal Attacks}
We test robustness against three classic removal attacks on CIFAR10. As shown in Figure \ref{fig: finetune_cifar10_wsr}, ComMark maintains a 98.10\% WSR under finetuning, outperforming prior methods, many of which fail significantly. Under 90\% pruning (Figure \ref{fig: prune_cifar10_wsr}), our WSR remains above 95\%, while others drop below 50\%. For quantization (Figure \ref{fig: quantization_cifar10_wsr}), ComMark survives until 2-bit precision, while most competitors fail at 3 or 4 bits. These results highlight the advantage of frequency-domain watermarking over pixel-level methods. Full results on other datasets are in our Appendix \ref{sec: more evaluation against watermark removal attacks}.

\subsection{Robustness against Evasion Attacks}

To bypass ownership verification, an adversary can use data preprocessing techniques to alter the query samples submitted by defender. We test the robustness of various watermarking methods against 11 input preprocessing attacks \cite{lukas2022sok} on CIFAR10, as shown in Figure \ref{fig: robustness against evasion attacks on cifar10}. Our method outperforms others, maintaining high watermark success rates across multiple attacks, while previous methods only show robustness against specific attacks. For instance, the Noise method resists geometric transformations like cropping and rotation but fails against Gaussian noise or JPEG compression. Our method's robustness stems from frequency-domain compression, which discards high-frequency information, embedding this behavior globally in the spatial domain. This makes it resilient to geometric transformations, noise, compression, and other preprocessing. Similar results are observed on GTSRB, CIFAR100, and VGGFace datasets, as shown in Figures \ref{fig: robustness against evasion attacks on gtsrb}, \ref{fig: robustness against evasion attacks on cifar100}, and \ref{fig: robustness against evasion attacks on vggface} of our Appendix.

\subsection{Covertness}

In practical scenarios, adversaries may use detection mechanisms, such as manual inspection or binary classifiers, to block ownership verification via query samples. To counter this, watermark samples must be highly covert, i.e., visually indistinguishable. Figure \ref{fig: visual comparison of covertness on cifar10} compares the visual differences between clean samples and watermark samples using different methods on the CIFAR10 task, as well as the residual between the two after being magnified by three times. Our ComMark method achieves excellent covertness, with differences imperceptible even when residuals are magnified three times. We further quantitatively assess covertness using PSNR, SSIM, and LPIPS metrics, as shown in Table \ref{tab: Quantitative comparison of covertness}. ComMark ranks second across all metrics, closely trailing the MAB method, aligning with the visual results in Figure \ref{fig: visual comparison of covertness on cifar10}. By discarding high-frequency information during watermark sample construction, our method effectively maintains imperceptible differences between watermark and clean samples. Notably, while BlindMark and MAB are also covert, their robustness proves to be poor, as detailed in previous sections.

\begin{table*}[t]
    \vspace{-0.0cm}
    \centering
    \footnotesize
    \caption{Different watermark success rate (WSR) performances (\%) of our watermarking method when resisting watermark evasion attacks and reducing false watermark triggering with (w/) and without (w/o) attack loss $L_{attk}$.}
    \label{tab: Ablation about attack loss}
    \resizebox{\textwidth}{!}{
    \begin{tabular}{
        m{1.71cm}
        m{0.75cm}<{\centering}
        m{0.75cm}<{\centering}
        m{0.75cm}<{\centering}
        m{0.75cm}<{\centering}
        m{0.75cm}<{\centering}
        m{0.75cm}<{\centering}
        m{0.75cm}<{\centering}
        m{0.75cm}<{\centering}
        m{0.65cm}<{\centering}
        m{0.65cm}<{\centering}
        m{0.65cm}<{\centering}
        m{0.65cm}<{\centering}
        m{0.65cm}<{\centering}
        m{0.65cm}<{\centering}
        m{0.65cm}<{\centering}
        m{0.65cm}<{\centering}}
        \toprule
        \multirow{3}{*}{\textbf{Method}} & \multicolumn{8}{c}{\textbf{Resisting Watermark Evasion Attacks (WSR$\uparrow$)}} & \multicolumn{8}{c}{\textbf{Reducing False Watermark Triggering (WSR$\downarrow$)}} \\
        \cmidrule(lr){2-9} \cmidrule(lr){10-17}
        & \multicolumn{2}{c}{GTSRB} & \multicolumn{2}{c}{CIFAR10} & \multicolumn{2}{c}{CIFAR100} & \multicolumn{2}{c}{VGGFace} & \multicolumn{2}{c}{GTSRB} & \multicolumn{2}{c}{CIFAR10} & \multicolumn{2}{c}{CIFAR100} & \multicolumn{2}{c}{VGGFace} \\
        \cmidrule(lr){2-3} \cmidrule(lr){4-5} \cmidrule(lr){6-7} \cmidrule(lr){8-9} \cmidrule(lr){10-11} \cmidrule(lr){12-13} \cmidrule(lr){14-15} \cmidrule(lr){16-17}
        & w/o & w/ & w/o & w/ & w/o & w/ & w/o & w/ & w/o & w/ & w/o & w/ & w/o & w/ & w/o & w/ \\
        \midrule
        Cropping & 98.59 & 99.93 & 98.72 & 99.88 & 99.63 & 99.95 & 99.59 & 99.92 & 7.72 & 6.58 & 17.26 & 10.33 & 1.87 & 1.05 & 5.38 & 1.75 \\
        Rotation & 99.75 & 99.86 & 99.23 & 99.98 & 99.53 & 99.98 & 99.66 & 99.95 & 12.38 & 8.24 & 12.98 & 9.71 & 1.68 & 1.32 & 5.60 & 2.37 \\
        Scaling & 99.23 & 99.65 & 98.32 & 99.99 & 99.35 & 100.00 & 99.67 & 99.97 & 7.10 & 4.35 & 13.35 & 10.09 & 1.54 & 0.89 & 1.69 & 1.55 \\
        Brightness & 99.23 & 99.92 & 92.68 & 100.00 & 99.01 & 100.00 & 99.49 & 100.00 & 17.82 &  12.37 & 15.97 &  9.70 & 14.46 &  1.08 & 12.95 &  4.17 \\
        Noise & 99.90 & 99.90 & 99.83 & 100.00 & 99.97 & 100.00 & 99.93 & 100.00 & 24.32 &  12.53 & 29.91 &  9.67 & 22.31 &  1.14 & 25.78 &  3.68 \\
        Blur & 97.85 & 99.39 & 83.19 & 98.77 & 88.95 & 99.76 & 99.35 & 99.73 & 7.97 &  4.29 & 15.74 &  10.33 & 1.26 &  0.88 & 1.14 &  0.80 \\
        Image Quant. & 100.00 & 100.00 & 100.00 & 100.00 & 100.00 & 100.00 & 100.00 & 100.00 & 31.54 &  5.03 & 34.94 &  16.49 & 7.40 &  3.67 & 18.52 &  13.03 \\
        Color Quant. & 100.00 & 100.00 & 99.91 & 100.00 & 99.05 & 99.93 & 99.89 & 100.00 & 19.87 &  12.33 & 16.23 &  9.81 & 3.57 &  0.93 & 3.94 &  1.74 \\
        JPEG2000 & 95.83 & 97.79 & 82.59 & 95.30 & 97.33 & 98.46 & 78.50 & 94.54 & 33.16 &  14.43 & 34.98 &  14.50 & 25.97 &  9.30 & 25.23 &  7.34 \\
        WEBP & 88.35 & 99.04 & 55.79 & 88.56 & 58.26 & 93.36 & 69.05 & 92.72 & 34.93 &  15.05 & 27.96 &  18.99 & 26.49 &  8.13 & 34.47 &  12.23 \\
        \bottomrule
    \end{tabular}
    }
    \vspace{-0.0cm}
\end{table*}

\begin{figure*}[t]
\centering
\includegraphics[width=0.99\textwidth]{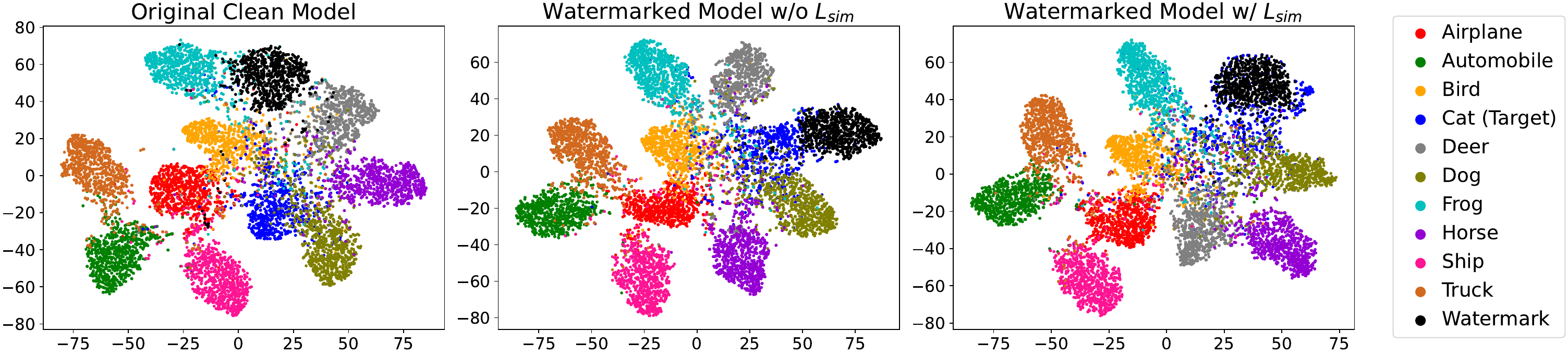}
\caption{Visualization of t-SNE in the feature space of models with (w/) and without (w/o) similarity loss $L_{sim}$.}
\label{fig: tsne}
\end{figure*}

\subsection{Ablation Study}

In this section, we perform ablation experiments to evaluate the impact of loss function designs and hyperparameters on ComMark's performance, focusing on attack loss, similarity loss, watermark sample rate, and compression quality factor. Due to space limitation, results for the latter two are detailed in Table \ref{fig: watermark_ratio} and Table \ref{fig: quality_factor} of our Appendix.

\subsubsection{Effect of Attack Loss}
We first evaluate the impact of attack loss, which generates attacked samples during training. The left half of Table \ref{tab: Ablation about attack loss} shows robustness against watermark evasion attacks, with attack loss improving watermark success rates by 0\% to 36\%, depending on the attack type. The right half of Table \ref{tab: Ablation about attack loss} indicates that attack loss reduces false triggering success rates by 0\% to 27\%. Overall, our method already exhibits strong resilience to watermark evasion attacks and false triggering, and attack loss further enhances these properties. It should be clear that, unlike watermark evasion attack where an adversary processes the defender's watermark samples to prevent ownership verification, false triggering is where an adversary uses some data processing techniques to try to forge watermark samples to cause ownership ambiguity.

\subsubsection{Effect of Similarity Loss}
Our similarity loss has two key goals: enhancing entanglement between watermark samples and target class samples, and stabilizing and accelerating model training. The t-SNE plots \cite{van2008visualizing} in Figure \ref{fig: tsne} show that similarity loss brings watermark samples closer to target class samples in the representation space, strengthening their correlation and improving watermark robustness. Figure \ref{fig: training_with_simloss} in the Appendix compares training progress with and without similarity loss, revealing faster convergence, stabilized training, and slight improvements in both model accuracy and watermark success rate when similarity loss is applied.

\section{Conclusion}

In this paper, we propose a novel watermarking approach using covert behavioral patterns as watermark triggers. Specifically, we introduce a black-box method based on frequency-domain compression, where high-frequency information is discarded through quantization to create covert watermark samples. This global compression behavior, when reverted to the spatial domain, enhances resistance to removal. Additionally, we employ two optimization techniques during training: attack simulation, which improves resistance to attacks and reduces false triggering, and similarity loss, which enhances robustness and stabilizes training by clustering features of same-label samples. Extensive experiments across diverse tasks and datasets demonstrate our method's superior covertness and robustness. We hope this work will help model owners protect their intellectual property.


\bibliographystyle{ACM-Reference-Format}
\bibliography{reference}

\appendix

\begin{table}[t]
    \centering
    \caption{Comparison of robustness (\%) against cross-dataset extraction attacks. The primary task is CIFAR10.}
    \label{tab: Comparison of robustness against cross-dataset extraction attacks}
    \resizebox{0.48\textwidth}{!}{
    \begin{tabular}{lcccccccc}
        \toprule
        \multirow{3}{*}{\textbf{Method}} & \multicolumn{8}{c}{\textbf{Query Set of Adversary}} \\
        \cmidrule(lr){2-9}
        & \multicolumn{2}{c}{CIFAR10} & \multicolumn{2}{c}{CIFAR100} & \multicolumn{2}{c}{GTSRB} & \multicolumn{2}{c}{VGGFace} \\
        \cmidrule(lr){2-3} \cmidrule(lr){4-5} \cmidrule(lr){6-7} \cmidrule(lr){8-9}
        & Acc & WSR$\uparrow$ & Acc & WSR$\uparrow$ & Acc & WSR$\uparrow$ & Acc & WSR$\uparrow$ \\
        \midrule
        Abstract & 65.13 & 72.00 & 43.31 & 74.00 & 16.55 & 33.00 & 11.08 & 16.00 \\
        Content	& 36.35 & 9.23 & 53.13 & 21.09 & 17.92 & 56.45 & 15.58 & \textbf{74.57} \\
        MEAD & 50.94 & \underline{95.36} & 35.74 & \underline{96.88} & 17.96 & \underline{62.04} & 17.18 & 67.38 \\
        Noise & 61.80 & 14.06 & 57.67 & 15.47 & 16.33 & 39.80 & 15.95 & 52.83 \\
        BlindMark & 51.21 & 17.40 & 32.33 & 18.20 & 14.50 & 9.60 & 16.60 & 12.40 \\
        MAB & 66.23 & 12.40 & 46.79 & 12.60 & 14.14 & 8.00 & 16.29 & 11.20 \\
        \textbf{ComMark} & \cellcolor{gray!20}77.57 & \cellcolor{gray!20}\textbf{98.13} & \cellcolor{gray!20}72.32 & \cellcolor{gray!20}\textbf{98.93} & \cellcolor{gray!20}22.91 & \cellcolor{gray!20}\textbf{69.09} & \cellcolor{gray!20}28.41 & \cellcolor{gray!20}\underline{71.42} \\
        \bottomrule
    \end{tabular}
    }
    \vspace{-0.0cm}
\end{table}

\begin{table}[t]
    \centering
    \caption{Comparison of robustness (\%) against cross-architecture extraction attacks. The model architecture of victim is ResNet18.}
    \label{tab: Comparison of robustness against cross-architecture extraction attacks}
    \resizebox{0.48\textwidth}{!}{
    \begin{tabular}{lcccccccc}
        \toprule
        \multirow{3}{*}{\textbf{Method}} & \multicolumn{8}{c}{\textbf{Model Architecture of Adversary}} \\
        \cmidrule(lr){2-9}
        & \multicolumn{2}{c}{ResNet18} & \multicolumn{2}{c}{ResNet34} & \multicolumn{2}{c}{VGG16} & \multicolumn{2}{c}{AlexNet} \\
        \cmidrule(lr){2-3} \cmidrule(lr){4-5} \cmidrule(lr){6-7} \cmidrule(lr){8-9}
        & Acc & WSR$\uparrow$ & Acc & WSR$\uparrow$ & Acc & WSR$\uparrow$ & Acc & WSR$\uparrow$ \\
        \midrule
        Abstract & 65.13 & 72.00 & 44.66 & 34.00 & 10.00 & 10.00 & 27.13 & 36.00 \\
        Content	& 36.35 & 9.23 & 51.02 & 10.77 & 18.97 & 1.54 & 29.15 & 19.27 \\
        MEAD & 50.94 & \underline{95.36} & 28.50 & \underline{90.80} & 19.78 & \underline{67.19} & 19.58 & \underline{71.94} \\
        Noise & 61.80 & 14.06 & 53.42 & 6.42 & 10.00 & 52.06 & 30.53 & 9.12 \\
        BlindMark & 51.21 & 17.40 & 62.09 & 19.00 & 28.72 & 13.20 & 36.59 & 12.60 \\
        MAB & 66.23 & 12.40 & 58.99 & 13.60 & 10.00 & 11.00 & 23.86 & 6.40 \\
        \textbf{ComMark} & \cellcolor{gray!20}88.12 & \cellcolor{gray!20}\textbf{98.13} & \cellcolor{gray!20}75.59 & \cellcolor{gray!20}\textbf{92.85} & \cellcolor{gray!20}27.55 & \cellcolor{gray!20}\textbf{76.81} & \cellcolor{gray!20}31.83 & \cellcolor{gray!20}\textbf{82.49} \\
        \bottomrule
    \end{tabular}
    }
    \vspace{-0.0cm}
\end{table}

\begin{table}[t]
    \centering
    \caption{Comparison of robustness (\%) against cross-dataset and cross-architecture extraction attacks. The victim model is ResNet18 trained on CIFAR10.}
    \label{tab: Comparison of robustness against cross-dataset and cross-architecture extraction attacks}
    \resizebox{0.48\textwidth}{!}{
    \begin{tabular}{lcccccccc}
        \toprule
        \multirow{4}{*}{\textbf{Method}} & \multicolumn{8}{c}{\textbf{Query Set and Model Architecture of Adversary}} \\
        \cmidrule(lr){2-9}
        & \multicolumn{2}{c}{CIFAR100} & \multicolumn{2}{c}{GTSRB} & \multicolumn{2}{c}{VGGFace} & \multicolumn{2}{c}{TinyImageNet} \\
        & \multicolumn{2}{c}{VGG16} & \multicolumn{2}{c}{AlexNet} & \multicolumn{2}{c}{ResNet34} & \multicolumn{2}{c}{DenseNet161} \\
        \cmidrule(lr){2-3} \cmidrule(lr){4-5} \cmidrule(lr){6-7} \cmidrule(lr){8-9}
        & Acc & WSR$\uparrow$ & Acc & WSR$\uparrow$ & Acc & WSR$\uparrow$ & Acc & WSR$\uparrow$ \\
        \midrule
        Abstract & 10.00 & 10.00 & 14.14 & 19.00 & 10.43 & 10.00 & 23.20 & 88.00 \\
        Content & 14.44 & 6.60 & 11.13 & 0.64 & 14.18 & \textbf{84.47} & 37.02 & 3.22 \\
        MEAD & 19.49 & \textbf{98.43} & 11.34 & \underline{46.91} & 12.54 & 79.83 & 17.16 & \textbf{99.74} \\
        Noise & 10.00 & 94.96 & 12.46 & 13.25 & 19.35 & 67.90 & 36.79 & 14.85 \\
        BlindMark & 18.67 & 11.80 & 12.03 & 9.80 & 11.27 & 11.40 & 20.94 & 13.20 \\
        MAB & 10.00 & 11.00 & 10.65 & 9.80 & 16.65 & 10.40 & 36.97 & 13.00 \\
        \textbf{ComMark} & \cellcolor{gray!20}17.74 & \cellcolor{gray!20}\underline{97.61} & \cellcolor{gray!20}14.27 & \cellcolor{gray!20}\textbf{53.11} & \cellcolor{gray!20}20.91 & \cellcolor{gray!20}\underline{83.27} & \cellcolor{gray!20}39.14 & \cellcolor{gray!20}\underline{99.59} \\
        \bottomrule
    \end{tabular}
    }
\end{table}

\begin{figure*}[t]
    \centering
    \vspace{-0.0cm}
    \begin{subfigure}{0.24\textwidth}
        \centering
        \includegraphics[width=1.0\textwidth]{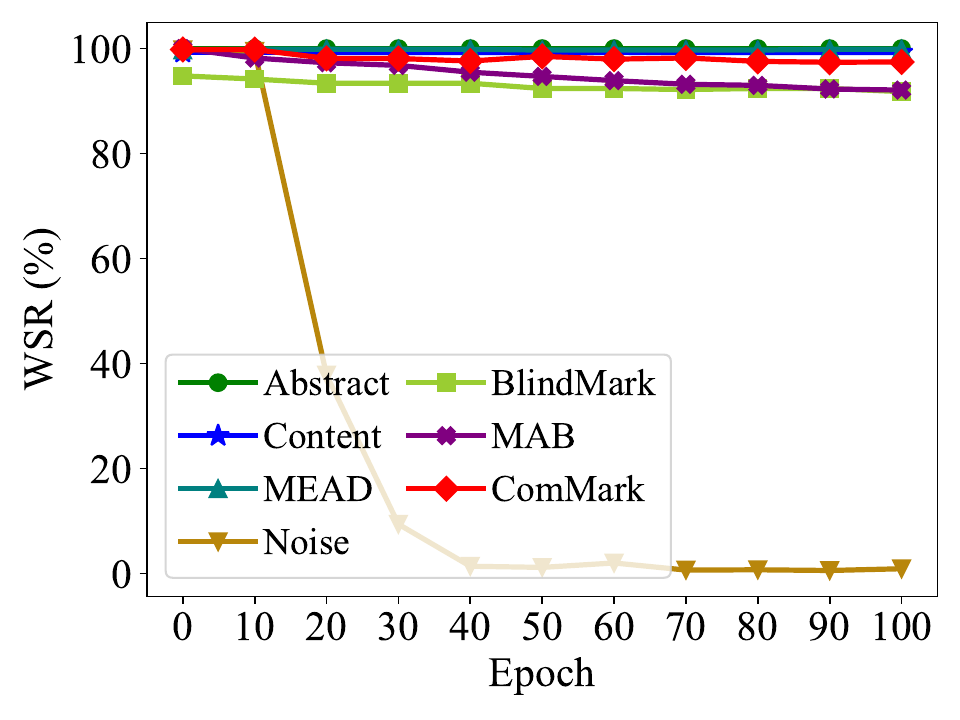}
        \caption{Model Finetuning Attack}
        \label{fig: finetune_gtsrb_wsr}
    \end{subfigure}
    \begin{subfigure}{0.24\textwidth}
        \centering
        \includegraphics[width=1.0\textwidth]{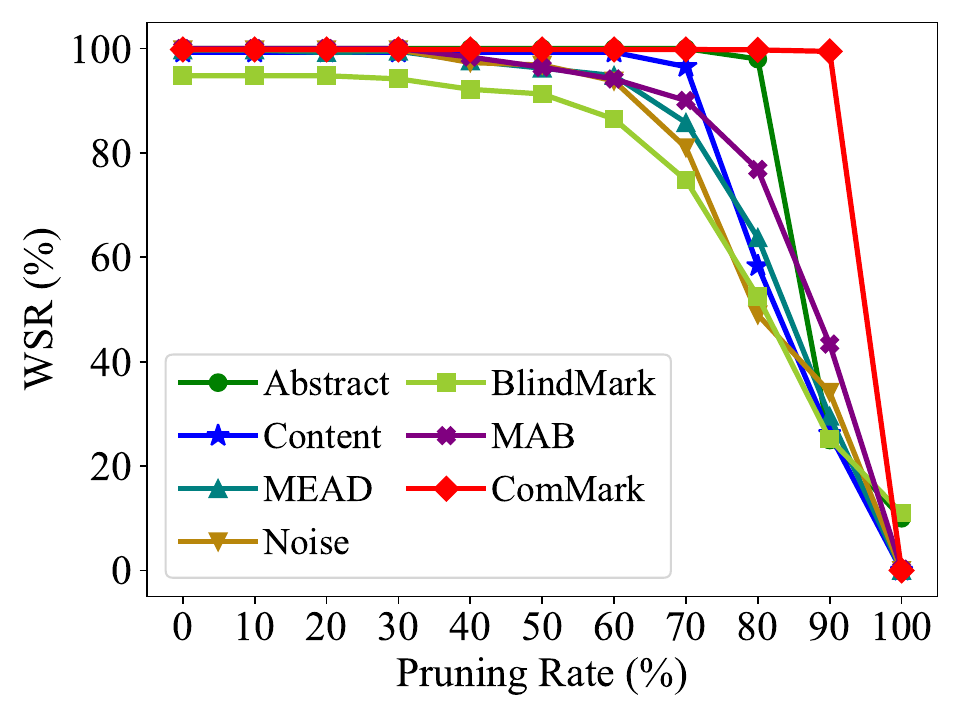}
        \caption{Model Pruning Attack}
        \label{fig: prune_gtsrb_wsr}
    \end{subfigure}
    \begin{subfigure}{0.24\textwidth}
        \centering
        \includegraphics[width=1.0\textwidth]{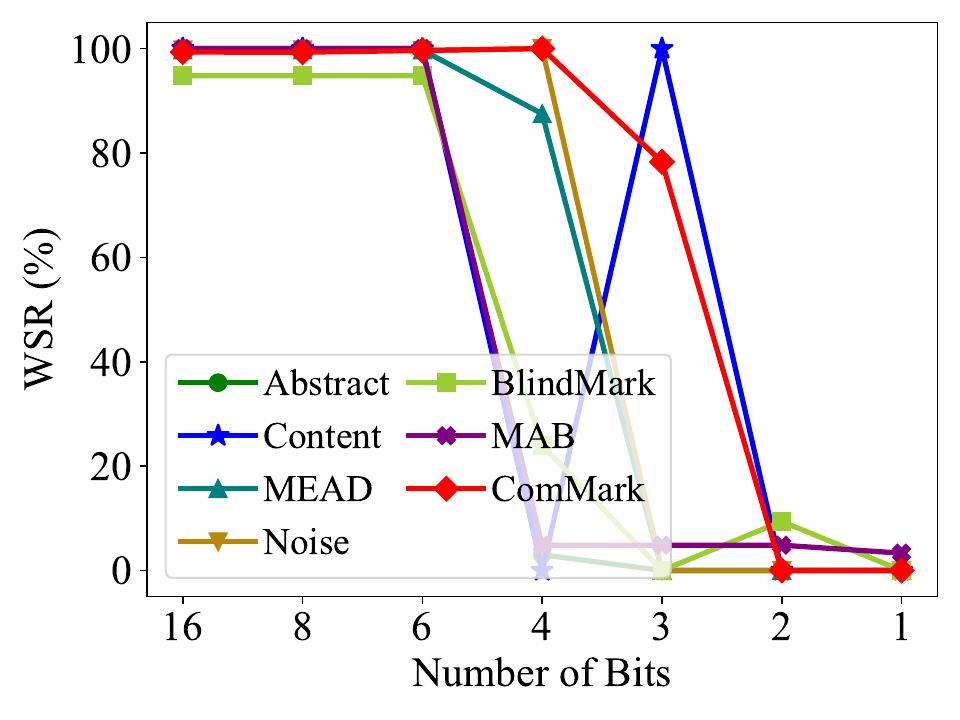}
        \caption{Model Quantization Attack}
        \label{fig: quantization_gtsrb_wsr}
    \end{subfigure}

    \caption{Comparison of robustness against watermark removal attacks. (On GTSRB)}
    \label{fig: robustness against removal attacks on gtsrb}
    \vspace{-0.0cm}
\end{figure*}

\begin{figure*}[t]
    \centering
    \vspace{-0.0cm}
    \begin{subfigure}{0.24\textwidth}
        \centering
        \includegraphics[width=1.0\textwidth]{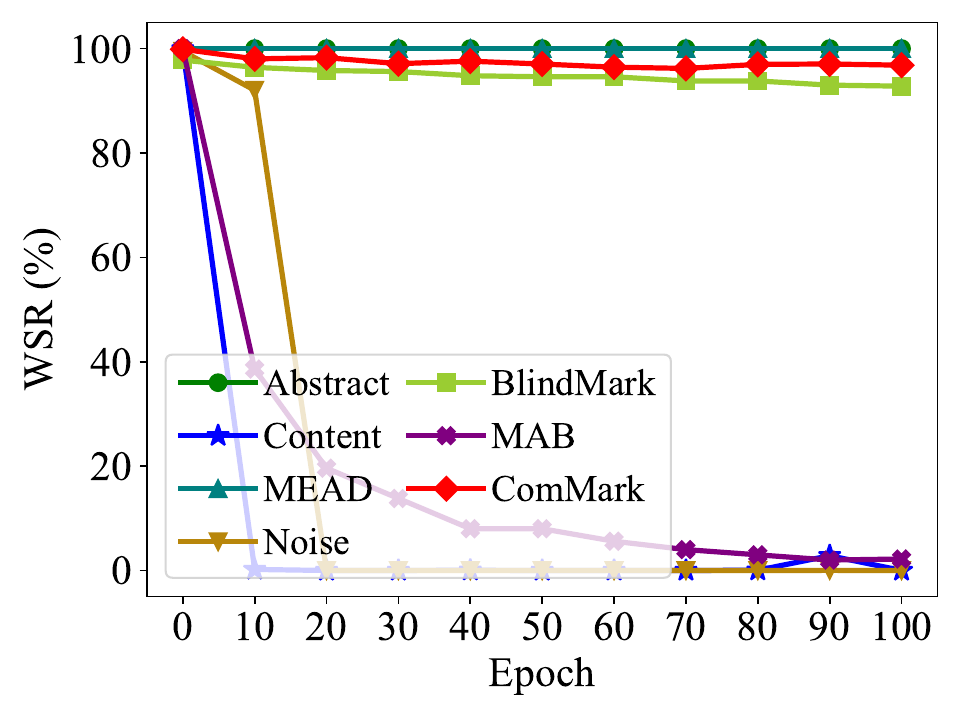}
        \caption{Model Finetuning Attack}
        \label{fig: finetune_cifar100_wsr}
    \end{subfigure}
    \begin{subfigure}{0.24\textwidth}
        \centering
        \includegraphics[width=1.0\textwidth]{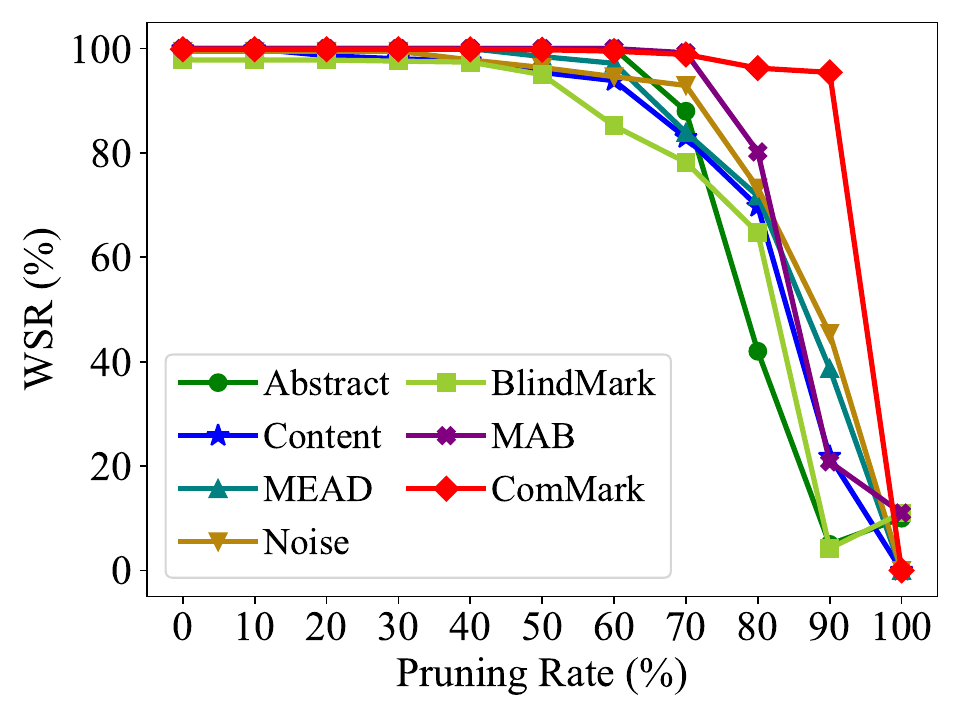}
        \caption{Model Pruning Attack}
        \label{fig: prune_cifar100_wsr}
    \end{subfigure}
    \begin{subfigure}{0.24\textwidth}
        \centering
        \includegraphics[width=1.0\textwidth]{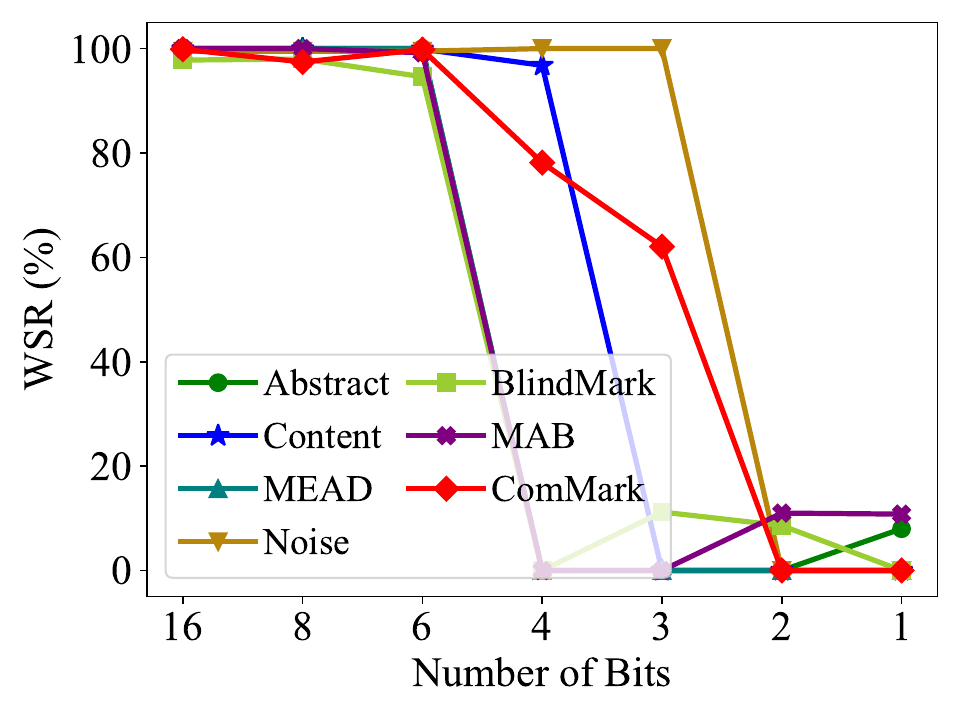}
        \caption{Model Quantization Attack}
        \label{fig: quantization_cifar100_wsr}
    \end{subfigure}

    \caption{Comparison of robustness against watermark removal attacks. (On CIFAR100)}
    \label{fig: robustness against removal attacks on cifar100}
    \vspace{-0.0cm}
\end{figure*}

\begin{figure*}[t]
    \centering
    \vspace{0.0cm}
    \begin{subfigure}{0.24\textwidth}
        \centering
        \includegraphics[width=1.0\textwidth]{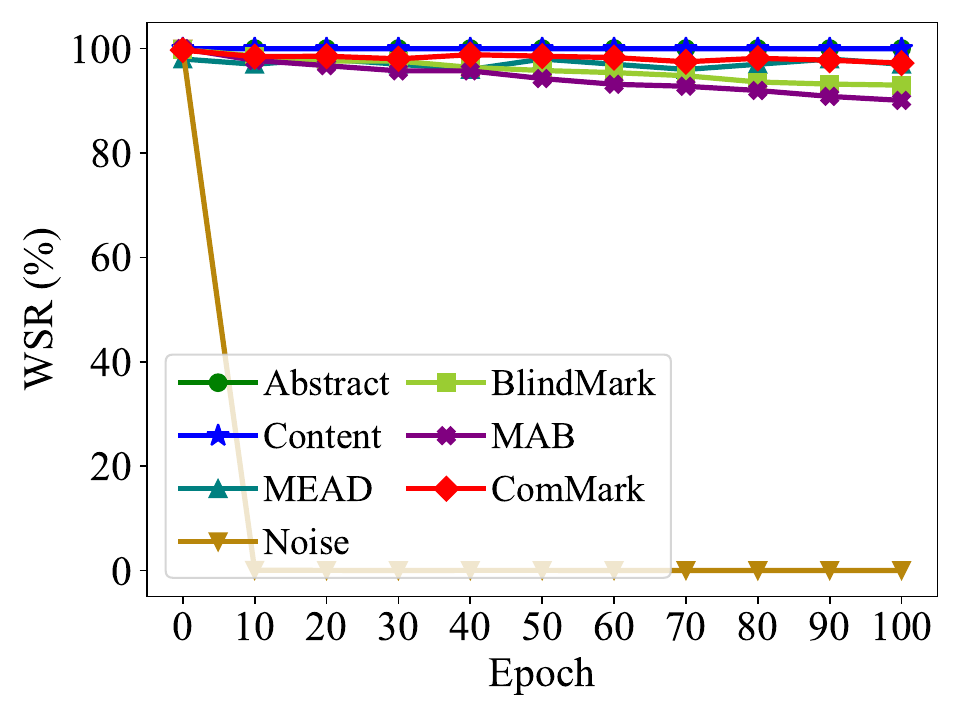}
        \caption{Model Finetuning Attack}
        \label{fig: finetune_vggface_wsr}
    \end{subfigure}
    \begin{subfigure}{0.24\textwidth}
        \centering
        \includegraphics[width=1.0\textwidth]{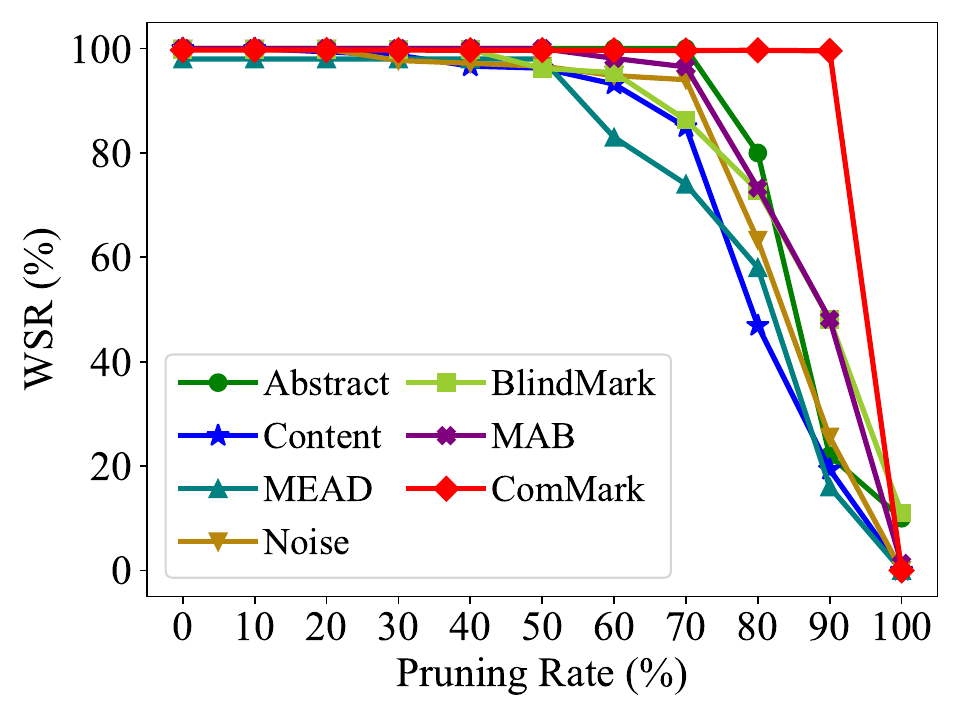}
        \caption{Model Pruning Attack}
        \label{fig: prune_vggface_wsr}
    \end{subfigure}
    \begin{subfigure}{0.24\textwidth}
        \centering
        \includegraphics[width=1.0\textwidth]{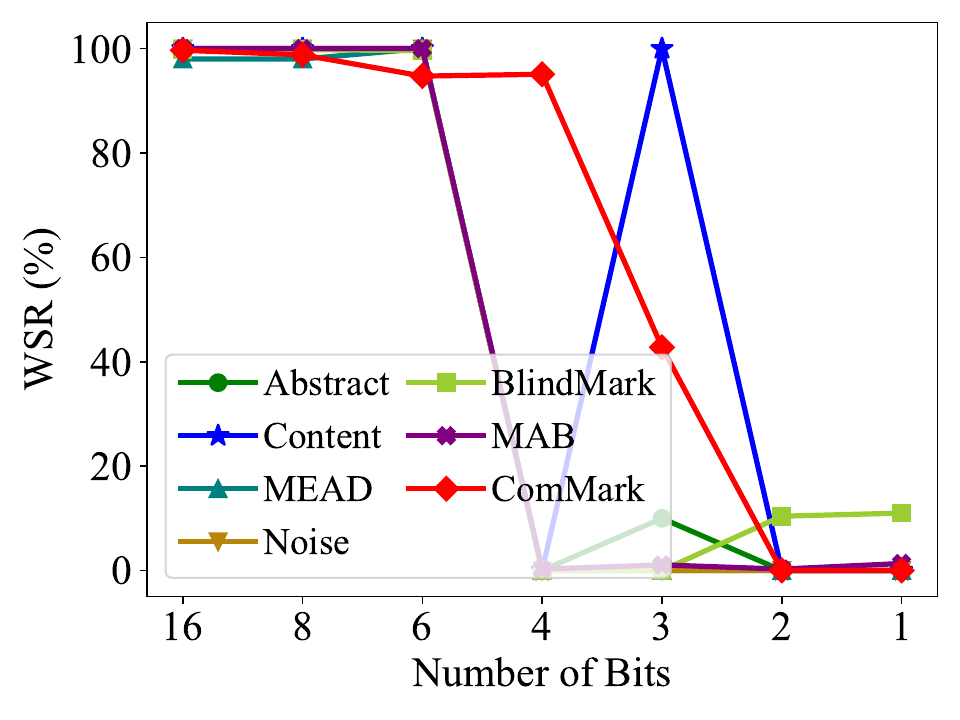}
        \caption{Model Quantization Attack}
        \label{fig: quantization_vggface_wsr}
    \end{subfigure}

    \caption{Comparison of robustness against watermark removal attacks. (On VGGFace)}
    \label{fig: robustness against removal attacks on vggface}
    \vspace{0.5cm}
\end{figure*}

\section{More Evaluation of Robustness against Model Extraction Attacks}
\label{sec: more evaluation against model extraction attacks}
\subsection{Model Extraction Attacks}
We briefly introduce the key ideas of popular extraction attacks here.

$\bullet$ Distillation \cite{hinton2015distilling}: This attack removes the hard loss term from the standard KL-divergence-based knowledge distillation loss, retaining only the soft loss term. In this scenario, the adversary has access to the original or similarly distributed primary task training data, and can query the victim model using these data while constraining the outputs of the victim and extracted models to be similar using KL divergence loss.

$\bullet$ JBDA \cite{papernot2017practical}: The adversary has access to a small set of original training data (seed samples) as the initial query set, and at each epoch, they expand the query set using Jacobian-based data augmentation technique for model extraction attack.

$\bullet$ Knockoff \cite{orekondy2019knockoff}: The adversary can only access publicly available substitute data for model queries. To improve query efficiency, the adversary uses a reinforcement learning strategy to construct a transfer set from a large pool of public data for the attack.

$\bullet$ Hard-Label: The three attack methods described above all operate in scenarios where the victim model outputs soft-label results. We adapt them to the hard-label output scenario to form three new attacks.

$\bullet$ Cross-Dataset: The adversary uses a query dataset with a different distribution from the victim's training set to perform the Knockoff attack.

$\bullet$ Cross-Architecture: The adversary uses an extracted model with a different model architecture from the victim model for the Knockoff attack.

$\bullet$ Cross-Dataset and Cross-Architecture: In this more challenging scenario, the adversary uses both a query dataset with a different distribution and an extracted model with a different architecture compared to the victim's.

\subsection{More Results}
The results in Table \ref{tab: Comparison of robustness against cross-dataset extraction attacks}, Table \ref{tab: Comparison of robustness against cross-architecture extraction attacks} and Table \ref{tab: Comparison of robustness against cross-dataset and cross-architecture extraction attacks} demonstrate the superior robustness of our proposed watermarking method even when it encounters the three challenging model extraction attacks, namely Cross-Dataset, Cross-Architecture, Cross-Dataset \& Cross-Architecture, respectively.

\section{More Evaluation of Robustness against Watermark Removal Attacks}
\label{sec: more evaluation against watermark removal attacks}
\subsection{Watermark Removal Attacks}
The principles of these attacks are as follows:

$\bullet$ Finetuning \cite{adi2018turning}: The adversary finetunes the watermarked model using a subset of labeled training data, but only updates the parameters of some layers of the network. In our evaluation, we finetune for 100 epochs and update only the parameters of the last layer.

$\bullet$ Pruning \cite{uchida2017embedding}: The adversary prunes a certain proportion of the smallest network weights and sets them to zero. We vary the pruning rate from 0\% to 100\%.

$\bullet$ Quantization \cite{lukas2022sok}: The adversary compresses all weights of the network into a lower bit representation to save storage resources. We decrease the number of bit from 16 to 1.

\subsection{More Results}
Figure \ref{fig: robustness against removal attacks on gtsrb}, Figure \ref{fig: robustness against removal attacks on cifar100} and Figure \ref{fig: robustness against removal attacks on vggface} show the results of our watermarking approach to resist removal attacks on three tasks, GTSRB, CIFAR100 and VGGFace, respectively. It can be seen that our embedded watermark is more difficult to be erased by current popular watermark removal attacks compared to previous black-box methods. Overall, our method has better robustness.

\section{Covertness against Watermark Detection Attacks}
In addition to demonstrating the covertness of our method through visual residuals and image similarity in the previous section, we now evaluate the covertness of various watermarking methods against watermark detection attacks. We adopt two popular anomaly detection methods, Local Outlier Factor (LOF) \cite{breunig2000lof} and Isolation Forest (IF) \cite{liu2008isolation}, to assess the covertness of watermarks. LOF identifies local outliers by comparing the density differences between data points and their neighbors, considering points with large density differences as anomalies. IF isolates data points by constructing multiple random trees, where points that are harder to isolate are considered anomalous.

The watermark detection results are shown in Table \ref{tab: Comparison of robustness against watermark detection attacks}. It can be seen that while maintaining an acceptable loss in primary task accuracy and false positive rates of clean samples, our watermarking method achieves a 0\% watermark detection rate across all datasets and anomaly detection methods, meaning that watermark samples are not detected as anomalous by the respective algorithms. Among previous watermarking methods, only Abstract achieves 0\% watermark detection rate across all scenarios, while other methods have some watermark samples detected as anomalous in certain scenarios.

\begin{table}[t]
    \centering
    \caption{Comparison of covertness (\%) against watermark detection attacks. AccLoss, FP, and DetW are abbreviations for the metrics Accuracy Loss, False Positive, and Detected Watermark, respectively.}
    \label{tab: Comparison of robustness against watermark detection attacks}
    \resizebox{0.48\textwidth}{!}{
    \begin{tabular}{llcccccc}
        \toprule
        \multirow{2}{*}{\textbf{Dataset}} & \multirow{2}{*}{\textbf{Method}} & \multicolumn{3}{c}{\textbf{LOF}} & \multicolumn{3}{c}{\textbf{IF}} \\
        \cmidrule(lr){3-5} \cmidrule(lr){6-8}
        & & AccLoss & FP & DetW$\downarrow$ & AccLoss & FP & DetW$\downarrow$ \\
        \midrule
        \multirow{7}{*}{GTSRB} & Abstract & 13.98 & 14.73 & \textbf{0.00} & 12.55 & 13.46 & \textbf{0.00} \\
        & Content & 13.91 & 14.73 & \textbf{0.00} & 12.71 & 13.46 & \textbf{0.00} \\
        & MEAD & 13.91 & 14.73 & 29.17 & 12.89 & 13.46 & \underline{10.69} \\
        & Noise	& 13.99 & 14.73 & \textbf{0.00} & 12.66 & 13.46 & \textbf{0.00} \\
        & BlindMark	& 13.86 & 14.77 & \underline{1.00} & 12.42 & 13.45 & \textbf{0.00} \\
        & MAB & 13.96 & 14.73 & 11.73 & 12.76 & 13.46 & 10.71 \\
        & \textbf{ComMark} & \cellcolor{gray!20}13.69 & \cellcolor{gray!20}14.73 & \cellcolor{gray!20}\textbf{0.00} & \cellcolor{gray!20}12.29 & \cellcolor{gray!20}13.46 & \cellcolor{gray!20}\textbf{0.00} \\
        \midrule
        \multirow{7}{*}{CIFAR10} & Abstract	& 8.53 & 10.22 & \textbf{0.00} & 10.94 & 13.43 & \textbf{0.00} \\
        & Content & 8.53 & 10.22 & 9.99 & 10.88 & 13.43 & \underline{12.13} \\
        & MEAD & 8.49 & 10.22 & \textbf{0.00} & 10.85 & 13.43 & \textbf{0.00} \\
        & Noise & 8.35 & 10.22 & 10.27 & 10.81 & 13.43 & 13.33 \\
        & BlindMark & 8.30 & 10.22 & \textbf{0.00} & 10.76 & 13.43 & \textbf{0.00} \\
        & MAB & 8.45 & 10.22 & \underline{9.80} & 10.97 & 13.43 & 13.40 \\
        & \textbf{ComMark} & \cellcolor{gray!20}8.13 & \cellcolor{gray!20}10.22 & \cellcolor{gray!20}\textbf{0.00} & \cellcolor{gray!20}10.60 & \cellcolor{gray!20}13.43 & \cellcolor{gray!20}\textbf{0.00} \\
        \midrule
        \multirow{7}{*}{CIFAR100} & Abstract & 5.66 & 7.64 & \textbf{0.00} & 9.58 & 12.30 & \textbf{0.00} \\
        & Content & 5.47 & 7.64 & \underline{7.23} & 9.25 & 12.30 & \underline{11.61} \\
        & MEAD & 5.38 & 7.64 & \textbf{0.00} & 9.22 & 12.30 & \textbf{0.00} \\
        & Noise	& 5.54 & 7.64 & 7.50 & 9.22 & 12.30 & 11.97 \\
        & BlindMark	& 5.60 & 7.64 & \textbf{0.00} & 9.21 & 12.30 & \textbf{0.00} \\
        & MAB & 5.56 & 7.64 & 9.80 & 9.29 & 12.30 & 14.40 \\
        & \textbf{ComMark} & \cellcolor{gray!20}5.23 & \cellcolor{gray!20}7.64 & \cellcolor{gray!20}\textbf{0.00} & \cellcolor{gray!20}9.02 & \cellcolor{gray!20}12.30 & \cellcolor{gray!20}\textbf{0.00} \\
        \midrule
        \multirow{7}{*}{VGGFace} & Abstract & 7.40 & 18.81 & \textbf{0.00} & 7.74 & 20.68 & \textbf{0.00} \\
        & Content & 7.04 & 18.81 & \textbf{0.00} & 7.38 & 20.68 & \textbf{0.00} \\
        & MEAD & 7.35 & 18.81 & 46.00 & 7.90 & 20.68 & 36.00 \\
        & Noise	& 6.93 & 18.81 & \textbf{0.00} & 7.51 & 20.68 & \textbf{0.00} \\
        & BlindMark & 7.13 & 18.81 & \textbf{0.00} & 7.66 & 20.68 & \underline{0.40} \\
        & MAB & 7.35 & 18.81 & \underline{13.10} & 7.85 & 20.68 & 12.30 \\
        & \textbf{ComMark} & \cellcolor{gray!20}6.95 & \cellcolor{gray!20}18.81 & \cellcolor{gray!20}\textbf{0.00} & \cellcolor{gray!20}7.46 & \cellcolor{gray!20}20.68 & \cellcolor{gray!20}\textbf{0.00} \\
        \bottomrule
    \end{tabular}
    }
\end{table}

\begin{figure}[t]
\centering
\includegraphics[width=0.4\textwidth]{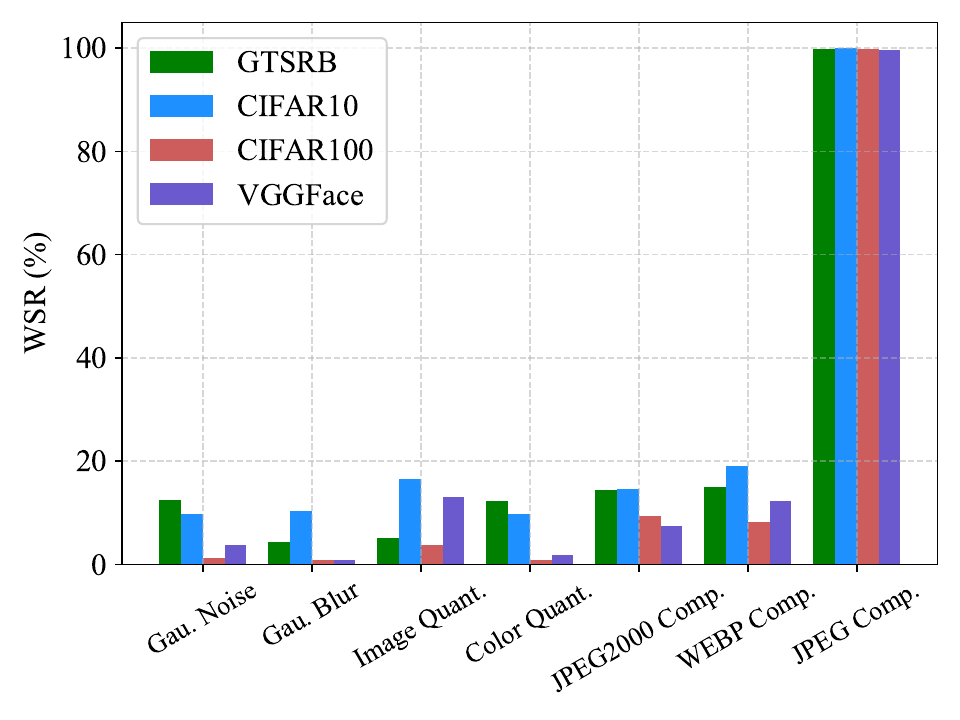}
\caption{Resistance to false watermark triggering conditions.}
\label{fig: uniqueness_false_trigger}
\vspace{-0.0cm}
\end{figure}

\begin{figure}[t]
\centering
\includegraphics[width=0.35\textwidth]{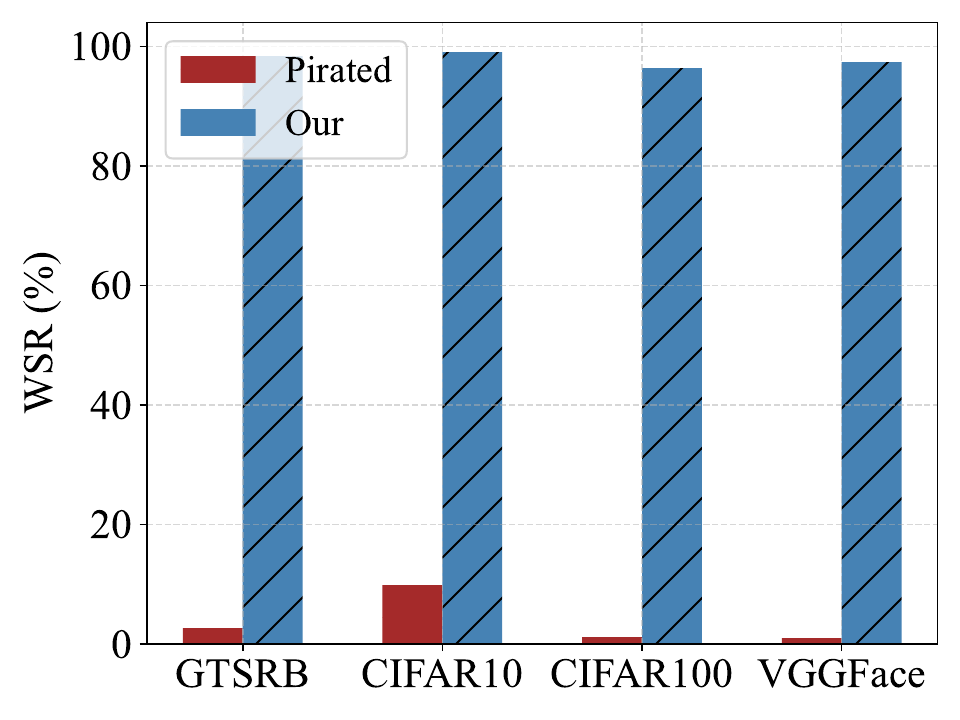}
\caption{Resistance to watermark ambiguity attacks.}
\label{fig: uniqueness_piracy_attack}
\end{figure}

\section{Uniqueness}
\label{sec: evaluation of uniqueness}
An ideal watermarking method, in addition to having sufficient effectiveness, covertness, and robustness, should also ensure that it is difficult for adversaries to trigger the watermark using other patterns, thereby preventing ownership disputes. We first investigate whether the watermarked model might mistakenly interpret some pixel-level artifacts as watermark triggering conditions. To explore this, we generate watermark samples using data preprocessing operations such as Gaussian Noise, Gaussian Blur, Image Quantization, Color Quantization, JPEG2000 Compression, and WEBP Compression, and test if these could lead to false watermark triggering. The results are shown in Figure \ref{fig: uniqueness_false_trigger}. We observe that only watermark samples generated using our JPEG Compression are able to trigger the watermark successfully, achieving over 99\% watermark success rate, while the success rate for other samples is very low, with the highest being only 18.99\%. This indicates that the watermarked model has learned the specific behavior condition of our JPEG compression, rather than pixel-level artifact features.

Additionally, we deploy a strong watermark ambiguity attack \cite{lv2024mea}, where the adversary first uses 20\% of the query data to embed their own watermark into the extracted model (with the watermark sample consisting of a fixed-colored square placed at the bottom right corner of clean samples), and then uses the remaining 80\% of the query data to perform Knockoff model extraction attack. We then inspect the survival of both the adversary's watermark and our watermark in the final extracted model. The results in Figure \ref{fig: uniqueness_piracy_attack} show that our watermark is still successfully verified with a very high success rate, while the adversary's pre-embedded watermark is easily removed after the model extraction attack.

In conclusion, our unique watermarking approach effectively resists false triggering and watermark ambiguity attacks, providing a stronger black-box technology for model copyright protection.

\begin{table*}[t]
    \centering
    \footnotesize
    \caption{Generalized performance of our watermarking method on various deep learning tasks. In the image generation and image caption tasks, we use SSIM and BLEU-4 metrics for evaluating the model performance, respectively, when the other tasks are evaluated using test accuracy.}
    \label{tab: Generalizability of the method}
    \resizebox{\textwidth}{!}{
    \begin{tabular}{
        m{2.8cm}
        m{1.3cm}<{\centering}
        m{2.0cm}<{\centering}
        m{1.6cm}<{\centering}
        m{1.2cm}<{\centering}
        m{1.9cm}<{\centering}
        m{1.6cm}<{\centering}
        m{1.6cm}<{\centering}
        m{1.7cm}<{\centering}}
        \toprule
        \multicolumn{2}{l}{\textbf{Task Domain} $\rightarrow$} & \textbf{Speech Command Recognition} & \textbf{Audio Scene Classification} & \textbf{Sentiment Analysis} & \textbf{News Topic Classification} & \textbf{Image Generation} & \textbf{Image Caption} & \textbf{Video Action Recognition} \\
        \midrule
        \multicolumn{2}{l}{\textbf{Data Modality} $\rightarrow$} & Audio & Audio & Text & Text & Image & Image+Text & Video \\
        \multicolumn{2}{l}{\textbf{Task Dataset} $\rightarrow$} & Speech Commands & UrbanSound8K & IMDB & AG News & Fashion MNIST & MSCOCO & UCF101 \\
        \multicolumn{2}{l}{\textbf{Model Architecture} $\rightarrow$} & M5 & VGG-like & GPT2 & Multi-Layer LSTM & AutoEncoder & ResNet50+LSTM & C3D \\
        \midrule
        \midrule
        \multirow{2}{*}{\textbf{Clean Model}} & Acc & 86.14\% & 80.26\% & 89.30\% & 88.82\% & 0.7358 & 0.1599 & 96.85\% \\
        & WSR & 2.57\% & 8.90\% & 48.80\% & 24.07\% & 0.2569 & 0.0711 & 0.85\% \\
        \midrule
        \multirow{2}{*}{\textbf{Watermarked Model}} & Acc & 83.04\% & 77.73\% & 88.00\% & 86.96\% & 0.7155 & 0.1512 & 92.07\% \\
        & WSR & 99.49\% & 97.82\%	& 99.40\% & 96.30\%	& 0.9707 & 0.9138 & 98.56\% \\
        \midrule
        \multirow{2}{*}{\textbf{Extracted Model}} & Acc & 80.51\% & 75.39\% & 85.70\% & 85.16\% & 0.7042 & 0.1386 & 88.45\% \\
        & WSR & 90.12\%	& 91.96\% & 72.70\%	& 73.00\% & 0.6831 & 0.7925 & 85.14\% \\
        \midrule
        \multirow{2}{*}{\textbf{Fine-tuned Model}} & Acc & 84.14\% & 81.63\% & 94.40\% & 89.14\%	& 0.7289 & 0.1453 & 96.70\% \\
        & WSR & 97.52\% & 95.71\% & 86.75\%	& 78.89\% & 0.6923 & 0.7915	& 94.15\% \\
        \midrule
        \textbf{Watermark Sample vs. Clean Sample} & Cosine Similarity & 0.9982 & 0.9905 & 0.9714 & 0.9588 & 0.9950 & 0.9936 & 0.9408 \\
        \bottomrule
    \end{tabular}
    }
    \vspace{-0.0cm}
\end{table*}

\begin{figure*}[t]
    \centering
    \vspace{-0.0cm}
    \begin{subfigure}{0.24\textwidth}
        \centering
        \includegraphics[width=1.0\textwidth]{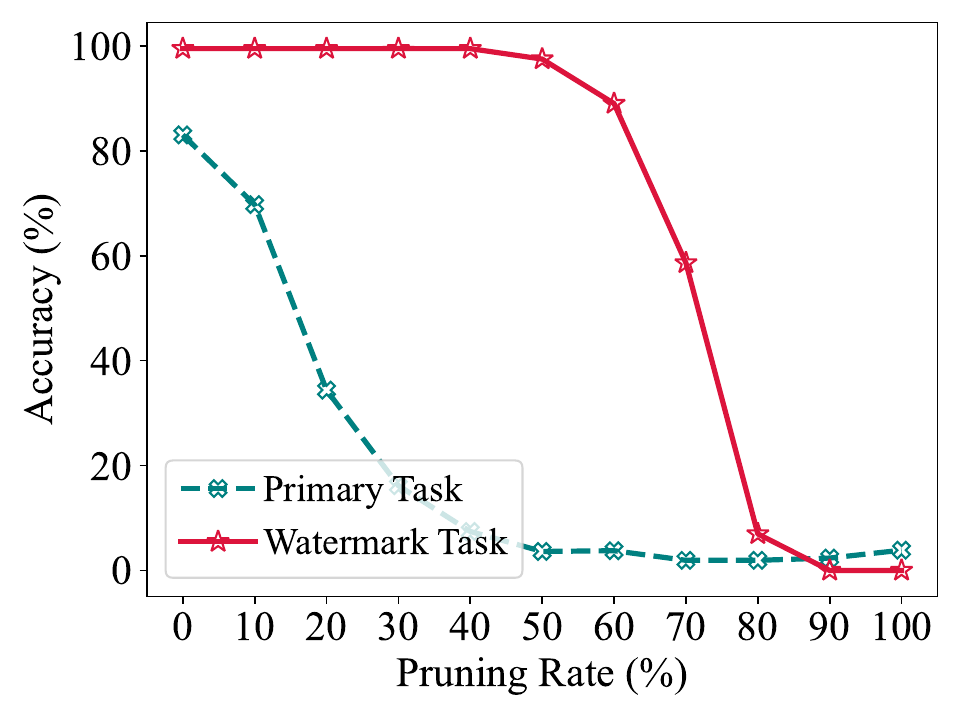}
        \caption{Speech Commands}
        \label{fig: prune_speech_commands}
    \end{subfigure}
    \begin{subfigure}{0.24\textwidth}
        \centering
        \includegraphics[width=1.0\textwidth]{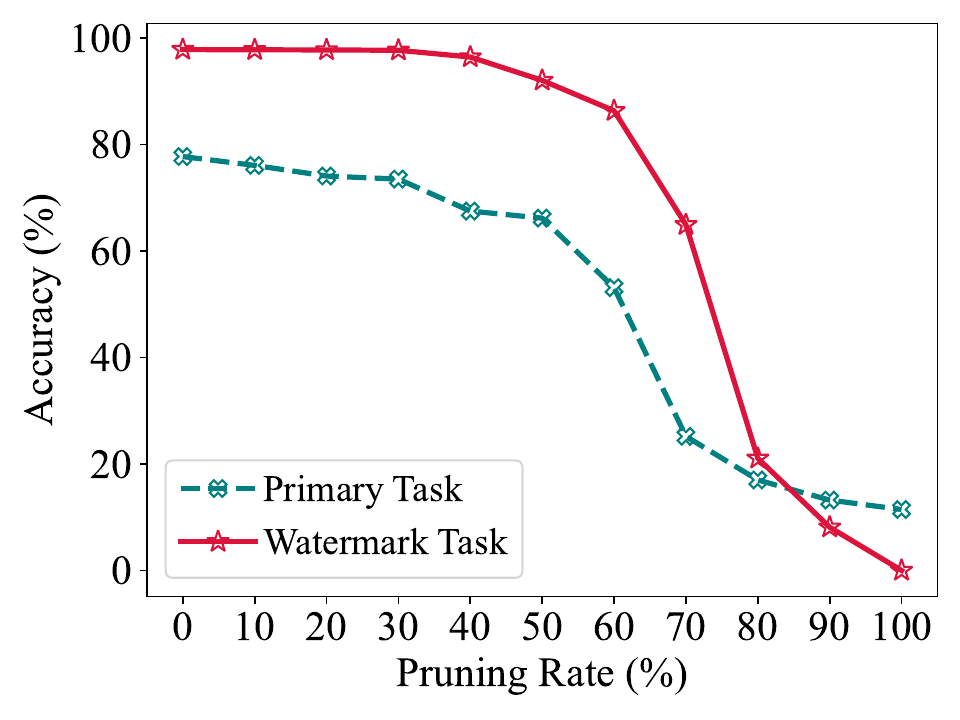}
        \caption{UrbanSound8K}
        \label{fig: prune_urbansound8K}
    \end{subfigure}
    \begin{subfigure}{0.24\textwidth}
        \centering
        \includegraphics[width=1.0\textwidth]{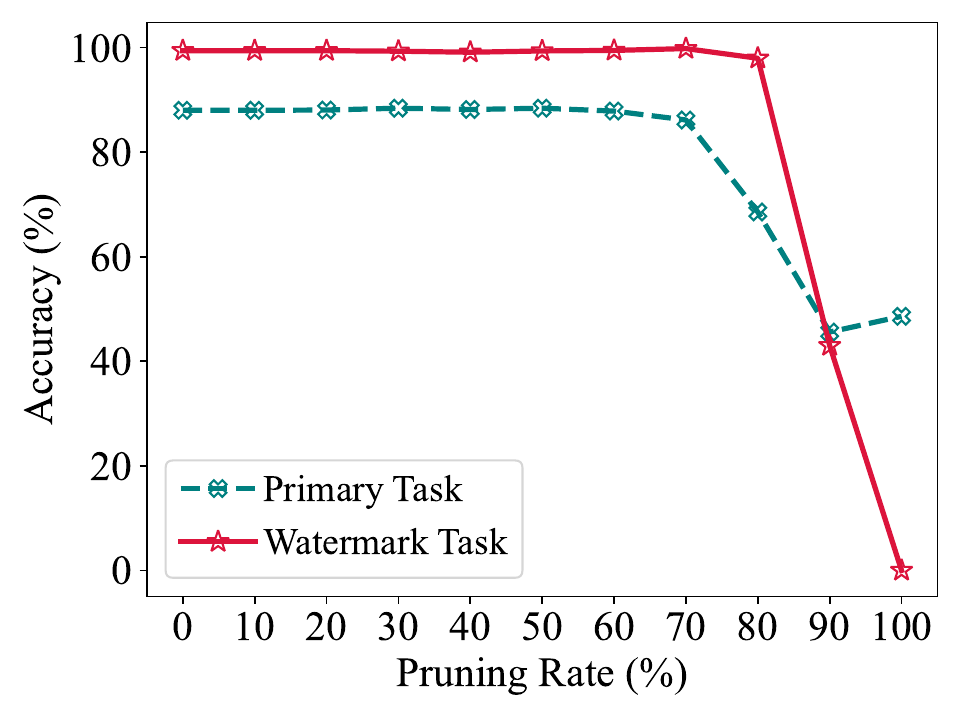}
        \caption{IMDB}
        \label{fig: prune_imdb}
    \end{subfigure}
    \begin{subfigure}{0.24\textwidth}
        \centering
        \includegraphics[width=1.0\textwidth]{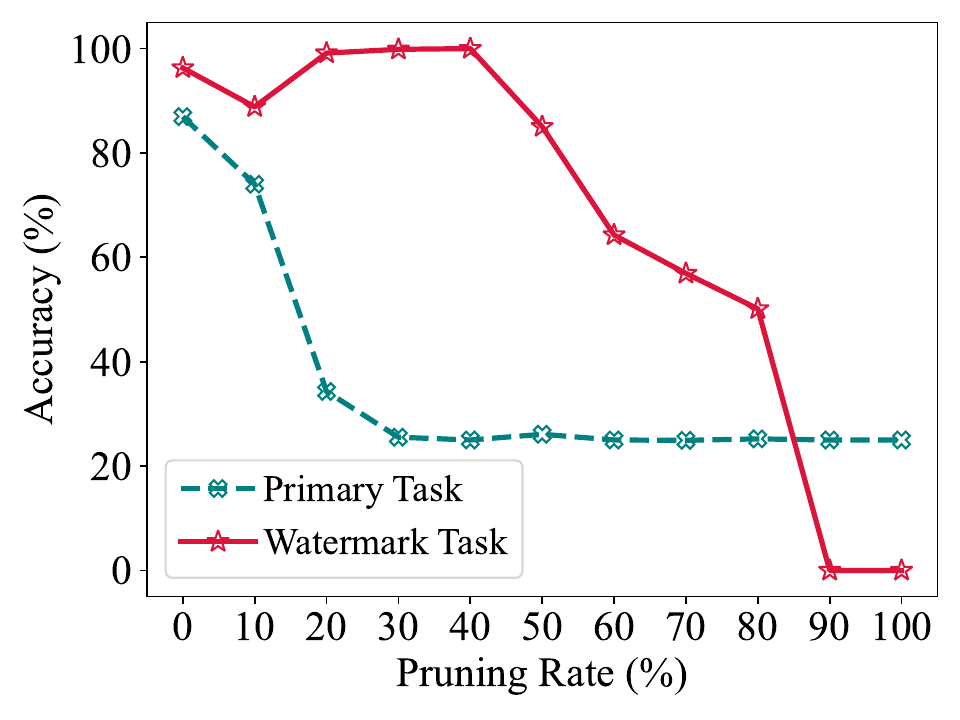}
        \caption{AG News}
        \label{fig: prune_ag_news}
    \end{subfigure}
    \vspace{0.2cm}

    \begin{subfigure}{0.24\textwidth}
        \centering
        \includegraphics[width=1.0\textwidth]{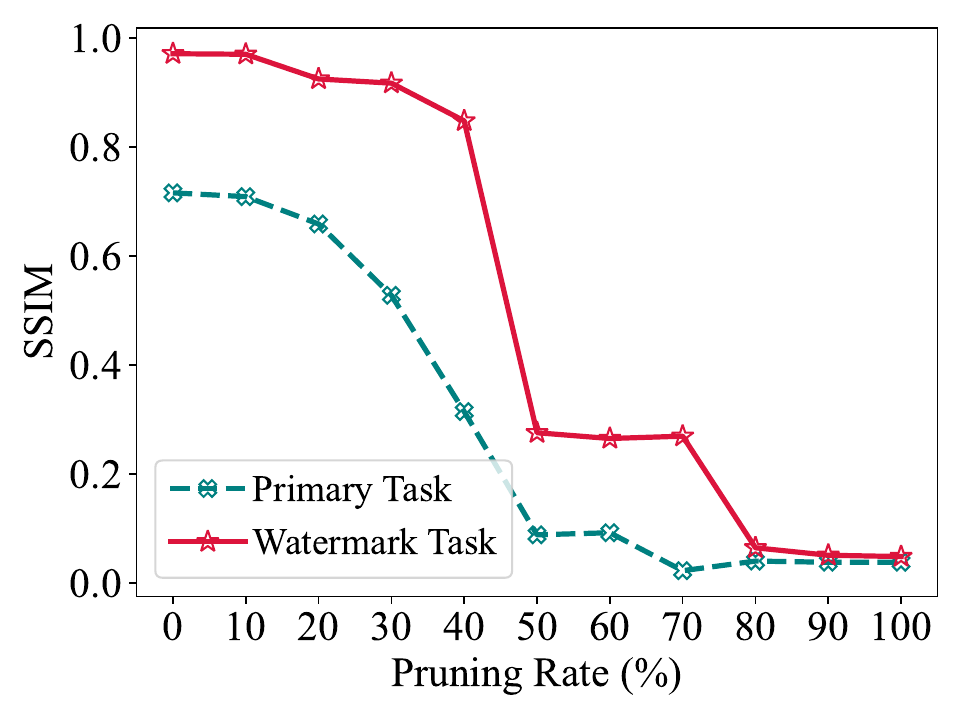}
        \caption{Fashion MNIST}
        \label{fig: prune_fashion_mnist}
    \end{subfigure}
    \begin{subfigure}{0.24\textwidth}
        \centering
        \includegraphics[width=1.0\textwidth]{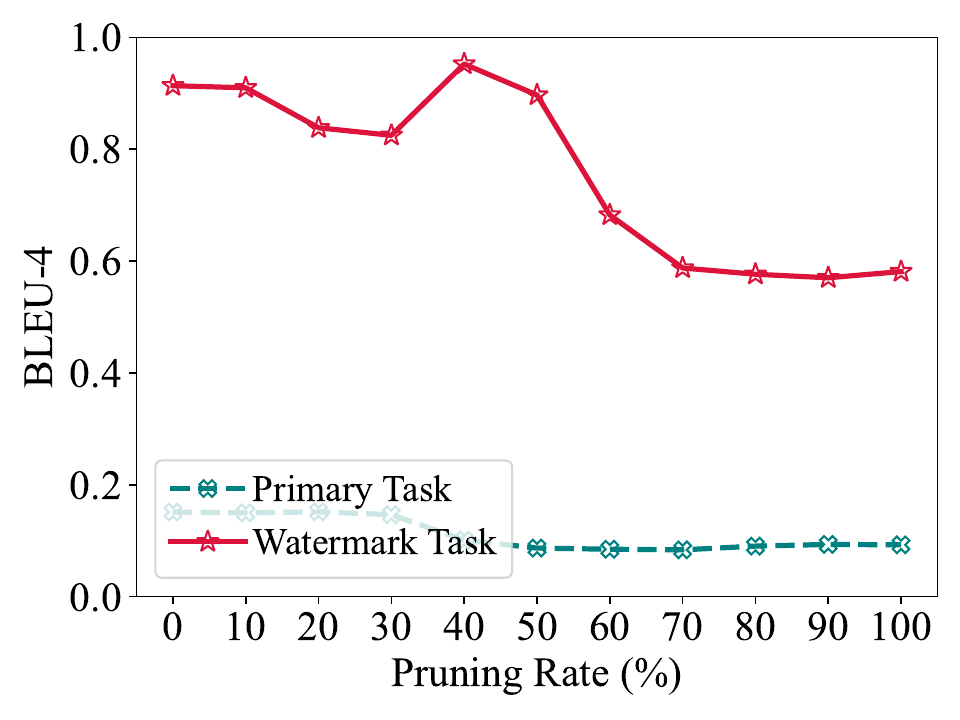}
        \caption{MSCOCO}
        \label{fig: prune_mscoco}
    \end{subfigure}
    \begin{subfigure}{0.24\textwidth}
        \centering
        \includegraphics[width=1.0\textwidth]{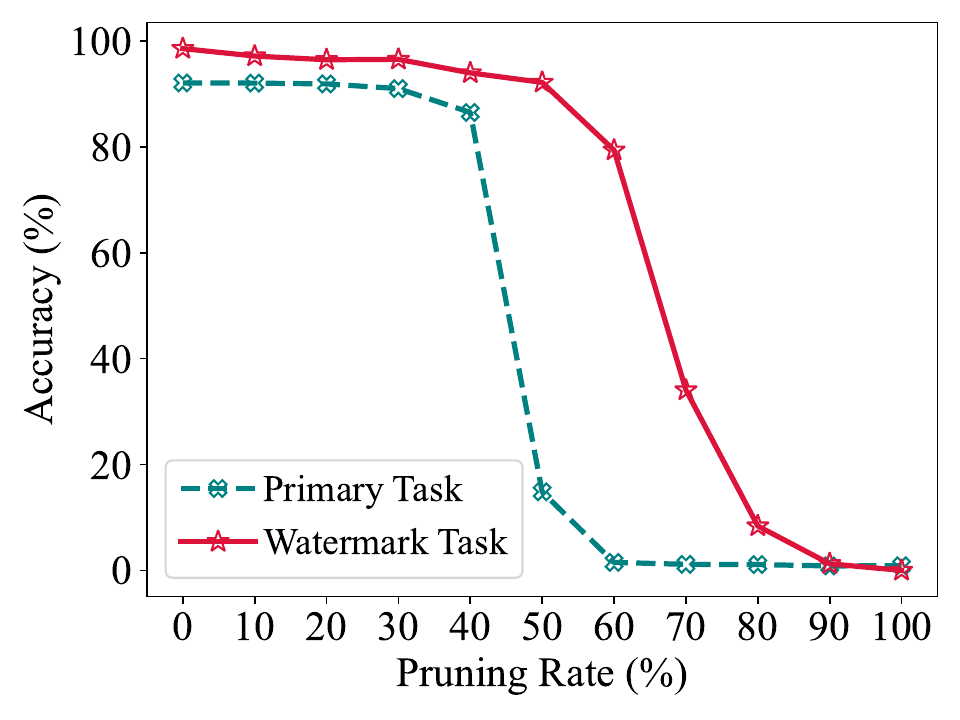}
        \caption{UCF101}
        \label{fig: prune_ucf101}
    \end{subfigure}
    \caption{Generalized robustness of our watermarking method against model pruning attacks on various deep learning tasks.}
    \label{fig: robustness against pruning on various tasks}
    \vspace{-0.0cm}
\end{figure*}

\section{Generalizability}
\label{sec: evaluation of generalizability}
Our watermarking method is generalizable and can be extended to various deep learning domains and data modalities. Specifically, when applying our method to tasks beyond image recognition, the main adjustment lies in selecting appropriate compression algorithms for different data modalities to construct the corresponding watermark samples. In this section, we extend our watermarking method to seven deep learning tasks: speech command recognition, audio scene classification, sentiment analysis, news topic classification, image generation, image captioning, and video action recognition. These tasks encompass four common data modalities: image, audio, video, and text. We first introduce the datasets and models, followed by the watermark sample construction methods for each data modality, and conclude with extensive experimental results.

\subsection{Datasets and Models}
The introduction of the datasets and models we used in this evaluation is as follows:

$\bullet$ Speech Commands \cite{warden2018speech}: A speech command recognition dataset containing 35 commands spoken by different individuals. The model we use is the M5 model from \cite{dai2017very}.

$\bullet$ UrbanSound8K \cite{salamon2014dataset}: An audio scene classification dataset consisting of 10 different environmental sounds from urban settings. The model we use is the VGG-like architecture (4Conv+1FC) from \cite{audio-scene-classification-model}.

$\bullet$ IMDB \cite{maas2011learning}: A sentiment analysis dataset containing a large collection of movie reviews classified as either positive or negative. We use the pre-trained GPT2 model \cite{radford2019language} as the initial architecture.

$\bullet$ AG News \cite{zhang2015character}: A news topic classification dataset containing four categories: World, Sports, Business, and Sci/Tec. We use a Multi-Layer LSTM \cite{news-topic-classification-model} as the model architecture.

$\bullet$ Fashion MNIST \cite{xiao2017fashion}: A dataset containing 10 categories of grayscale fashion images. This dataset serves as the target for our image generation task, with an AutoEncoder \cite{image-generation-model} as the generative model.

$\bullet$ MSCOCO \cite{lin2014microsoft}: A benchmark image captioning dataset where each image is paired with five different captions describing prominent entities and events. We use this dataset for the multimodal task of image captioning, and employ a model combination \cite{image-caption-model} of ResNet50 and LSTM.

$\bullet$ UCF101 \cite{soomro2012ucf101}: A video action recognition dataset consisting of 101 categories of human action video clips. We use C3D model \cite{video-action-recognition-model}, a 3D convolutional neural network designed for video data processing.

\subsection{Watermark Sample Construction Methods}
Image, audio, and video data are continuous, allowing for transformations from spatial to frequency domains, and there are numerous mature compression algorithms based on frequency domain transformations. Therefore, for these three types of data, we adopt the core steps of the JPEG \cite{wallace1991jpeg} (image compression), MP3 \cite{brandenburg1999mp3} (audio compression), and H264 \cite{richardson2004h} (video compression) algorithms to construct watermark samples. Specifically, for these data types, we first preprocess them by converting from the spatial to the frequency domain, then apply quantization-based compression in the frequency domain, and finally perform post-processing to restore the compressed data back to the spatial domain. This results in watermark samples that are both covert and robust, as they discard high-frequency information that is less perceptible to human eye and this discarding behavior is located in the global region of the image in the spatial domain.

However, text data is discrete and unsuitable for frequency-domain compression. To address this, we implement an approximate sample compression method. Specifically, we remove determiners, auxiliary verbs, coordinating conjunctions, and adjectives from the text through part-of-speech analysis \cite{asahara2000extended}, and the resulting text becomes the watermark sample. These words typically occupy only a small portion of the text and do not contribute significantly to the semantic core of the text. Therefore, their removal does not noticeably affect the original meaning, making this a covert form of approximate compression for text data.

\subsection{Evaluation Results}
Our experimental results are presented in Table \ref{tab: Generalizability of the method}, which includes the primary task accuracy and watermark success rates for clean models, watermarked models, extracted models, and fine-tuned models across various datasets. The performance of the watermarked models reflects the effectiveness and harmlessness of our method, while the performance of the extracted and fine-tuned models measures the robustness of the watermark. We observe that, even with the addition of our watermarking method, the accuracy drop is only 4.78\% on the UCF101 dataset, which has the largest impact on the primary task. Watermark success rates across all datasets exceed 90\%. In the case of model extraction attacks, the watermark success rate on the IMDB dataset, which is most severely affected by watermark removal, remains 72.70\%. During fine-tuning attacks, the most damaged watermark occurs in the Fashion MNIST image generation task, but it still retains an SSIM value of 0.6923, which is sufficient to assert model ownership. Furthermore, we compute the cosine similarity between clean samples and watermark samples to quantify watermark covertness. The cosine similarities across all datasets exceed 0.94, indicating a high similarity between watermark and original samples.

For robustness evaluation, we also observe the effect of our watermark under model pruning attacks, as shown in Figure \ref{fig: robustness against pruning on various tasks}. We find that only when the pruning rate is very high is our watermark completely removed, at which point the model's primary task performance is completely lost, rendering the model unusable. Therefore, it is infeasible for an adversary to remove our watermark without affecting model usability.

In summary, all the aforementioned results demonstrate that our method maintains consistently superb effectiveness, harmlessness, robustness, and covertness across different task domains and data modalities, providing strong evidence for the generalizability of our watermarking method.

\begin{figure*}[h]
    \centering
    \vspace{-0.0cm}
    \begin{subfigure}{0.24\textwidth}
        \centering
        \includegraphics[width=1.0\textwidth]{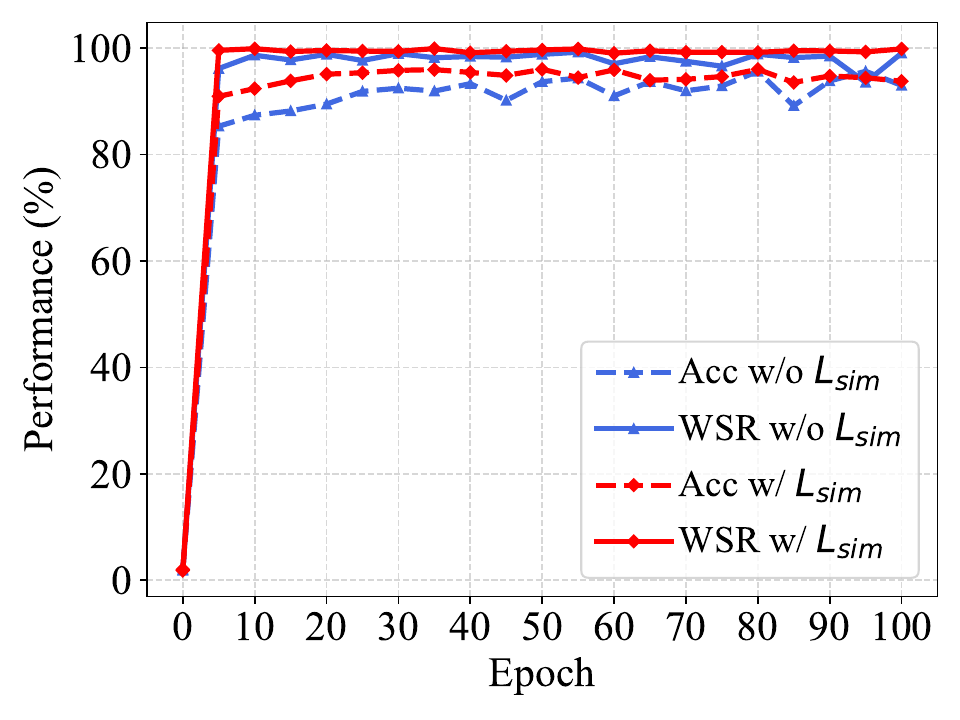}
        \caption{GTSRB}
        \label{fig: training_with_simloss_gtsrb}
    \end{subfigure}
    \begin{subfigure}{0.24\textwidth}
        \centering
        \includegraphics[width=1.0\textwidth]{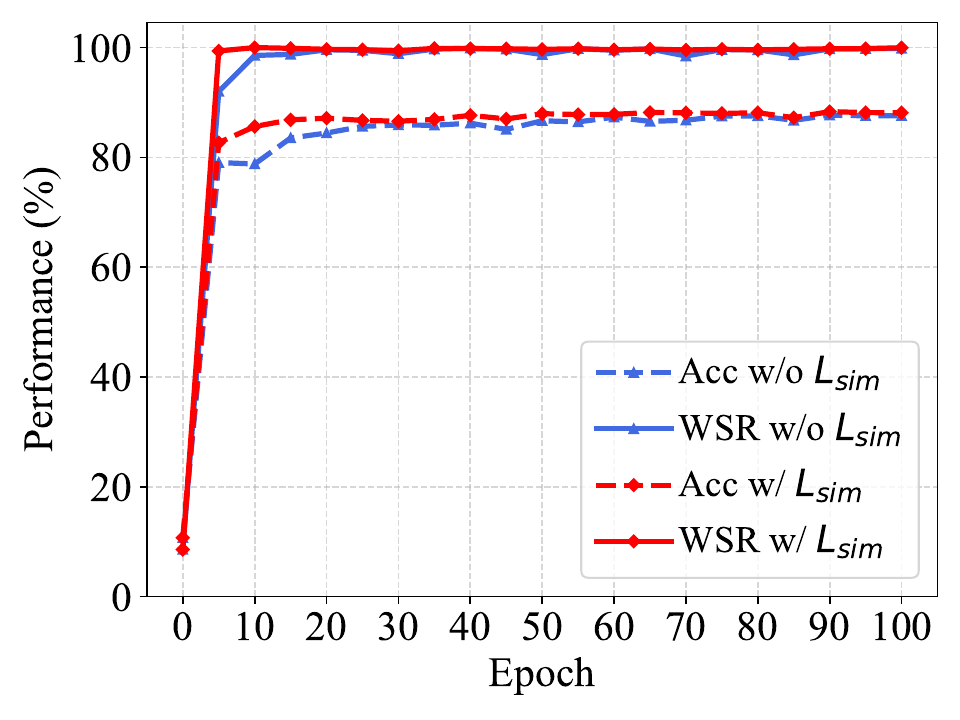}
        \caption{CIFAR10}
        \label{fig: training_with_simloss_cifar10}
    \end{subfigure}
    \begin{subfigure}{0.24\textwidth}
        \centering
        \includegraphics[width=1.0\textwidth]{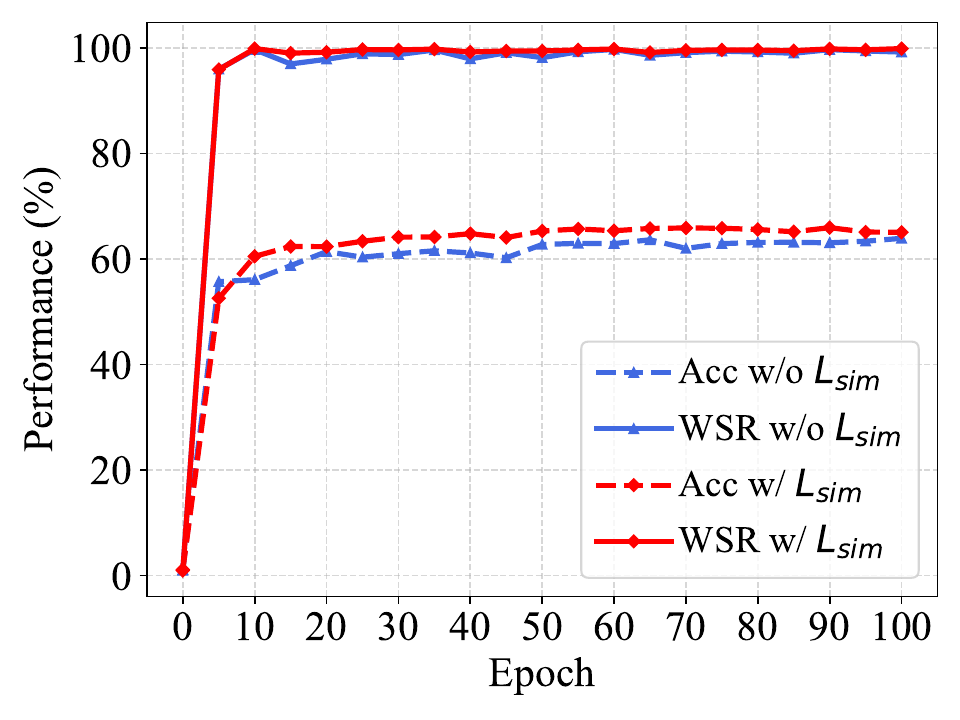}
        \caption{CIFAR100}
        \label{fig: training_with_simloss_cifar100}
    \end{subfigure}
    \begin{subfigure}{0.24\textwidth}
        \centering
        \includegraphics[width=1.0\textwidth]{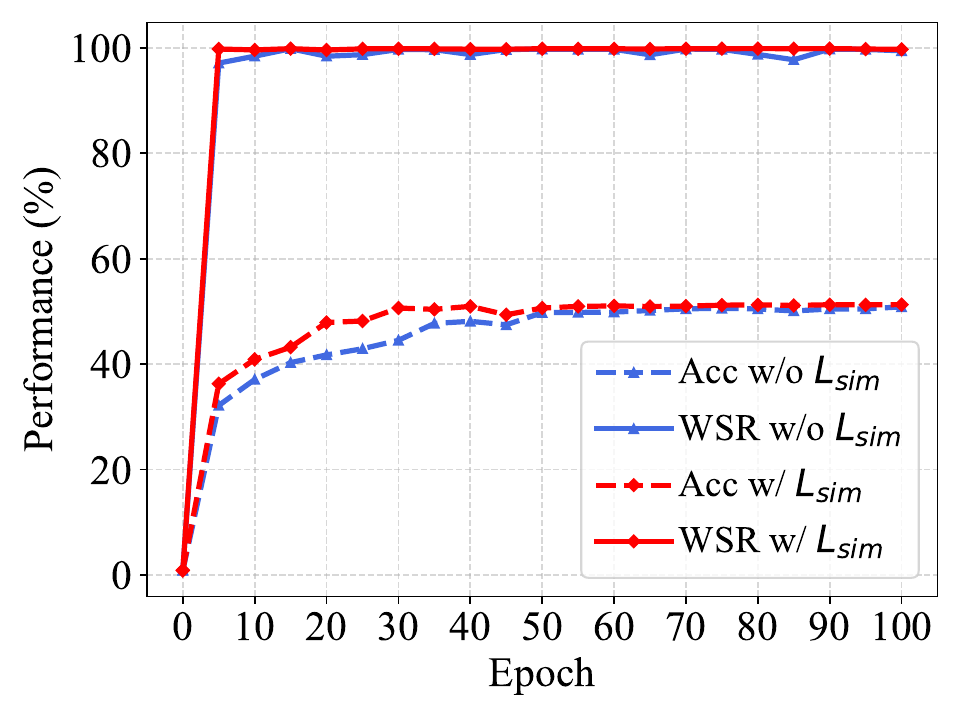}
        \caption{VGGFace}
        \label{fig: training_with_simloss_vggface}
    \end{subfigure}
    \caption{The change in model performance with (w) and without (w/o) similarity loss $L_{sim}$ as the training epochs increase.}
    \label{fig: training_with_simloss}
\end{figure*}

\begin{figure*}[t]
    \centering
    \vspace{-0.0cm}
    \begin{subfigure}{0.24\textwidth}
        \centering
        \includegraphics[width=1.0\textwidth]{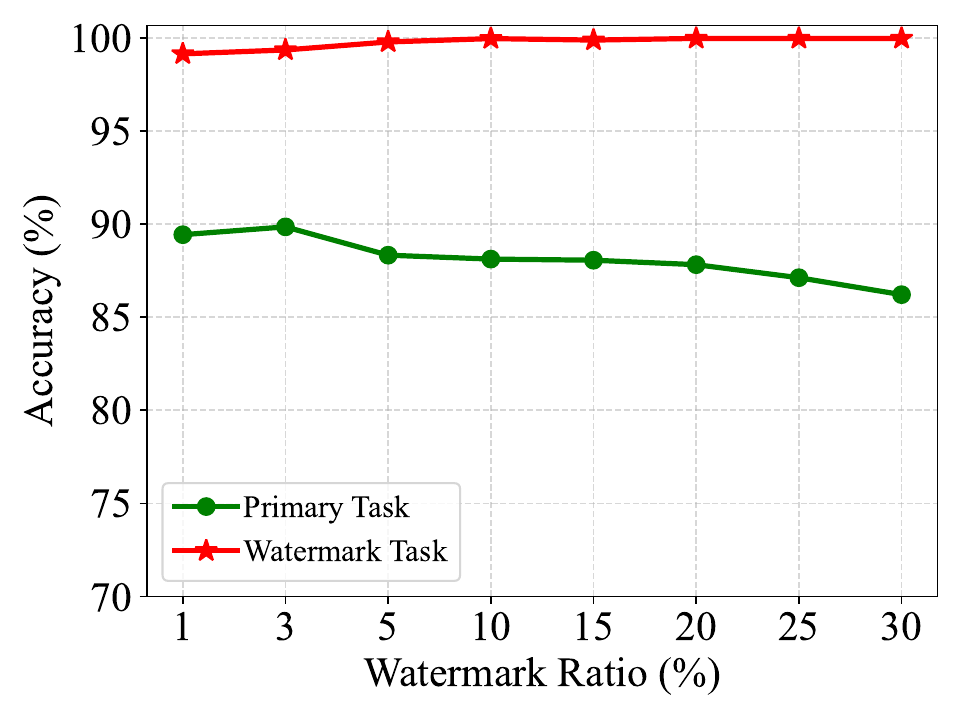}
        \caption{Effectiveness}
        \label{fig: watermark_ratio_effectiveness}
    \end{subfigure}
    \begin{subfigure}{0.24\textwidth}
        \centering
        \includegraphics[width=1.0\textwidth]{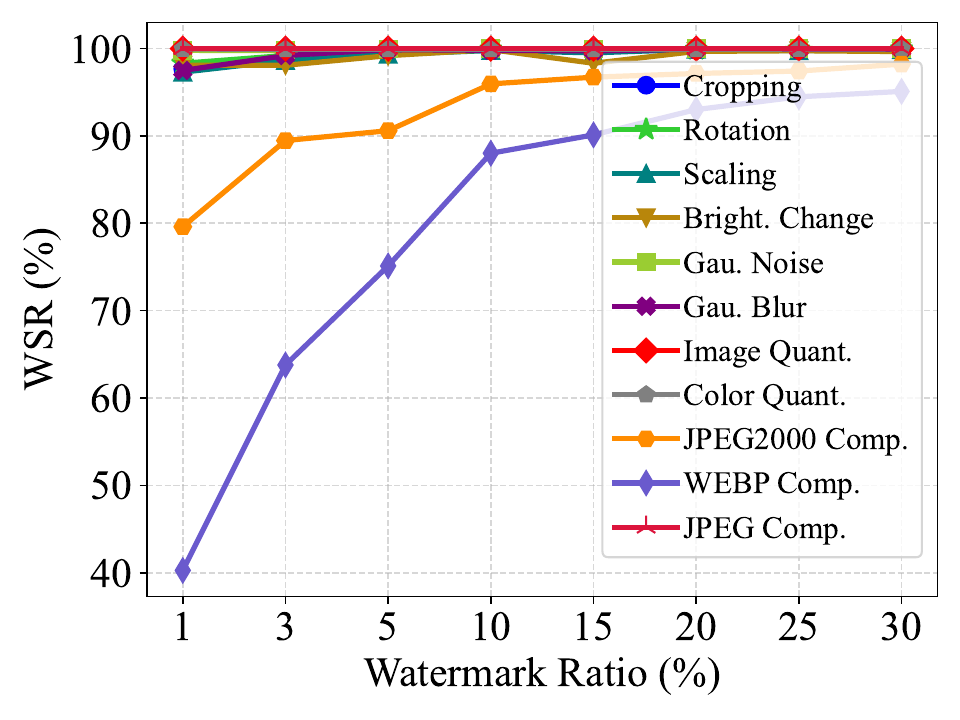}
        \caption{Robustness}
        \label{fig: watermark_ratio_robustness}
    \end{subfigure}
    \begin{subfigure}{0.24\textwidth}
        \centering
        \includegraphics[width=1.0\textwidth]{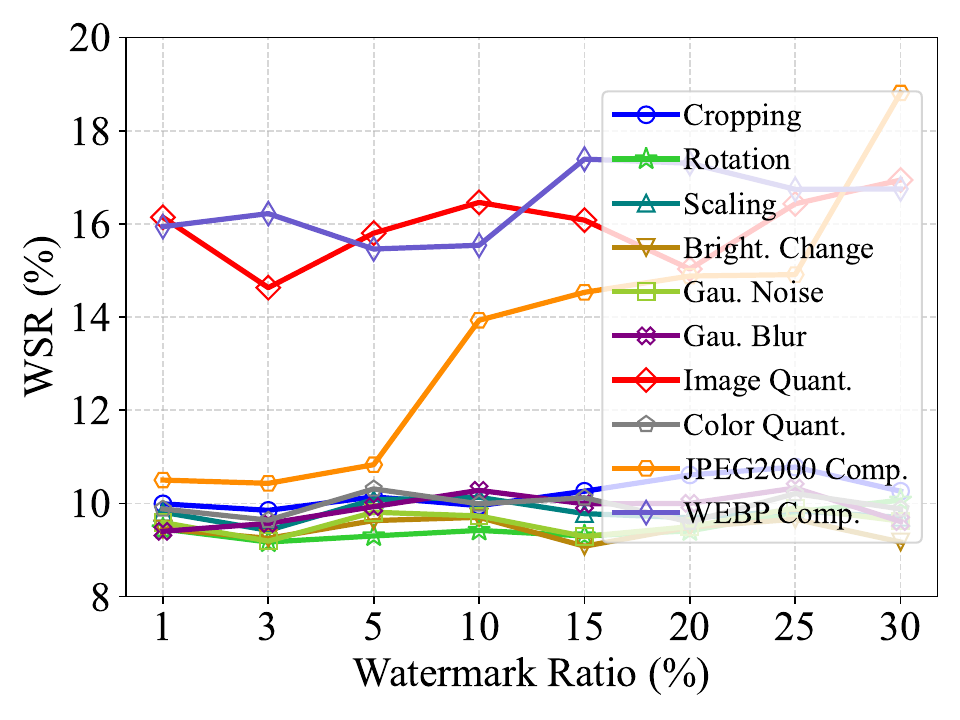}
        \caption{Uniqueness}
        \label{fig: watermark_ratio_uniqueness}
    \end{subfigure}
    \caption{The effect of different watermark sample rates.}
    \label{fig: watermark_ratio}
\end{figure*}

\begin{figure*}[t]
    \centering
    \vspace{-0.0cm}
    \begin{subfigure}{0.24\textwidth}
        \centering
        \includegraphics[width=1.0\textwidth]{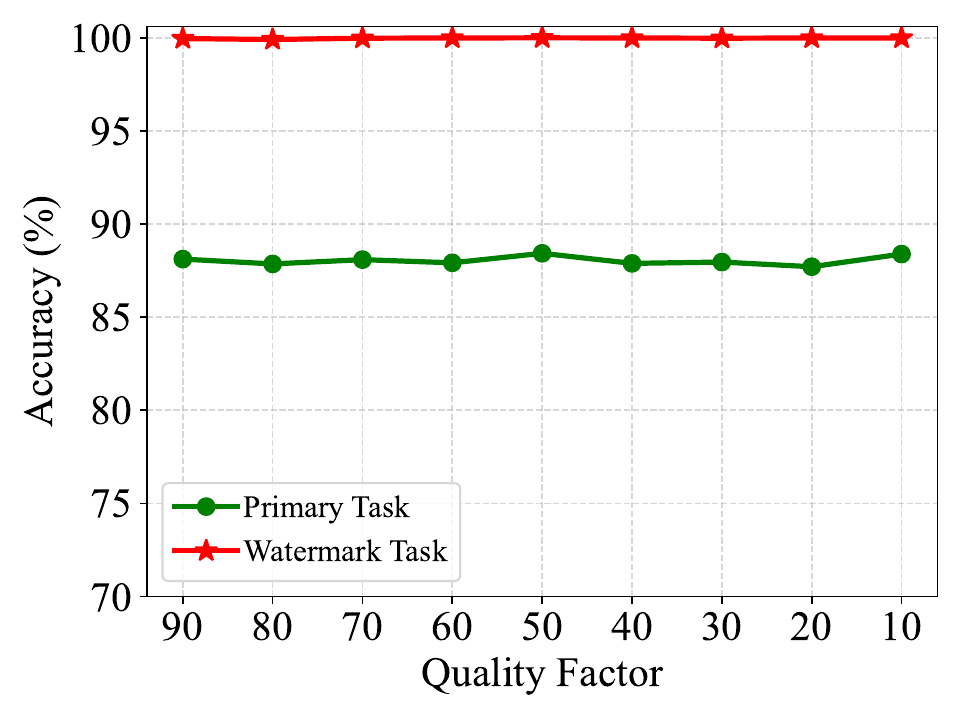}
        \caption{Effectiveness}
        \label{fig: quality_factor_effectiveness}
    \end{subfigure}
    \begin{subfigure}{0.24\textwidth}
        \centering
        \includegraphics[width=1.0\textwidth]{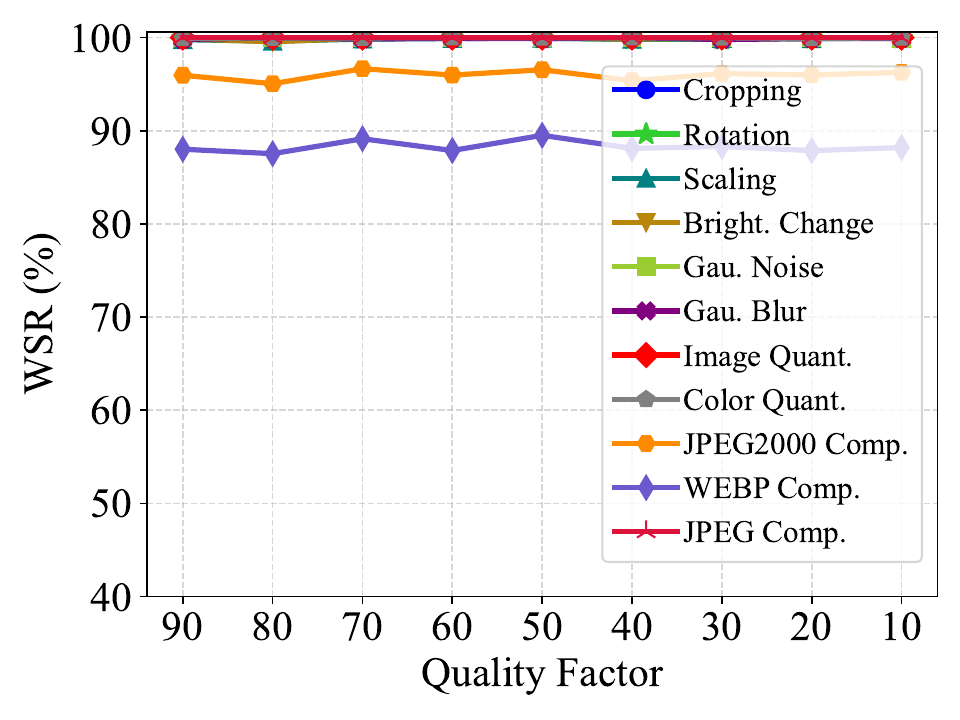}
        \caption{Robustness}
        \label{fig: quality_factor_robustness}
    \end{subfigure}
    \begin{subfigure}{0.24\textwidth}
        \centering
        \includegraphics[width=1.0\textwidth]{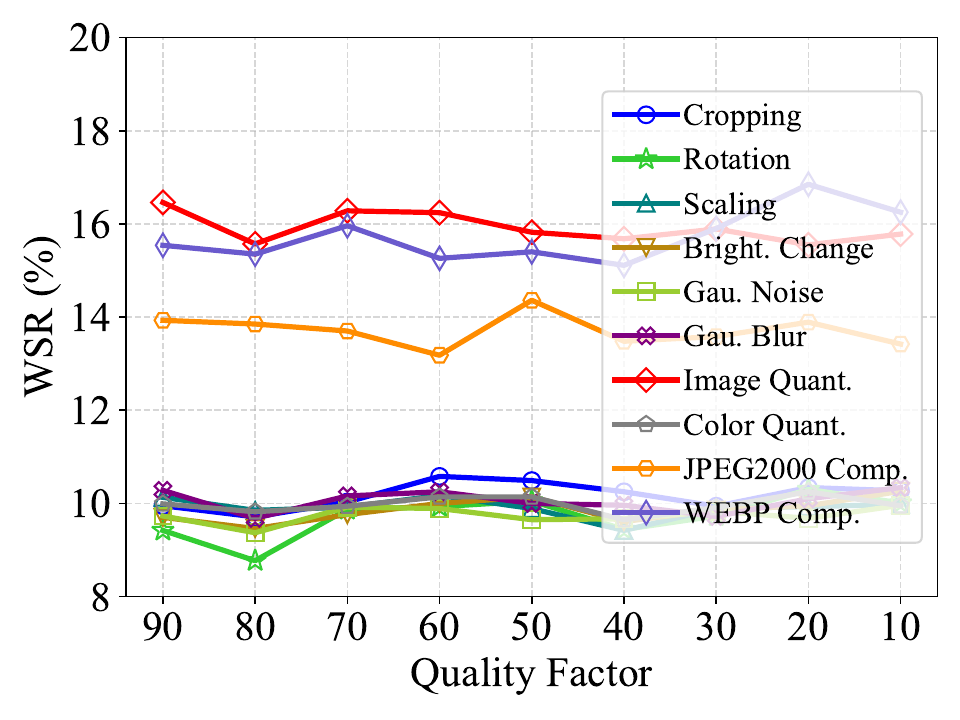}
        \caption{Uniqueness}
        \label{fig: quality_factor_uniqueness}
    \end{subfigure}
    \caption{The effect of different compression quality factors.}
    \label{fig: quality_factor}
\end{figure*}

\section{Additional Ablation Studies}
\label{sec: additional ablation studies}

\subsection{Effect of Watermark Sample Rate}
The watermark sample rate refers to the proportion of normal samples that are transformed into watermark samples. In Figure \ref{fig: watermark_ratio}, we demonstrate the performance of the watermarked model and its ability to resist evasion attacks and false triggering at different watermark sample rates. We observe that as the watermark sample rate increases, the primary task performance of the watermarked model gradually decreases, while the watermark success rate slightly increases. Meanwhile, the model's resistance to watermark evasion attacks improves significantly, with little change in its resistance to false triggering. These trends align with our expectations. Considering the balance between model's primary task performance and watermark robustness, we ultimately set the watermark sample rate to 10\%.

\subsection{Effect of Compression Quality Factor}
We also evaluate the impact of the JPEG compression quality factor in our watermarking algorithm on the final model performance, as well as its ability to resist evasion attacks and false triggering, with the results summarized in Figure \ref{fig: quality_factor}. We observe that as the compression quality factor decreases (i.e., as the degree of compression increases), there are no significant changes in the model's primary task accuracy, watermark success rate, or resistance to evasion attacks and false triggering. These results suggest that even slight compression can successfully embed the watermark, indicating that our watermarked model learns a compression behavior in frequency domain rather than specific pixel-level quantitative changes. A higher compression quality factor leads to a smaller difference between the watermark sample and the original clean sample, and considering covertness, we set the compression quality factor to 90.

\begin{figure*}[t]
    \centering
    \vspace{-0.0cm}
    \begin{subfigure}{1.0\textwidth}
        \centering
        \includegraphics[width=0.95\textwidth]{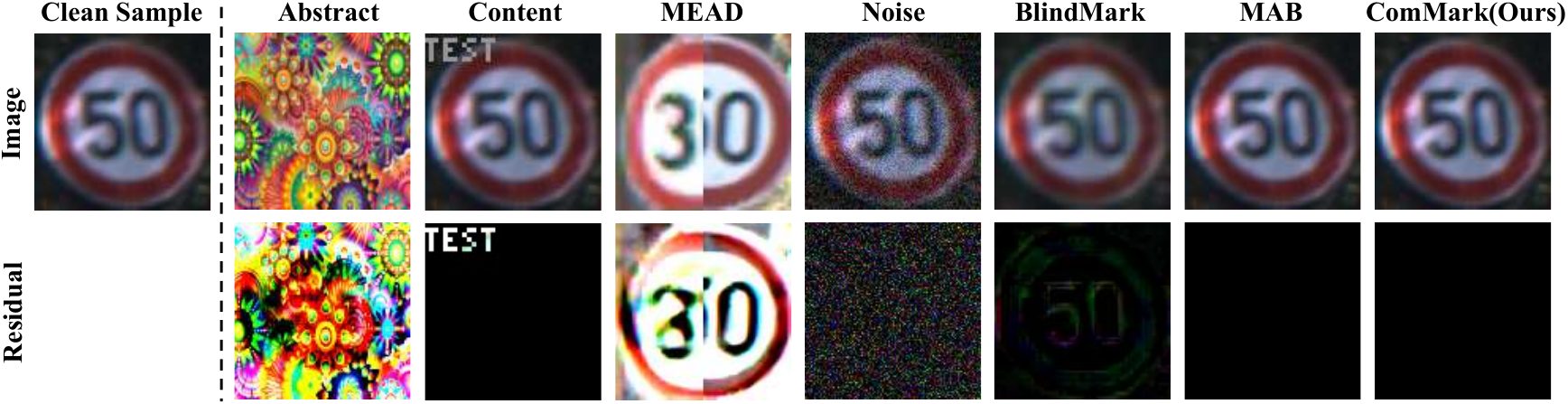}
        \caption{GTSRB}
        \label{fig: visual covertness gtsrb}
    \end{subfigure}
    \vspace{0.2cm}

    \begin{subfigure}{1.0\textwidth}
        \centering
        \includegraphics[width=0.95\textwidth]{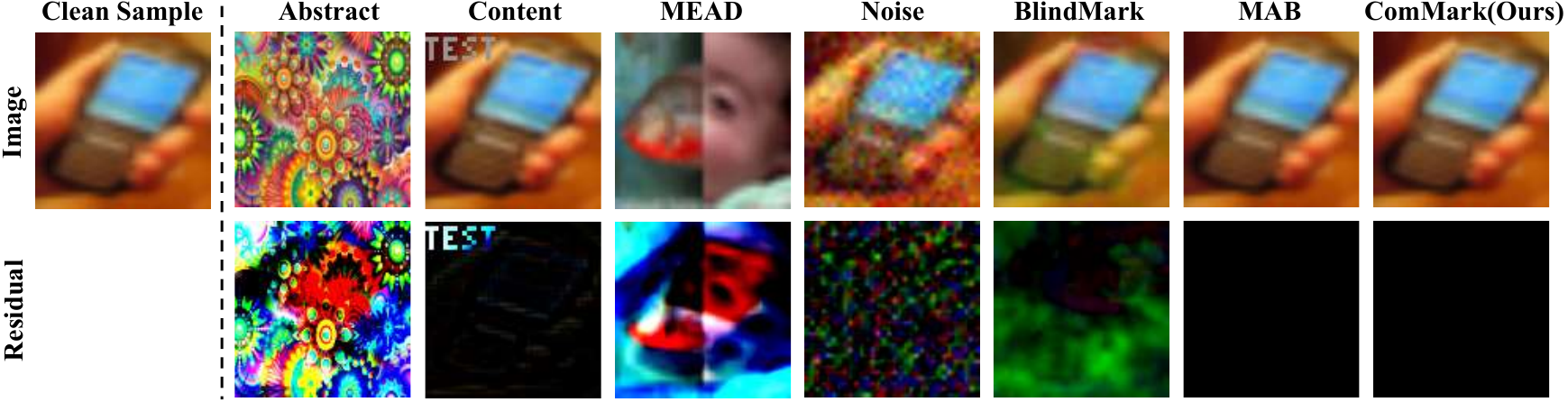}
        \caption{CIFAR100}
        \label{fig: visual covertness cifar100}
    \end{subfigure}
    \vspace{0.2cm}

    \begin{subfigure}{1.0\textwidth}
        \centering
        \includegraphics[width=0.95\textwidth]{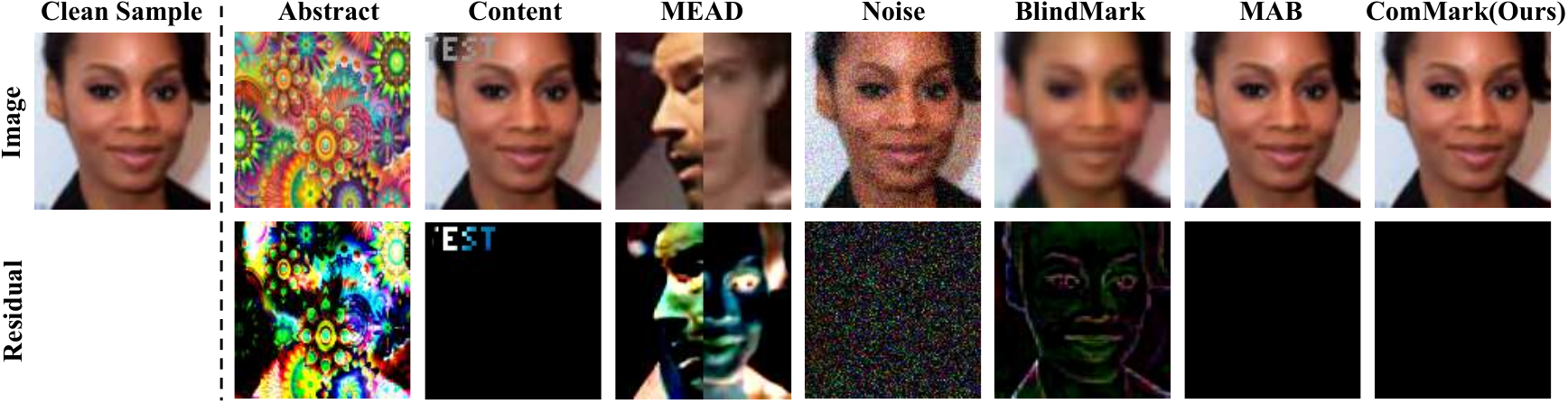}
        \caption{VGGFace}
        \label{fig: visual covertness vggface}
    \end{subfigure}
    \caption{Visual comparison of covertness of different watermarking methods. (On GTSRB, CIFAR100 and VGGFace)}
    \label{fig: visual comparison of covertness on more datasets}
    \vspace{-0.0cm}
\end{figure*}

\begin{figure*}[t]
    \centering
    \vspace{-0.0cm}
    \begin{subfigure}{0.24\textwidth}
        \centering
        \includegraphics[width=1.0\textwidth]{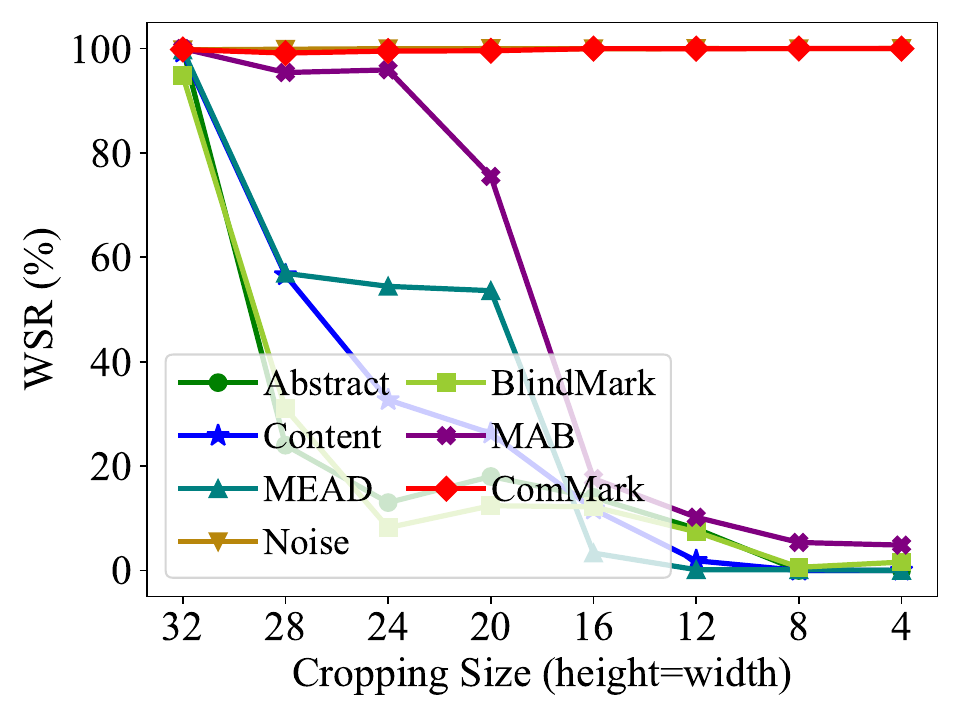}
        \caption{Cropping}
        \label{fig: crop_gtsrb_wsr}
    \end{subfigure}
    \begin{subfigure}{0.24\textwidth}
        \centering
        \includegraphics[width=1.0\textwidth]{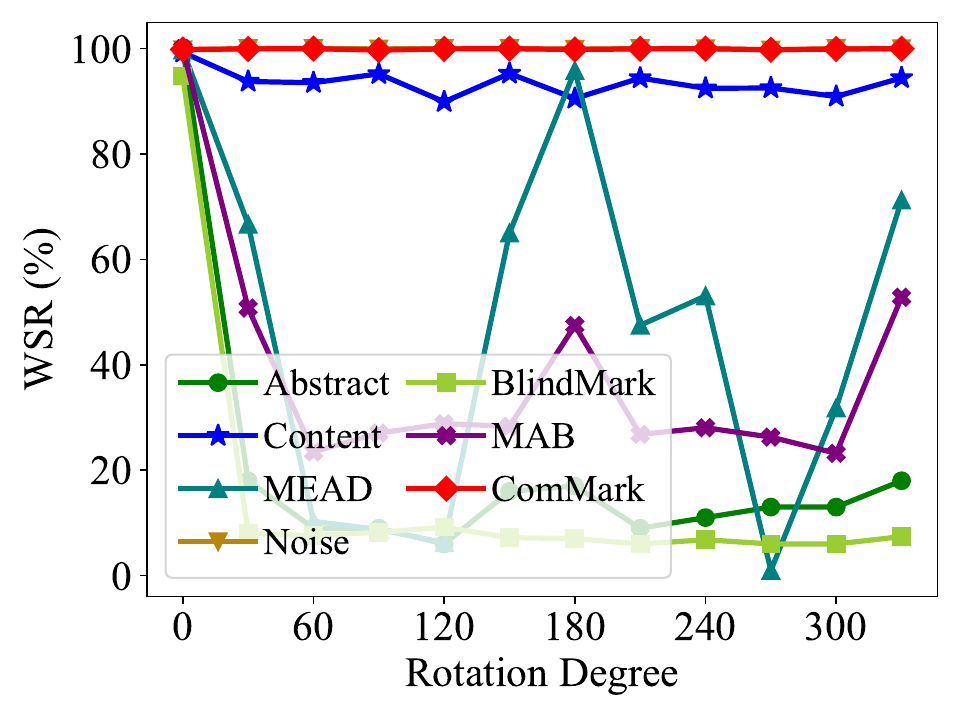}
        \caption{Rotation}
        \label{fig: rotate_gtsrb_wsr}
    \end{subfigure}
    \begin{subfigure}{0.24\textwidth}
        \centering
        \includegraphics[width=1.0\textwidth]{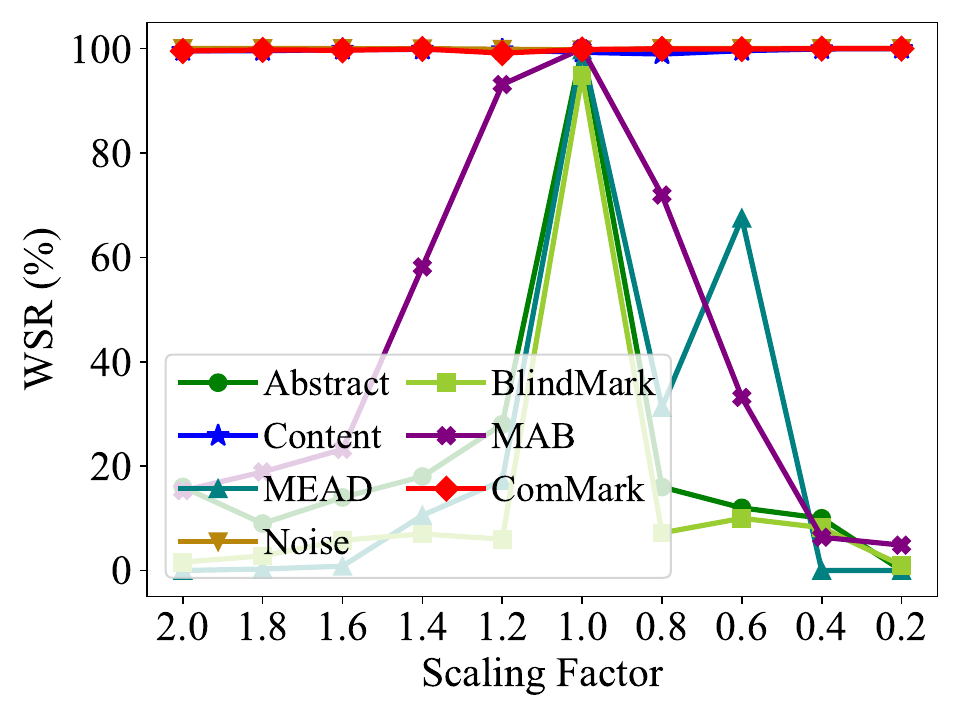}
        \caption{Scaling}
        \label{fig: scale_gtsrb_wsr}
    \end{subfigure}
    \begin{subfigure}{0.24\textwidth}
        \centering
        \includegraphics[width=1.0\textwidth]{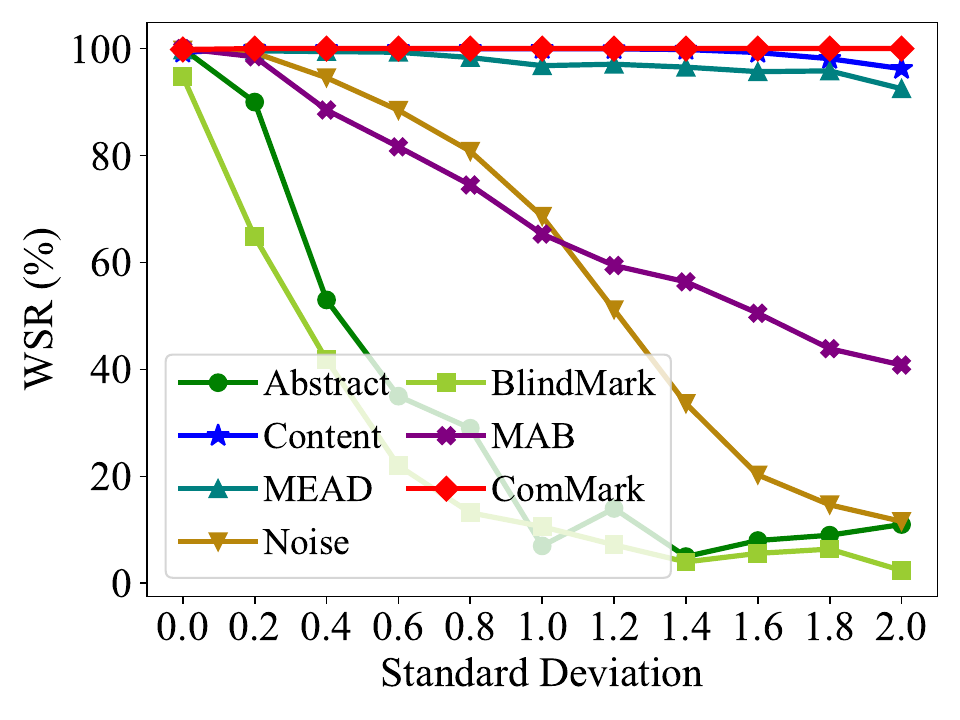}
        \caption{Gaussian Noise}
        \label{fig: gaussian_noise_gtsrb_wsr}
    \end{subfigure}
    \vspace{0.3cm}

    \begin{subfigure}{0.24\textwidth}
        \centering
        \includegraphics[width=1.0\textwidth]{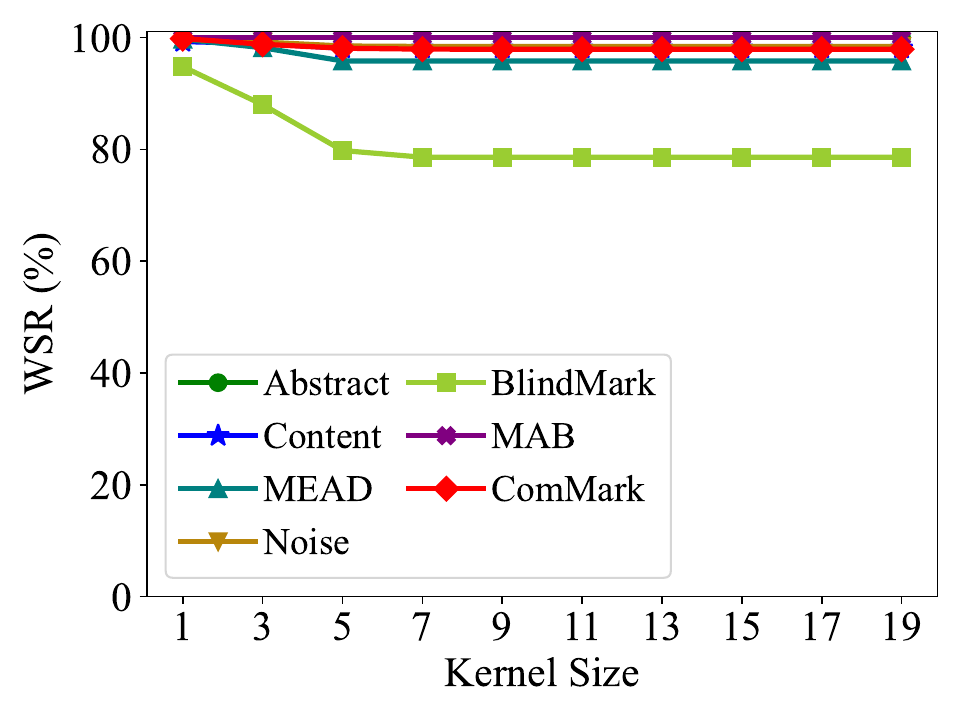}
        \caption{Gaussian Blur}
        \label{fig: gaussian_blur_gtsrb_wsr}
    \end{subfigure}
    \begin{subfigure}{0.24\textwidth}
        \centering
        \includegraphics[width=1.0\textwidth]{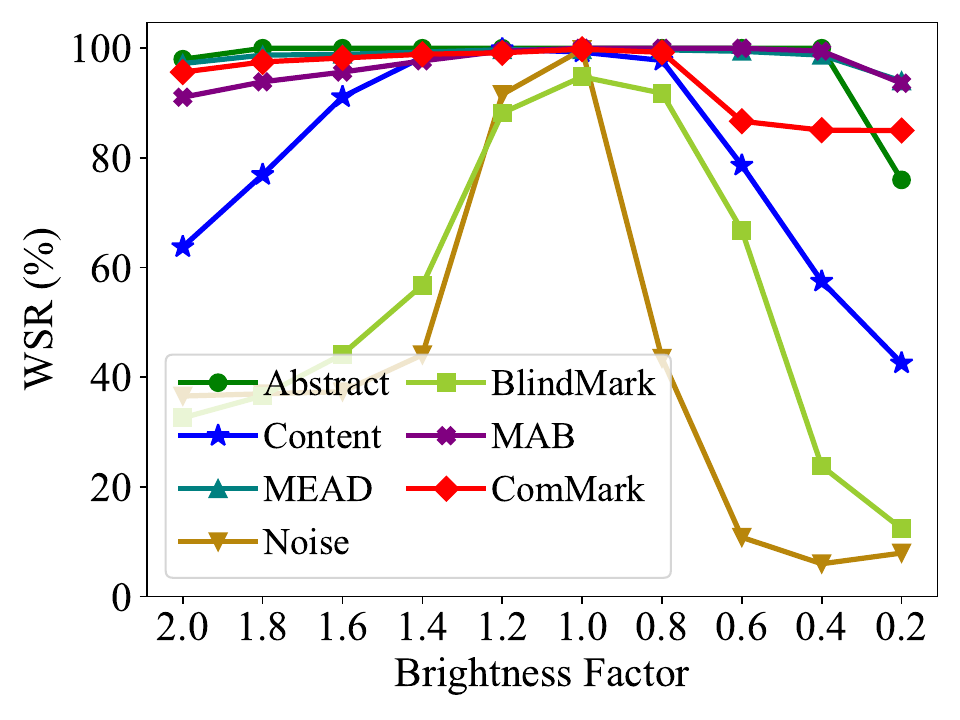}
        \caption{Brightness Change}
        \label{fig: brightness_change_gtsrb_wsr}
    \end{subfigure}
    \begin{subfigure}{0.24\textwidth}
        \centering
        \includegraphics[width=1.0\textwidth]{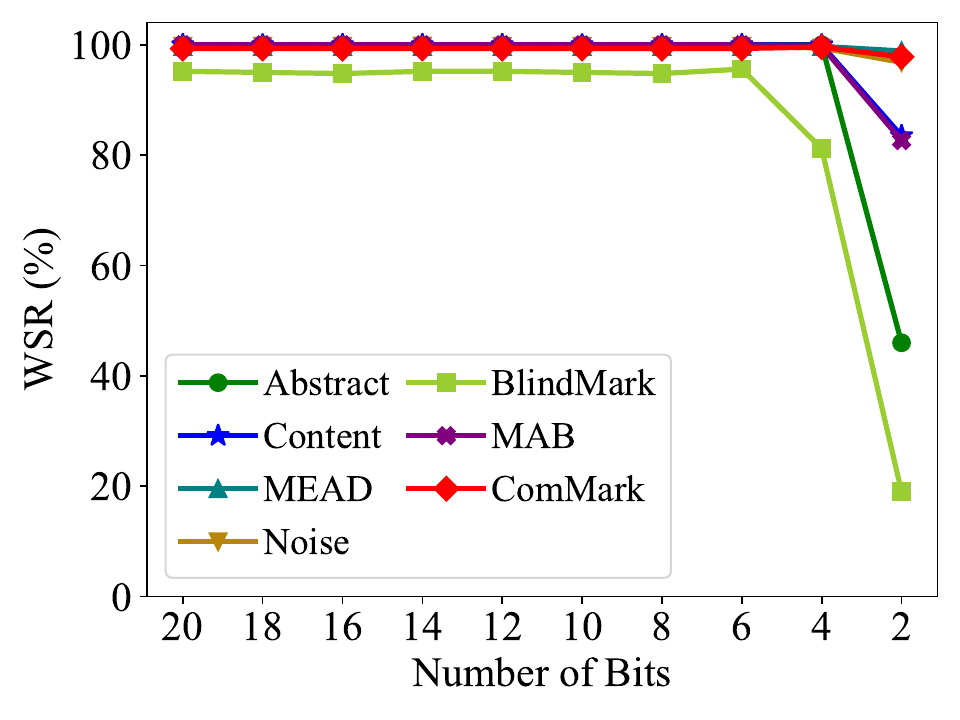}
        \caption{Image Quantization}
        \label{fig: image_quantization_gtsrb_wsr}
    \end{subfigure}
    \begin{subfigure}{0.24\textwidth}
        \centering
        \includegraphics[width=1.0\textwidth]{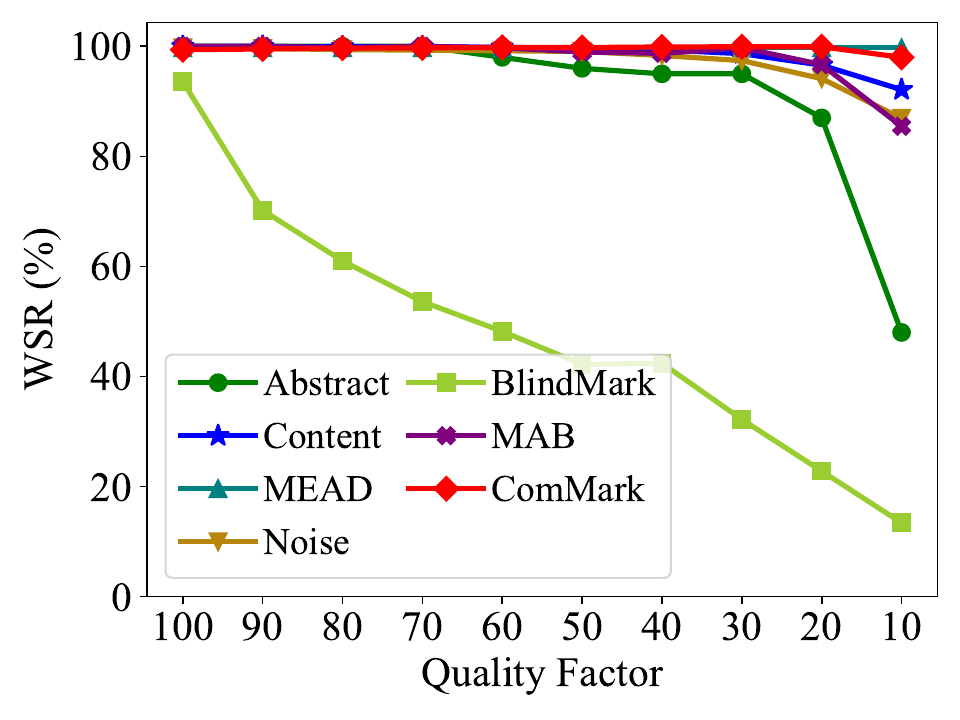}
        \caption{JPEG Compression}
        \label{fig: jpeg_compression_gtsrb_wsr}
    \end{subfigure}
    \vspace{0.3cm}

    \begin{subfigure}{0.24\textwidth}
        \centering
        \includegraphics[width=1.0\textwidth]{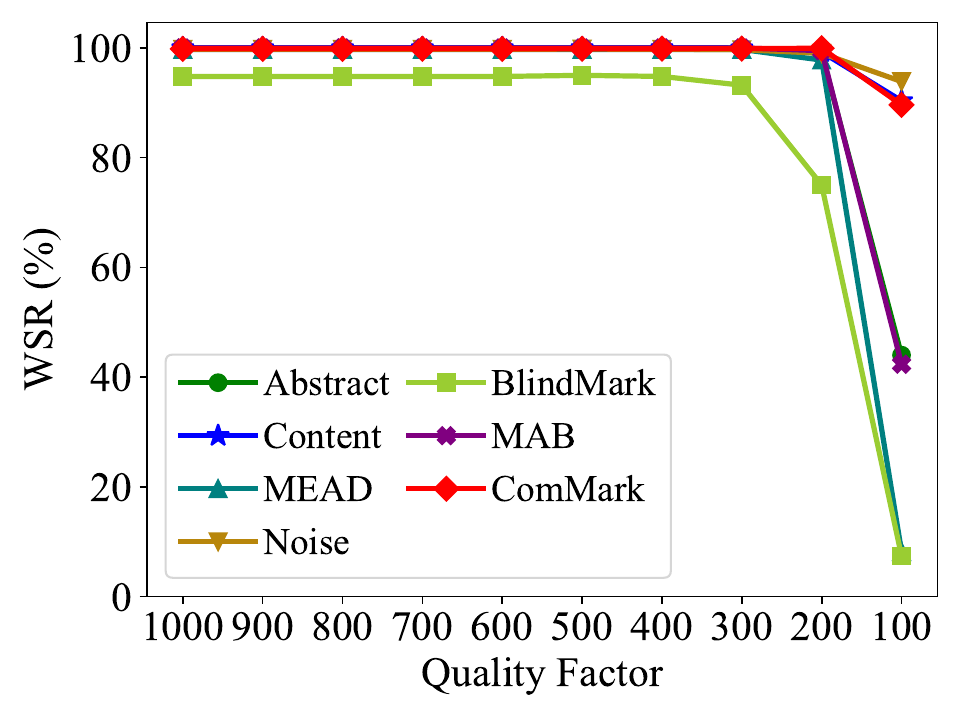}
        \caption{JPEG2000 Compression}
        \label{fig: jpeg2000_compression_gtsrb_wsr}
    \end{subfigure}
    \begin{subfigure}{0.24\textwidth}
        \centering
        \includegraphics[width=1.0\textwidth]{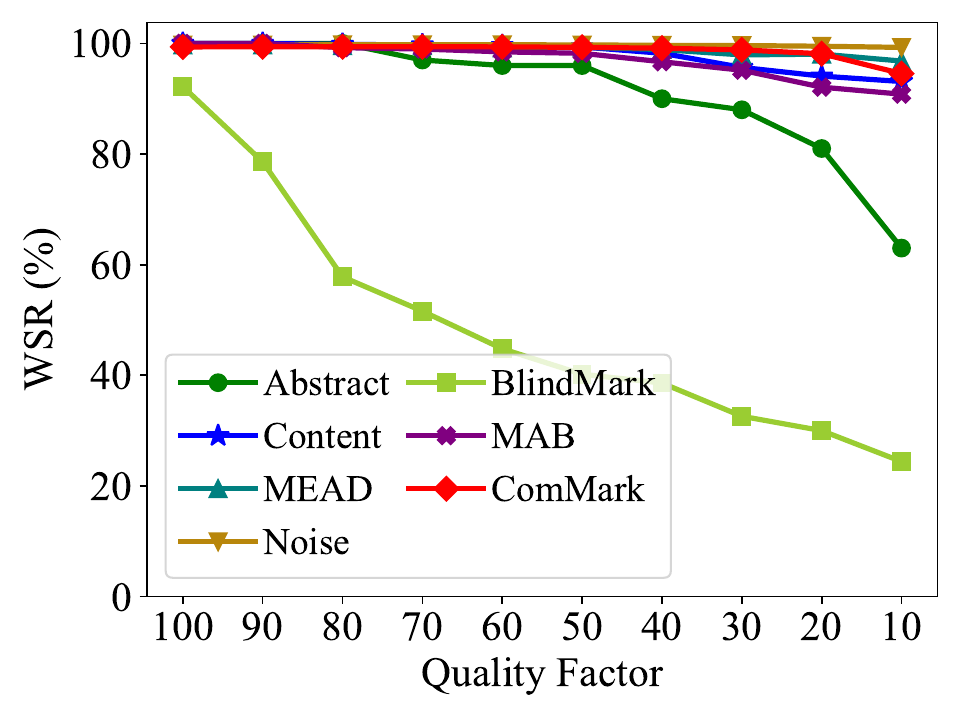}
        \caption{WEBP Compression}
        \label{fig: webp_compression_gtsrb_wsr}
    \end{subfigure}
    \begin{subfigure}{0.24\textwidth}
        \centering
        \includegraphics[width=1.0\textwidth]{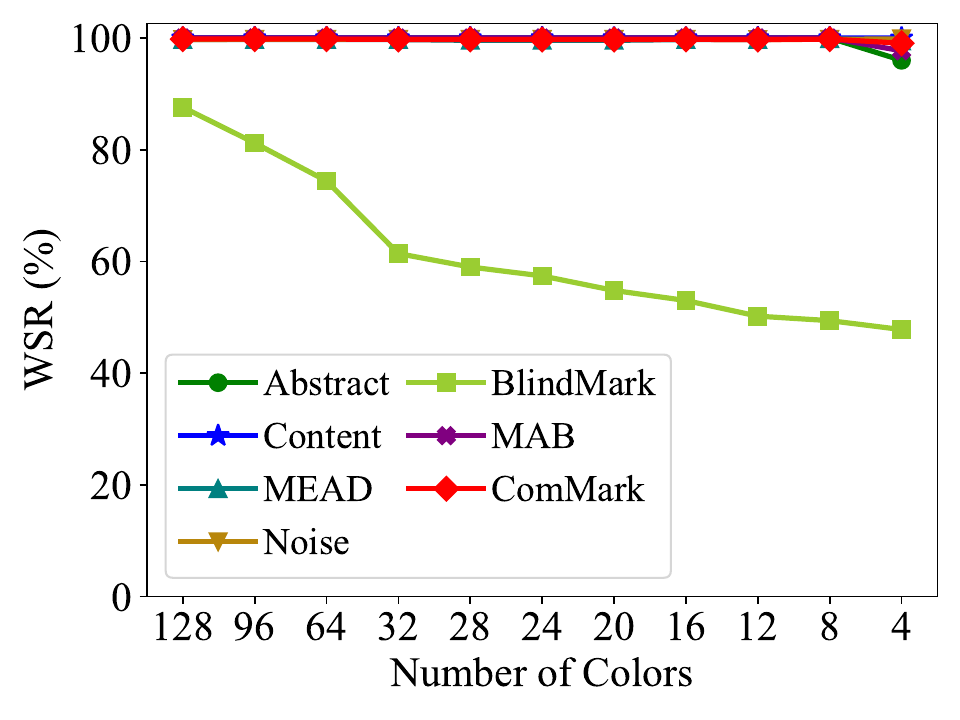}
        \caption{Color Quantization}
        \label{fig: color_quantization_gtsrb_wsr}
    \end{subfigure}
    \vspace{-0.0cm}
    \caption{Comparison of robustness against watermark evasion attacks. (On GTSRB)}
    \label{fig: robustness against evasion attacks on gtsrb}
    \vspace{-0.0cm}
\end{figure*}

\begin{figure*}[t]
    \centering
    \vspace{-0.0cm}
    \begin{subfigure}{0.24\textwidth}
        \centering
        \includegraphics[width=1.0\textwidth]{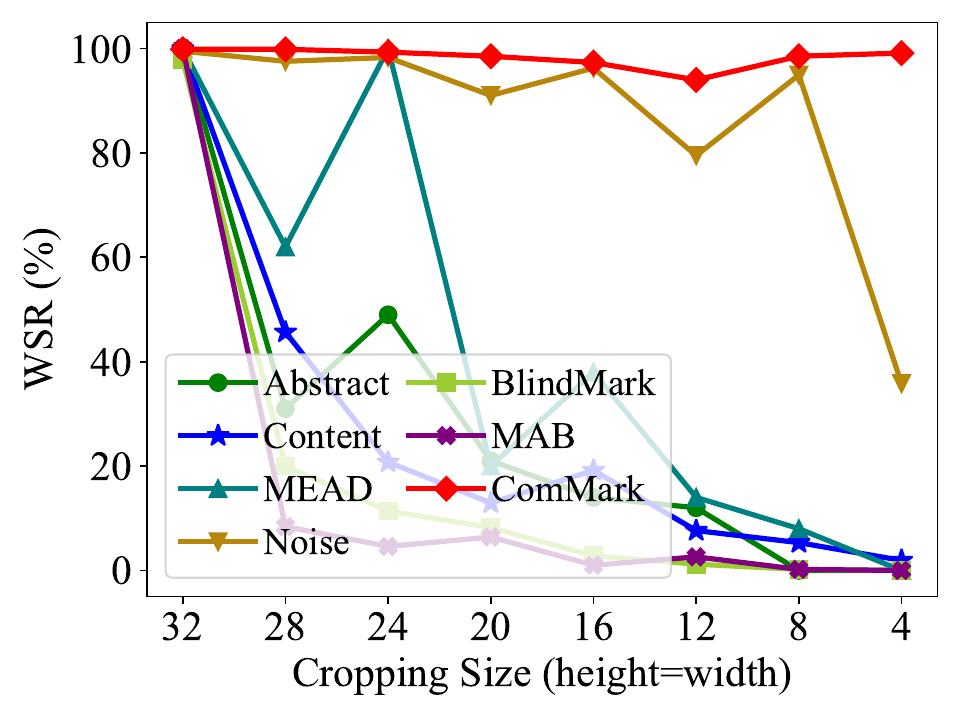}
        \caption{Cropping}
        \label{fig: crop_cifar100_wsr}
    \end{subfigure}
    \begin{subfigure}{0.24\textwidth}
        \centering
        \includegraphics[width=1.0\textwidth]{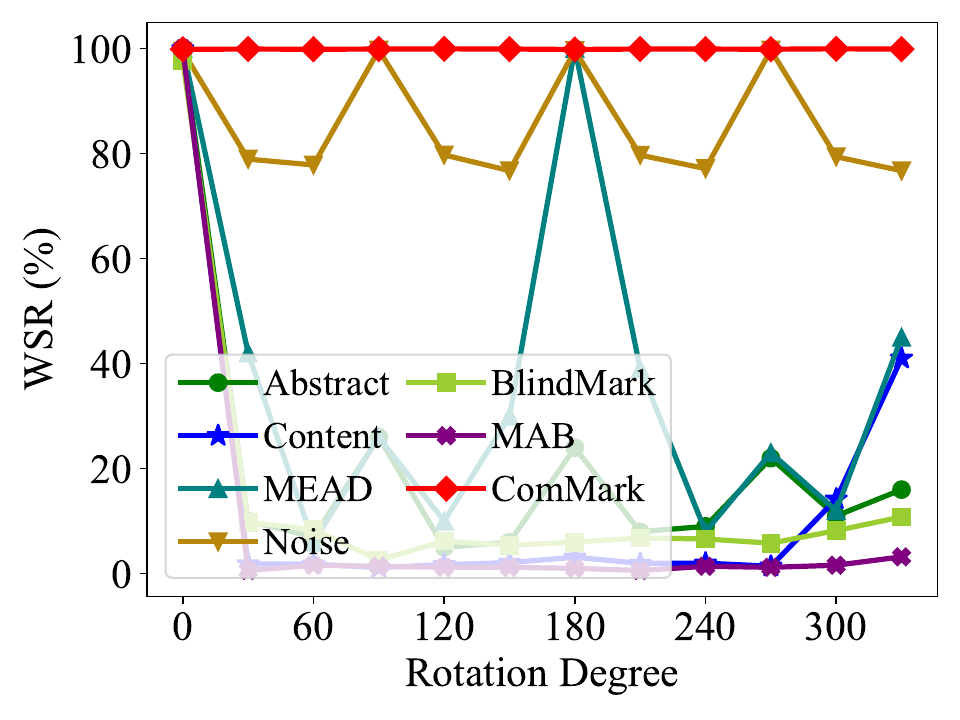}
        \caption{Rotation}
        \label{fig: rotate_cifar100_wsr}
    \end{subfigure}
    \begin{subfigure}{0.24\textwidth}
        \centering
        \includegraphics[width=1.0\textwidth]{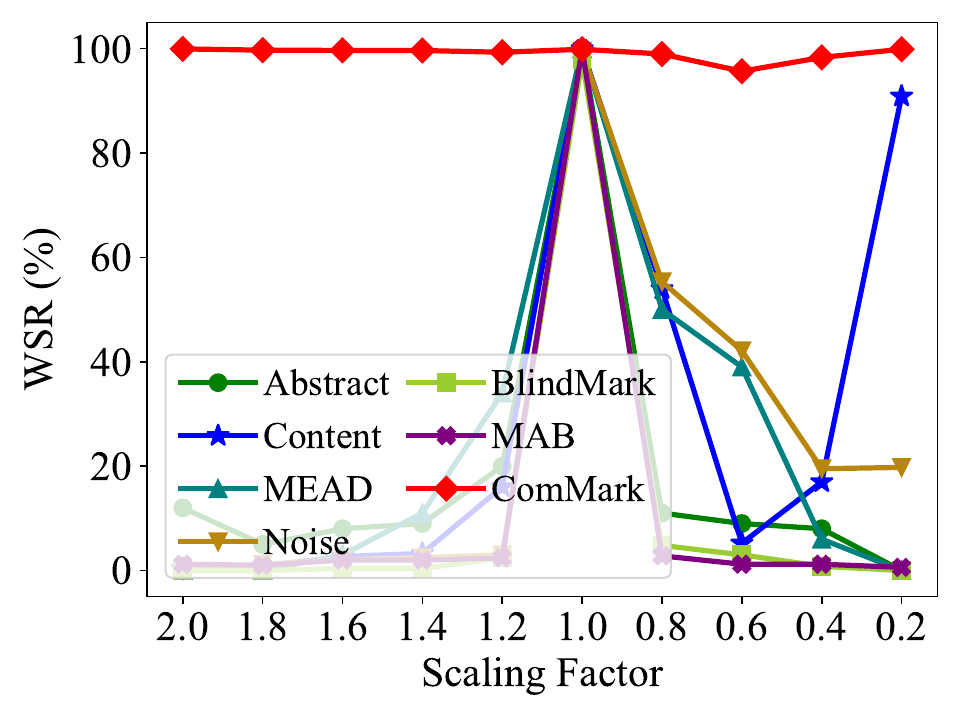}
        \caption{Scaling}
        \label{fig: scale_cifar100_wsr}
    \end{subfigure}
    \begin{subfigure}{0.24\textwidth}
        \centering
        \includegraphics[width=1.0\textwidth]{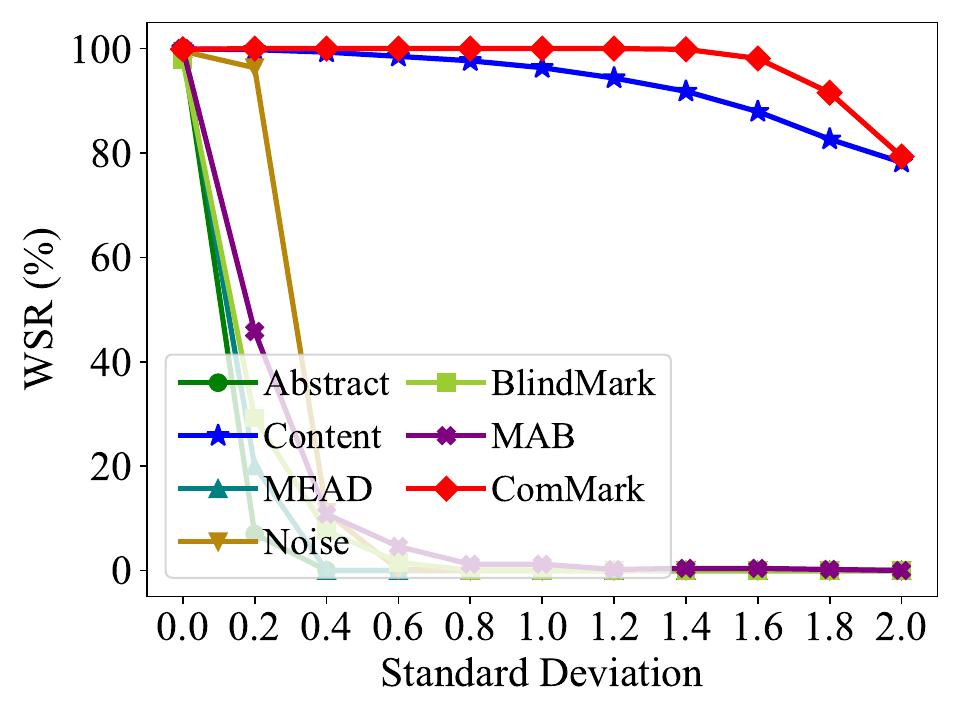}
        \caption{Gaussian Noise}
        \label{fig: gaussian_noise_cifar100_wsr}
    \end{subfigure}
    \vspace{0.3cm}

    \begin{subfigure}{0.24\textwidth}
        \centering
        \includegraphics[width=1.0\textwidth]{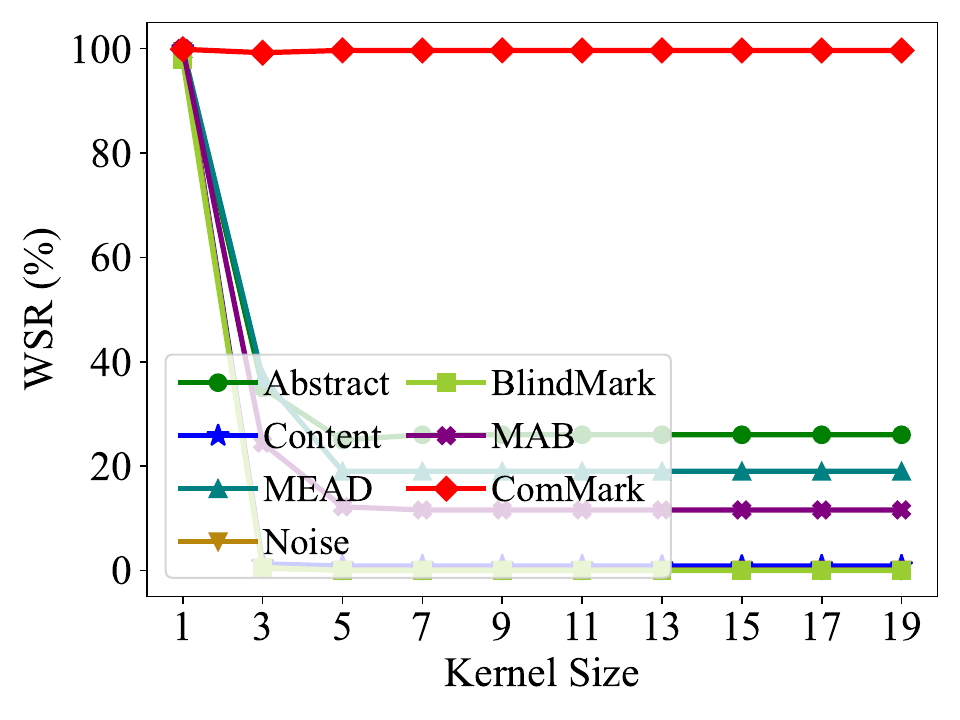}
        \caption{Gaussian Blur}
        \label{fig: gaussian_blur_cifar100_wsr}
    \end{subfigure}
    \begin{subfigure}{0.24\textwidth}
        \centering
        \includegraphics[width=1.0\textwidth]{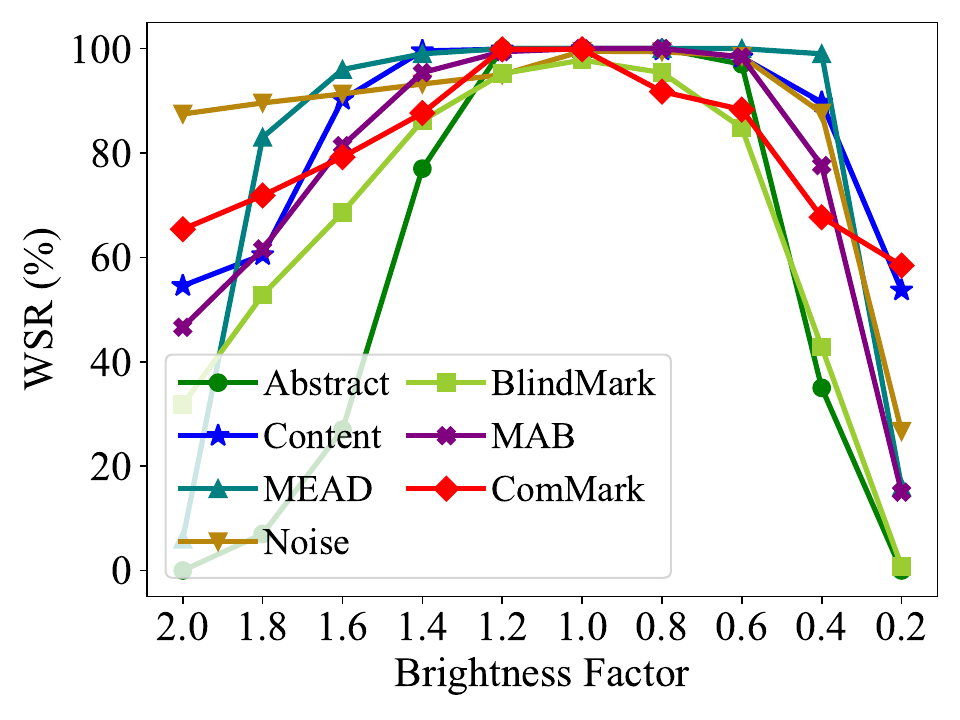}
        \caption{Brightness Change}
        \label{fig: brightness_change_cifar100_wsr}
    \end{subfigure}
    \begin{subfigure}{0.24\textwidth}
        \centering
        \includegraphics[width=1.0\textwidth]{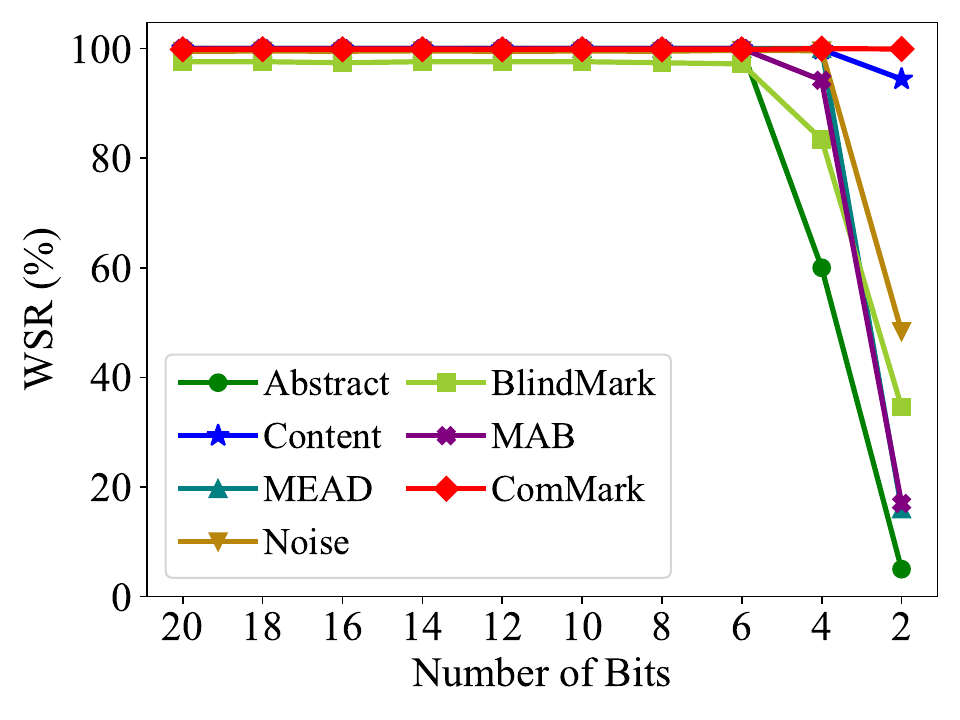}
        \caption{Image Quantization}
        \label{fig: image_quantization_cifar100_wsr}
    \end{subfigure}
    \begin{subfigure}{0.24\textwidth}
        \centering
        \includegraphics[width=1.0\textwidth]{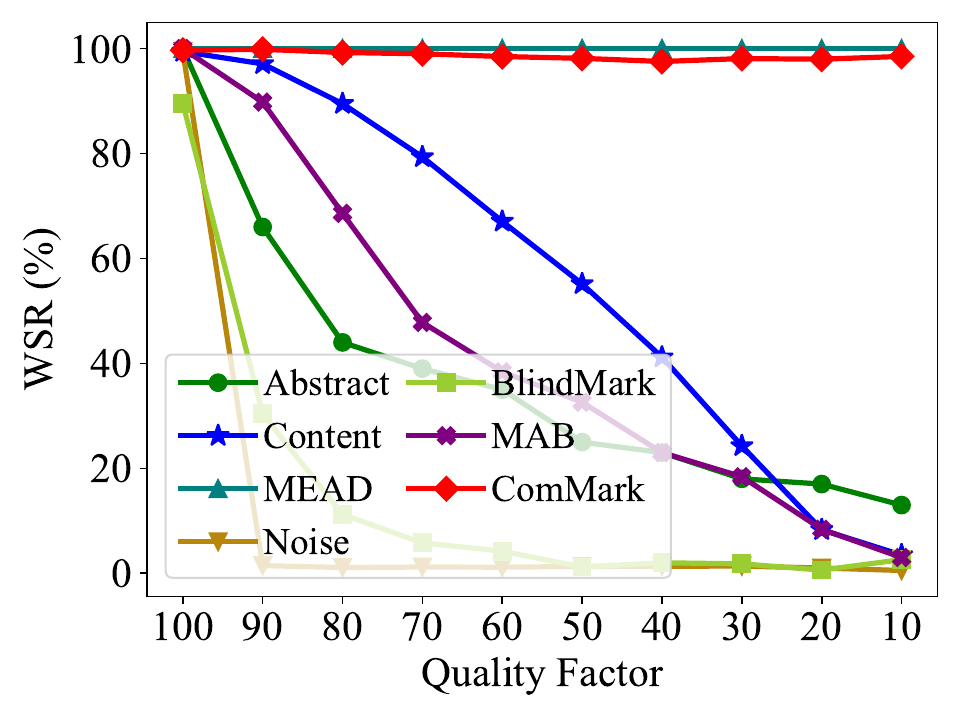}
        \caption{JPEG Compression}
        \label{fig: jpeg_compression_cifar100_wsr}
    \end{subfigure}
    \vspace{0.3cm}

    \begin{subfigure}{0.24\textwidth}
        \centering
        \includegraphics[width=1.0\textwidth]{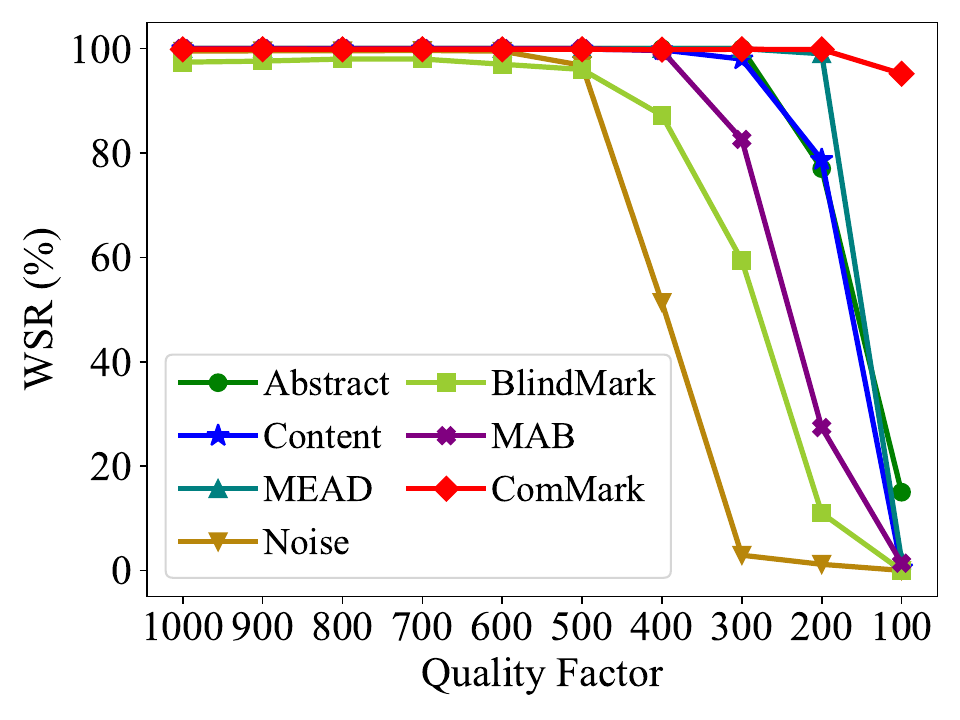}
        \caption{JPEG2000 Compression}
        \label{fig: jpeg2000_compression_cifar100_wsr}
    \end{subfigure}
    \begin{subfigure}{0.24\textwidth}
        \centering
        \includegraphics[width=1.0\textwidth]{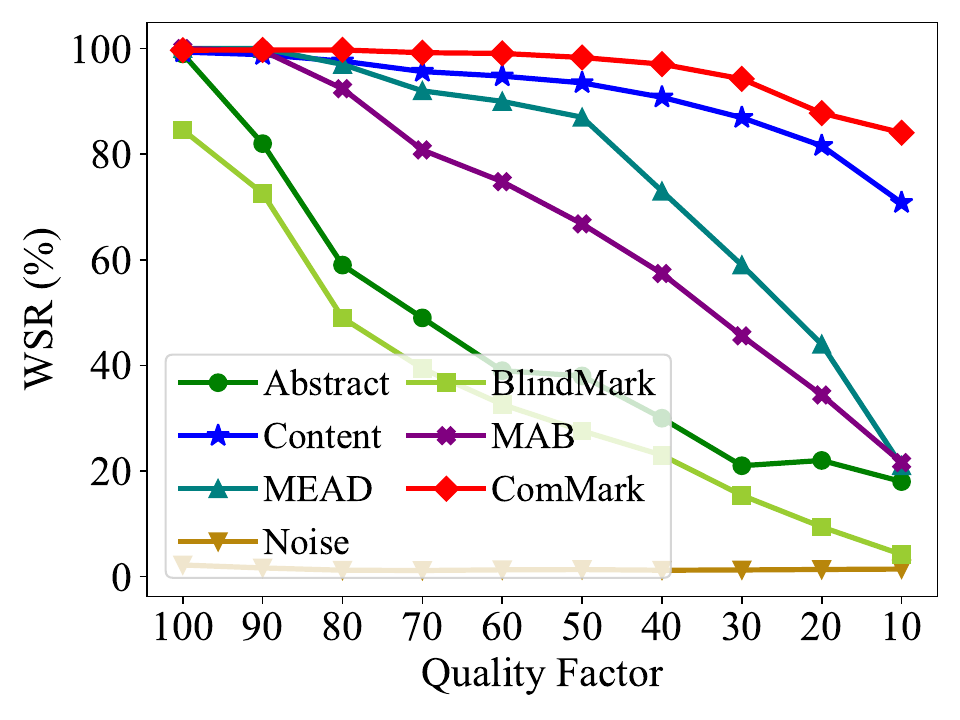}
        \caption{WEBP Compression}
        \label{fig: webp_compression_cifar100_wsr}
    \end{subfigure}
    \begin{subfigure}{0.24\textwidth}
        \centering
        \includegraphics[width=1.0\textwidth]{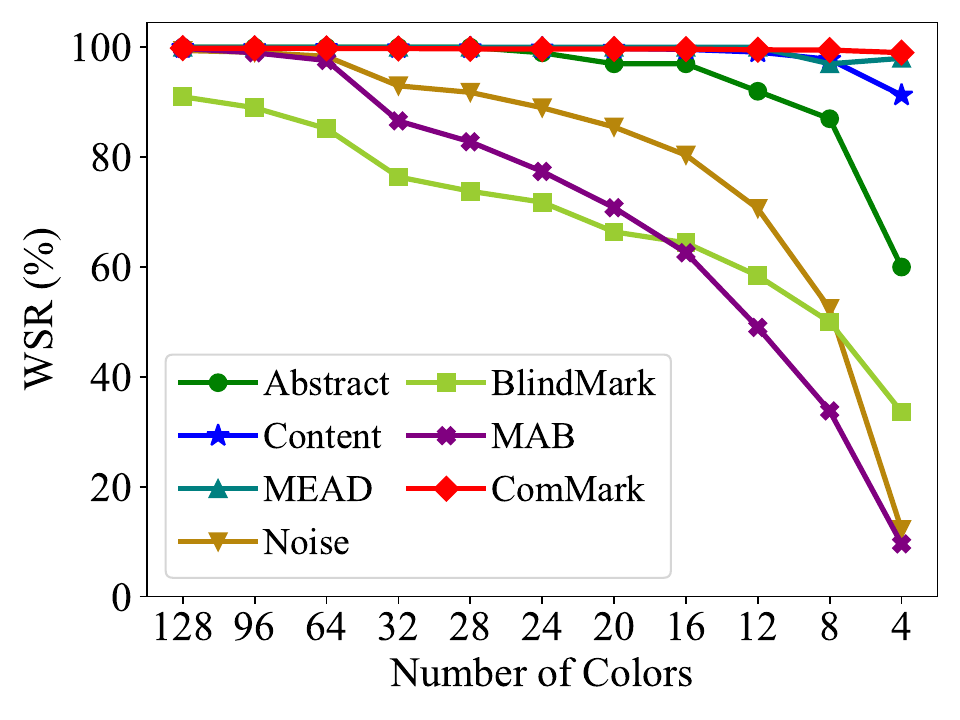}
        \caption{Color Quantization}
        \label{fig: color_quantization_cifar100_wsr}
    \end{subfigure}
    \vspace{-0.0cm}
    \caption{Comparison of robustness against watermark evasion attacks. (On CIFAR100)}
    \label{fig: robustness against evasion attacks on cifar100}
    \vspace{-0.0cm}
\end{figure*}

\begin{figure*}[t]
    \centering
    \vspace{-0.0cm}
    \begin{subfigure}{0.24\textwidth}
        \centering
        \includegraphics[width=1.0\textwidth]{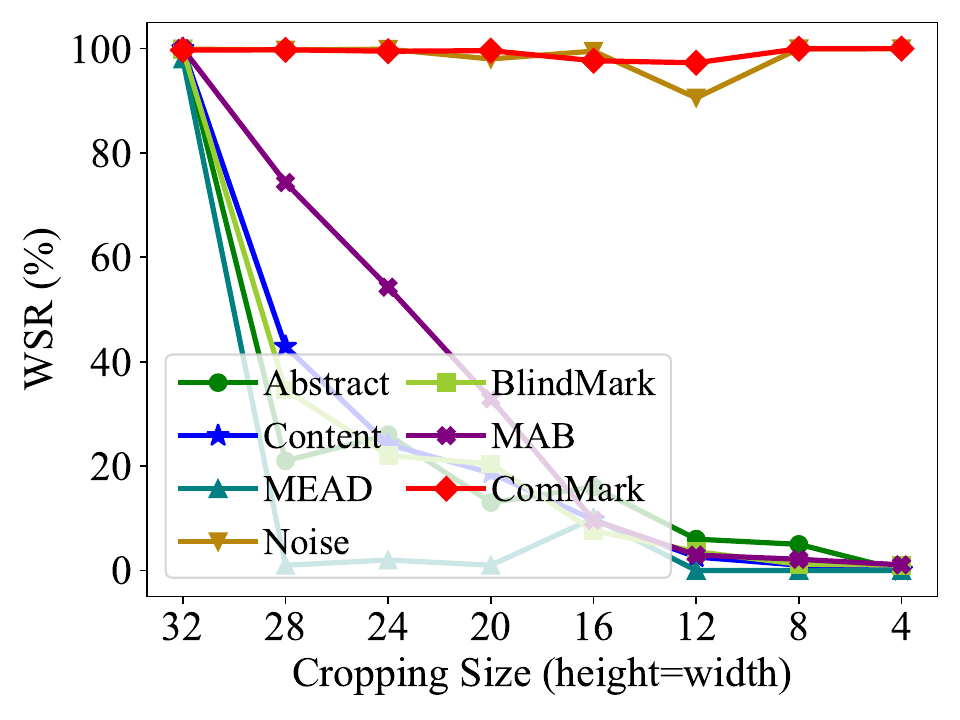}
        \caption{Cropping}
        \label{fig: crop_vggface_wsr}
    \end{subfigure}
    \begin{subfigure}{0.24\textwidth}
        \centering
        \includegraphics[width=1.0\textwidth]{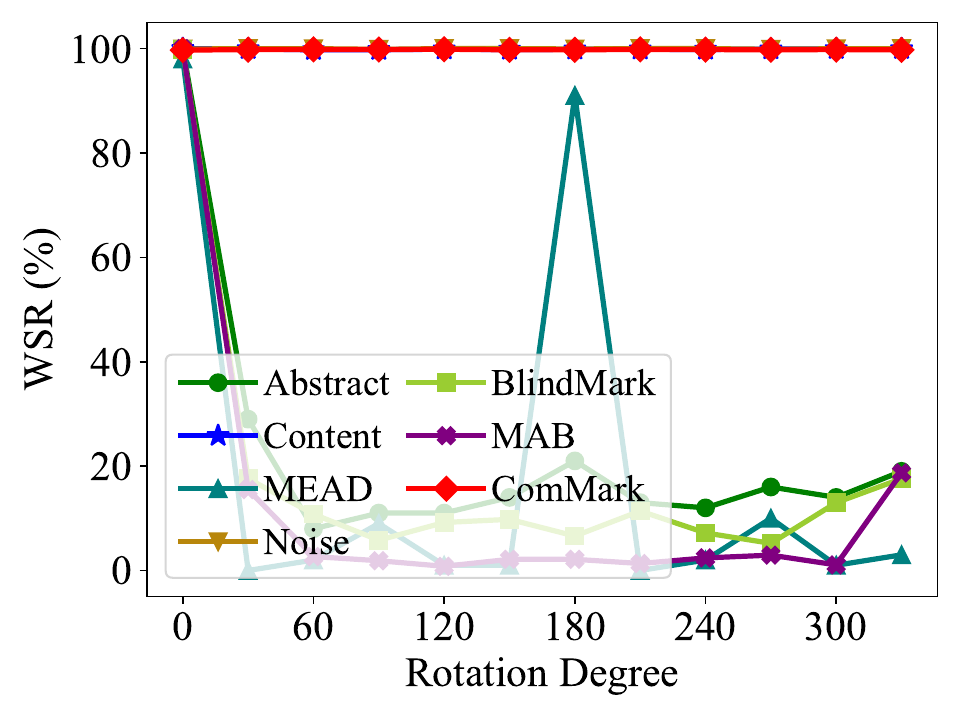}
        \caption{Rotation}
        \label{fig: rotate_vggface_wsr}
    \end{subfigure}
    \begin{subfigure}{0.24\textwidth}
        \centering
        \includegraphics[width=1.0\textwidth]{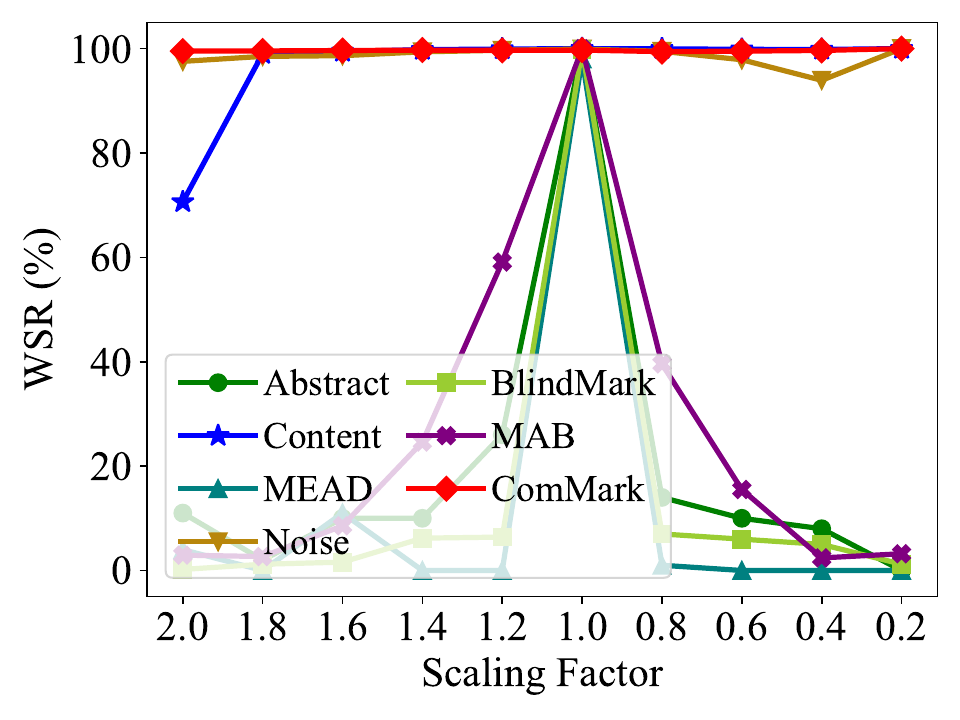}
        \caption{Scaling}
        \label{fig: scale_vggface_wsr}
    \end{subfigure}
    \begin{subfigure}{0.24\textwidth}
        \centering
        \includegraphics[width=1.0\textwidth]{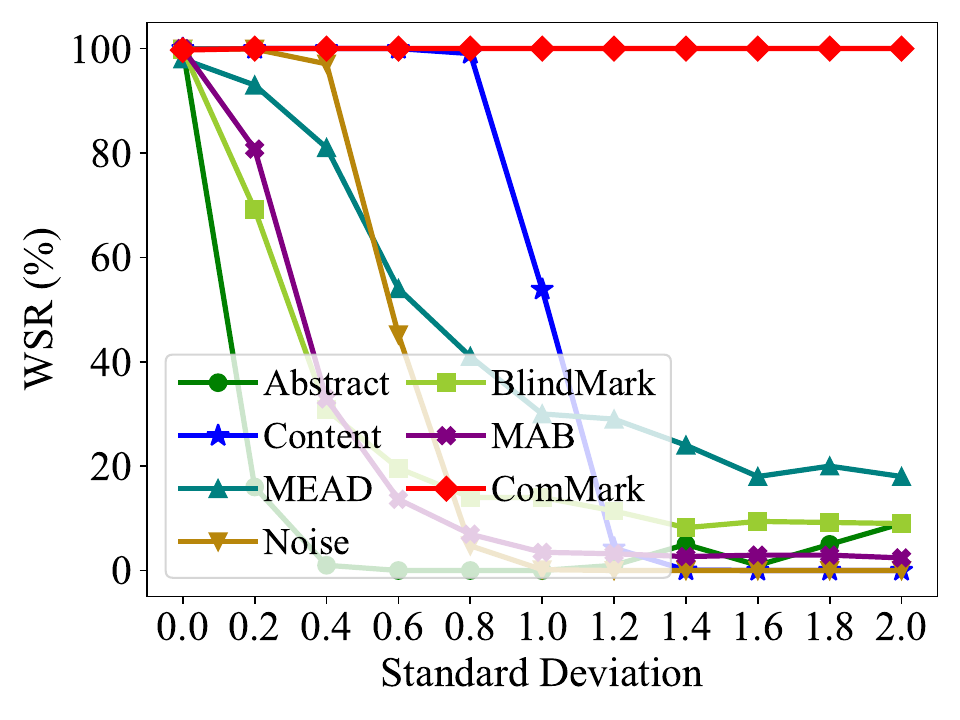}
        \caption{Gaussian Noise}
        \label{fig: gaussian_noise_vggface_wsr}
    \end{subfigure}
    \vspace{0.3cm}

    \begin{subfigure}{0.24\textwidth}
        \centering
        \includegraphics[width=1.0\textwidth]{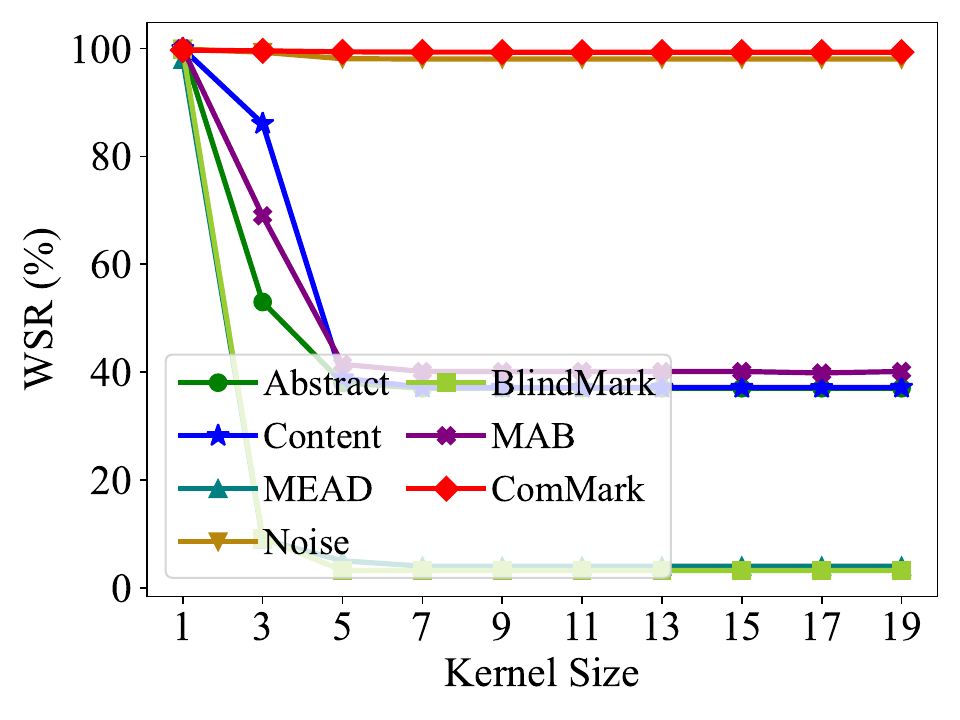}
        \caption{Gaussian Blur}
        \label{fig: gaussian_blur_vggface_wsr}
    \end{subfigure}
    \begin{subfigure}{0.24\textwidth}
        \centering
        \includegraphics[width=1.0\textwidth]{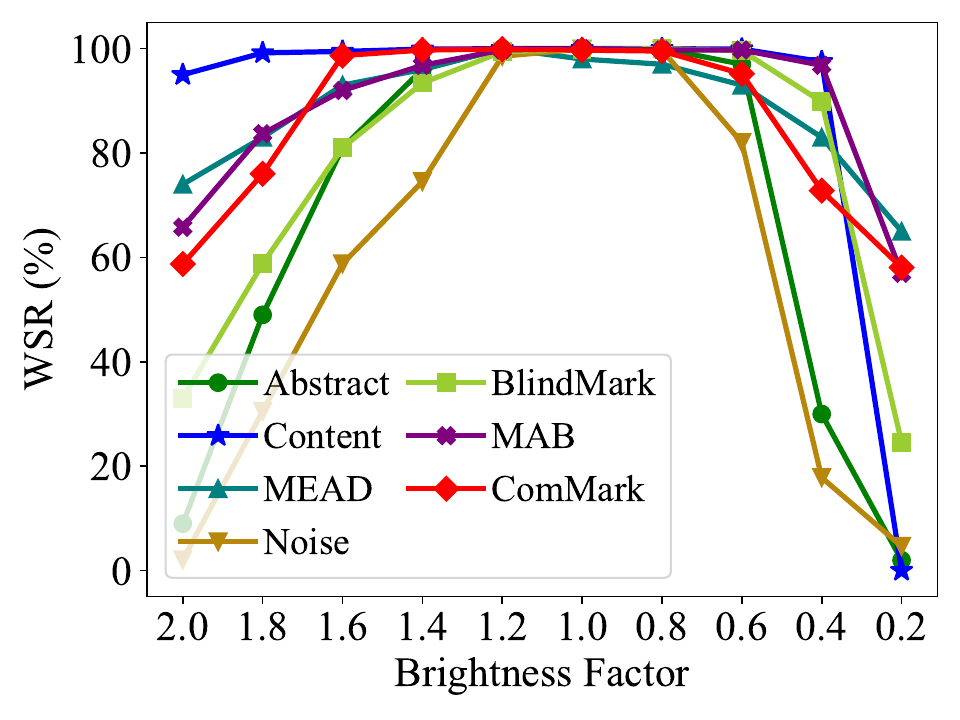}
        \caption{Brightness Change}
        \label{fig: brightness_change_vggface_wsr}
    \end{subfigure}
    \begin{subfigure}{0.24\textwidth}
        \centering
        \includegraphics[width=1.0\textwidth]{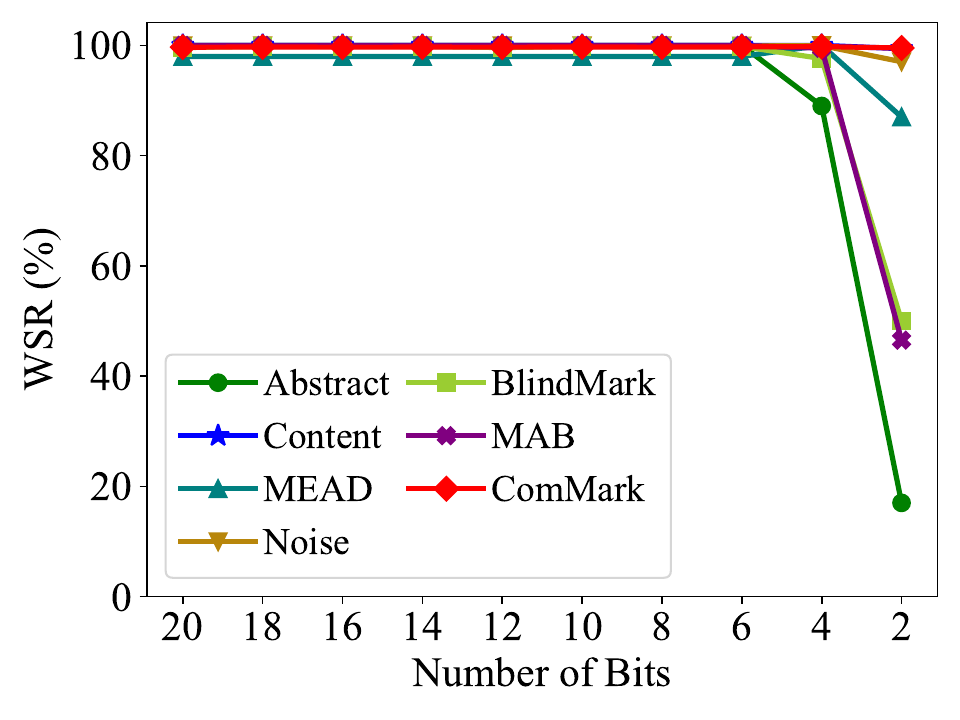}
        \caption{Image Quantization}
        \label{fig: image_quantization_vggface_wsr}
    \end{subfigure}
    \begin{subfigure}{0.24\textwidth}
        \centering
        \includegraphics[width=1.0\textwidth]{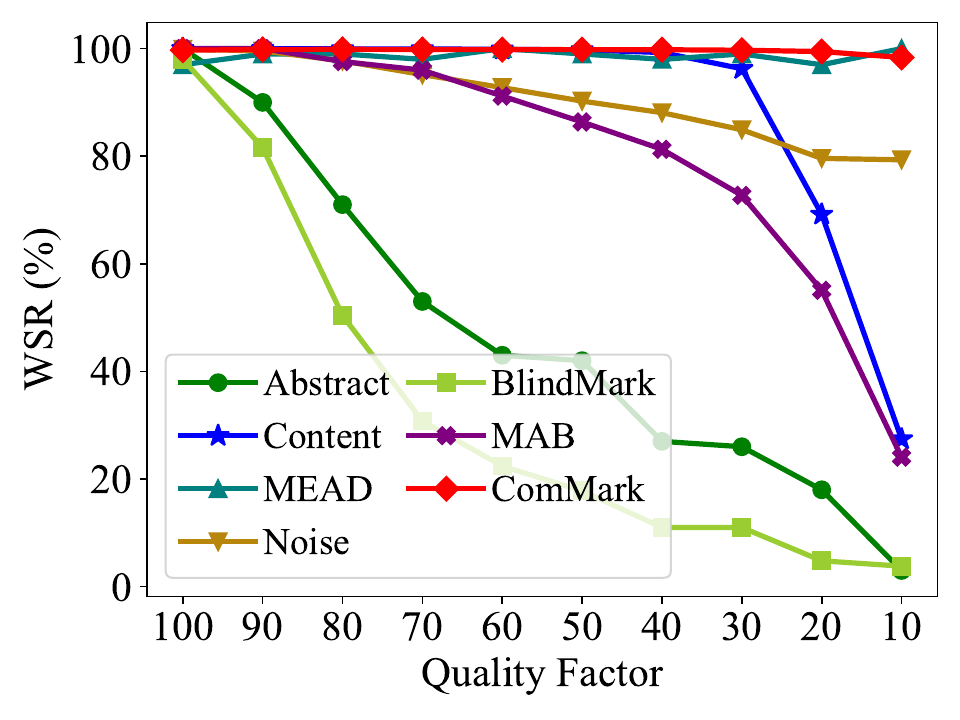}
        \caption{JPEG Compression}
        \label{fig: jpeg_compression_vggface_wsr}
    \end{subfigure}
    \vspace{0.3cm}

    \begin{subfigure}{0.24\textwidth}
        \centering
        \includegraphics[width=1.0\textwidth]{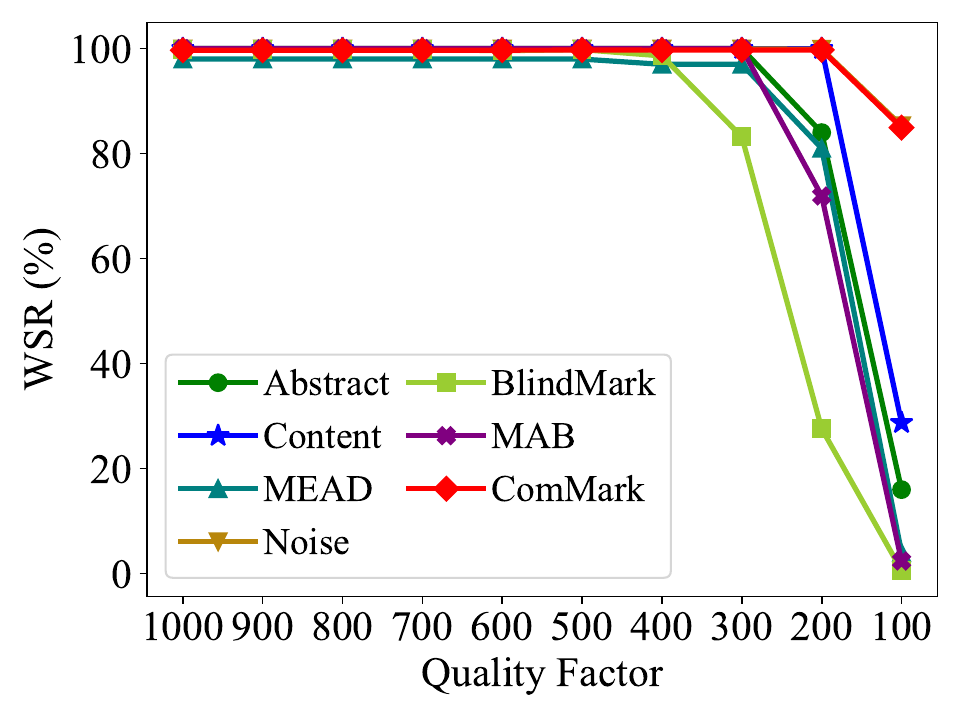}
        \caption{JPEG2000 Compression}
        \label{fig: jpeg2000_compression_vggface_wsr}
    \end{subfigure}
    \begin{subfigure}{0.24\textwidth}
        \centering
        \includegraphics[width=1.0\textwidth]{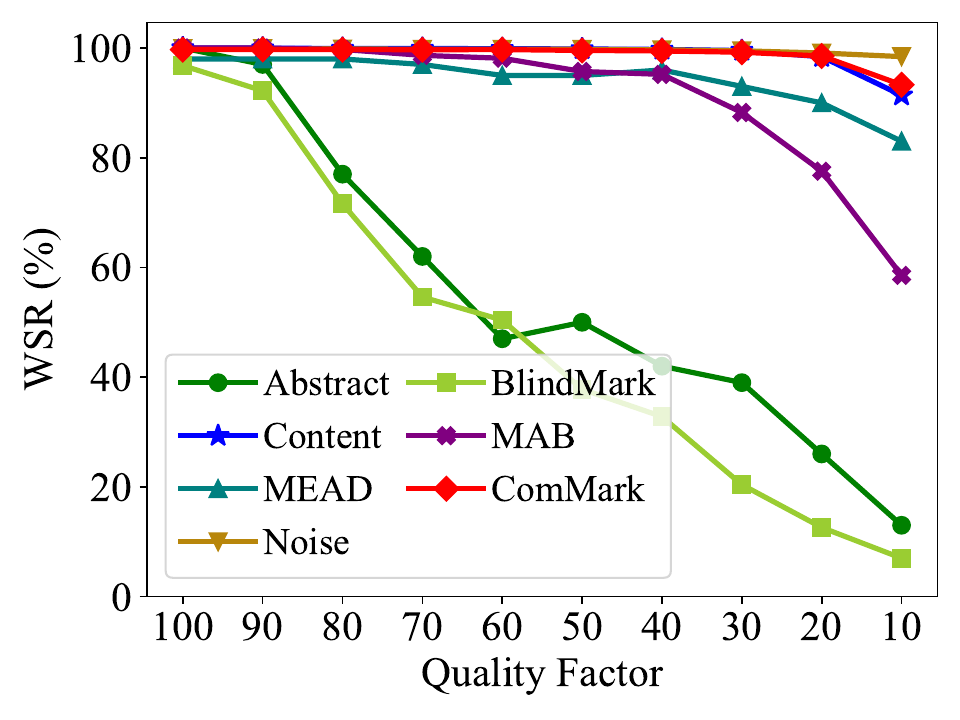}
        \caption{WEBP Compression}
        \label{fig: webp_compression_vggface_wsr}
    \end{subfigure}
    \begin{subfigure}{0.24\textwidth}
        \centering
        \includegraphics[width=1.0\textwidth]{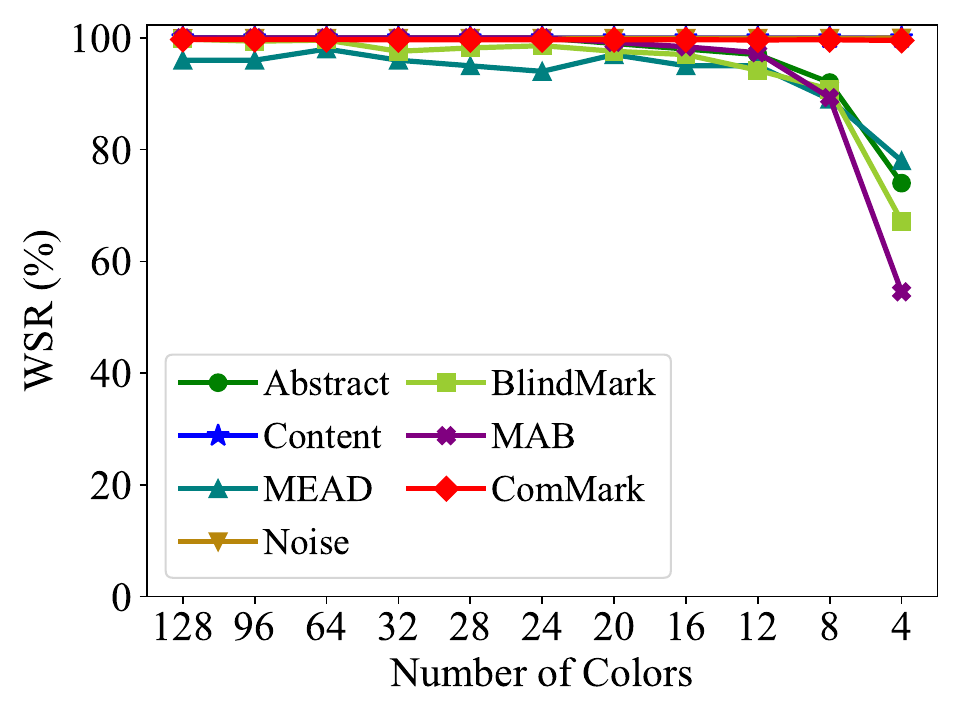}
        \caption{Color Quantization}
        \label{fig: color_quantization_vggface_wsr}
    \end{subfigure}
    \vspace{-0.0cm}
    \caption{Comparison of robustness against watermark evasion attacks. (On VGGFace)}
    \label{fig: robustness against evasion attacks on vggface}
    \vspace{-0.0cm}
\end{figure*}

\end{document}